\documentclass{aa}

\usepackage[T1]{fontenc} 
\usepackage[varg]{txfonts}
\usepackage{amsmath}
\usepackage{amssymb}
\usepackage{mathtools}
\usepackage{bm}
\usepackage{geometry}
\usepackage{enumitem}
\usepackage{graphicx}
\usepackage{multirow}
\usepackage{tikz}
\usepackage{makecell}
\usepackage{siunitx}
\DeclareSIPrefix\micro{\text{\textmu}}{-3} 
\usepackage{easyReview}
\usepackage{booktabs}
\usepackage{comment}

\usepackage{caption}
\usepackage{subcaption}

\usepackage[ruled,vlined]{algorithm2e}

\SetCommentSty{mycommfont}
\SetKwComment{tcp}{}{}

\DeclareMathAlphabet{\mathSpeCal}{OMS}{rsfs}{m}{n}

\newcommand{\norm}[1]{\big\lVert #1 \big\rVert}
\newcommand{\transp}{^\textsf{T}}
\newcommand{\W}{\mathrm{W}}
\newcommand{\C}{\mathrm{C}}
\newcommand{\CostFunc}{\mathSpeCal{C}}
\newcommand{\LogLike}{\mathSpeCal{L}}

\DeclareMathOperator*{\argmax}{arg\,max}

\newcommand{\SNR}{\mathcal{S}/\mathcal{N}\xspace}
\DeclareMathOperator{\Var}{Var}
\DeclareMathOperator{\Cov}{Cov}
\DeclareMathOperator{\E}{\mathbb{E}}
\newcommand{\PACO}{\texttt{PACO}\xspace}
\newcommand{\PACOs}{\texttt{PACO}'s\xspace}
\newcommand{\PACOME}{\texttt{PACOME}\xspace}
\newcommand{\PACOMEs}{\texttt{PACOME}'s\xspace}

\newcommand*{\V}[1]{\boldsymbol{#1}}   
\newcommand*{\M}[1]{\mathbf{#1}}       

\bibpunct{(}{)}{;}{a}{}{,}

\title{PACOME: Optimal multi-epoch combination of direct imaging observations for joint exoplanet detection and orbit estimation}
\titlerunning{The PACOME algorithm}

\authorrunning{Dallant et al.}
\author{Jules Dallant\inst{\ref{CRAL}}
             \and Maud Langlois\inst{\ref{CRAL}}
             \and Olivier Flasseur\inst{\ref{CRAL}}
             \and Éric Thiébaut\inst{\ref{CRAL}}
             }
        
\institute{Université Lyon 1, ENS de Lyon, CNRS, Centre de Recherche Astrophysique de Lyon UMR 5574, 69230 Saint-Genis-Laval, France \\\email{jules.dallant@univ-lyon1.fr}\label{CRAL}}

\date{Received March 17, 2023 / Accepted August 29, 2023}

\abstract
{
Exoplanet detections and characterizations via direct imaging require high contrast and high angular resolution. These requirements are typically pursued by combining (i) cutting-edge instrumental facilities equipped with extreme adaptive optics and coronagraphic systems, (ii) optimized differential imaging to introduce a diversity between the signals of the sought-for objects and that of the star, and (iii) dedicated (post-)processing algorithms to further eliminate  the residual stellar leakages.
}
{
With respect to the third technique, substantial efforts have been undertaken over this last decade on the design of more efficient post-processing algorithms. The whole data collection and retrieval processes currently allow to detect massive exoplanets at angular separations greater than a few tenths of au. The performance remains upper-bounded at shorter angular separations due to the the lack of diversity induced by the processing of each epoch of observations individually. We aim to propose a new algorithm that is able to combine several observations of the same star by accounting for the Keplerian orbital motion across epochs for the sought-for exoplanets in order to constructively co-add  their weak signals.
}
{
The proposed algorithm, \PACOME, integrates an exploration of the plausible orbits of the sought-for objects within an end-to-end statistical detection and estimation formalism. The latter is extended to a multi-epoch combination of the maximum likelihood framework of \PACO, which is a post-processing algorithm of single-epoch observations. From this, we derived a reliable multi-epoch detection criterion, interpretable both in terms of probability of detection and of false alarm. In addition, \PACOME is able to produce a few plausible estimates of the  orbital elements of the detected sources and provide their local error bars.}
{
We tested the proposed algorithm on several datasets obtained from the VLT/SPHERE instrument with IRDIS and IFS using the pupil tracking mode of the telescope. By resorting to injections of synthetic exoplanets, we show that \PACOME is able to detect sources remaining undetectable by the most advanced post-processing of each individual epoch. The gain in detection sensitivity scales as high as the square root of the number of epochs. We also applied \PACOME on a set of observations from the HR 8799 star hosting four known exoplanets, which can be detected by our algorithm, with very high signal-to-noise ratios.} 
{
\PACOME is an algorithm for combining multi-epoch high-contrast observations of a given star. Its sensitivity and the reliability of its astrophysical outputs permits the detection of new candidate companions at a statistically grounded confidence level. In addition, its implementation is efficient, fast, and fully automatized.
}


\begin{document} 

\maketitle

\section{Introduction}
\label{sec:introduction}

Direct imaging is an observational method that is particularly adapted to the detection and characterization of young giant exoplanets orbiting nearby stars \citep{traub2010direct}. It requires us to reach a high contrast and a high angular resolution through a combination of (i) cutting-edge observational facilities, (ii) custom observational techniques, and (iii) advanced (post)-processing algorithms. Concerning instrumental aspects, the main ground-based observatories are now equipped with dedicated instruments (e.g., GEMINI/GPI, \cite{macintosh2008gemini}; KECK/NIRC2, \cite{xuan2018characterizing}; MAGELLAN/MagAO-X, \cite{close2018optical}; VLT/SPHERE, \cite{beuzit2019sphere}; SUBARU/SCExAO, \cite{jovanovic2015subaru}) that integrate an extreme adaptive system and a coronagraphic mask to cancel out most of the stellar light. In spite of these cutting-edge facilities, the observations stay dominated by a strong and spatially correlated stellar nuisance, which is an obstacle to the detection of objects of interest. This nuisance component is formed by the additive contribution of the so-called speckles and of other sources of noise (i.e., thermal background, detector readout, and photon noise). Speckles are stellar leakages impacted by diffraction effects in the presence of residual (uncorrected) aberrations and currently remain the main source of limitation to the achievable contrast \citep{Soummer2007, Bailey2016}. In that context, observations are usually conducted with custom techniques in order to bring additional sources of diversity (e.g., temporal, spectral, polarimetric) to unmix the contribution of the sought-for objects from that of the nuisance. In this paper, we focus on angular differential imaging (ADI, \cite{marois2006angular}) in combination with spectral differential imaging (SDI, \cite{racine1999speckle}). The ADI process brings on a temporal diversity by using the pupil-tracking mode of the telescope, so that speckles remain quasi-static across exposures, while the objects of interest follow a circular rotation with respect to the star. This apparent motion is deterministic and depends solely on the experienced parallactic rotation angles. Then, SDI guarantees spectral diversity by simultaneously recording  images in several spectral channels using an integral field spectrograph (IFS) or a dual band imager (IRDIS) with less spectral leverage. A large variety of post-processing algorithms has been developed in the last decade to extract the relevant information from 3D (respectively, 4D) datasets recorded by ADI (respectively, ASDI), see e.g., \cite{pueyo2018direct, Cantalloube2020} for reviews. Among these methods, the \PACO algorithm \citep{Flasseur2018,Flasseur2020,Flasseur2020robust} has been shown to be particularly efficient for the processing of A(S)DI observations. It statistically captures  the spatio-temporo-spectral correlations of the data with a weighted multi-variate Gaussian model whose parameters are estimated, in a data-driven fashion, at the scale of a patch of a few tens of pixels. With VLT/SPHERE observations where the typical spatial correlation scale of speckles lies in a few tens of pixels, \cite{Flasseur2018, Flasseur2020, Flasseur2020robust, Cantalloube2020} and  \cite{Chomez2023} showed that this parameter-free method provides an improved detection sensitivity with respect to the baseline processing methods of the field (i.e., cADI \cite{marois2006angular}, TLOCI \cite{marois2014gpi} and KLIP \cite{soummer2012detection}, as well as \cite{amara2012pynpoint}). It also provides reliable estimates of the detection confidence and of the astro-photometry with the associated uncertainties. 

Whatever the chosen post-processing algorithm, detection performance stay upper-bounded by the lack of diversity at short angular separation, where (i) the speckle field is dominant and displays the largest temporal fluctuations and (ii) the apparent displacement of the objects of interest induced by A(S)DI is not sufficient to extract their signals without bias. From a data science point of view, two main avenues are currently investigated to mitigate these limitations. The first category of methods targets a better elimination of the nuisance component. It usually consists of building a finer model of the stellar (on-axis) point spread function (PSF). This model can be built from several datasets, resulting from the observation of different stars, in which the objects of interest are not expected to be co-localized. This is the general principle of approaches based on reference differential imaging  \citep{ruane2019reference, wahhaj2021search, sanghi2022efficiently, xie2022reference}. The second category of methods targets a better combination of the signal of the objects of interest. It consists of combining several epochs\footnote{In the following, we use the term "epochs" to refer to multiple datasets resulting from the observations of the same star at different times.} to constructively co-add  their signal by taking into account their proper Keplerian orbital motion. In the following, we focus on this second category of approaches, as it is more relevant to the algorithm presented in this paper.

The idea of exploiting several observations of the same star is a quite standard approach (e.g., in planetology for the detection of faint asteroid’s satellites \citep{Marchis2005, Berdeu2022} and it is not novel in the field exoplanet characterization by direct imaging. As an illustration, the evaluation (e.g., based on system stability criteria) of the exoplanet orbits are routinely performed by fitting their previously extracted astrometry at each individual epoch with dedicated algorithms \citep{blunt2020orbitize}. In the same vein, \cite{skemer2011sirius} proposed  "de-orbitizing" detection maps from different epochs, namely, to transform plus re-scale each map by compensating for the orbital motion of known companions based on ephemeris calculus. This pragmatic approach, based on a prior information about the source orbits, allows to re-detect them with an improved signal-to-noise ratio (S/N), which is useful for their characterization but remains blind to unknown candidates. From a theoretical point of view, \cite{males2013direct} studied the effect of the orbital motion on the detection sensitivity. They showed that orbital motion is difficult to exploit to reach $10^{-6}$ to $10^{-7}$ contrasts at small separations (typically, 0.1'' to 0.5'')  currently needed in  the search for Jupiter-like exoplanets by a multi-epoch combination of the data from the existing ten meters class telescopes. Proper orbital modeling will be even more crucial in the context of the quest of Neptune-like and Earth-like exoplanets with the thirty meters class telescopes (e.g., ELT, TMT, GMT). Indeed, the deep exploration of the inner environment (typically located at a few au) of the nearby solar-type stars will require contrasts of up to $10^{-8}$ to $10^{-9}$, implying long exposure times of possibly several tens of hours, which can only be achieved by conducting several observations split over several days, weeks, or months. For such observations, the orbital motion of exoplanets is not negligible at timescales of a few days or weeks. Combining the resulting multi-epoch observations without compensating for the Keplerian orbital motion will lead to a drastic degradation of the detection confidence, thus strongly limiting the achievable contrast, even for epochs separated from few weeks to few days. In that context, \cite{males2015orbital} suggested that orbital differential imaging (ODI), namely, the combination of multi-epoch observations with a proper compensation of the orbital motion of point-like sources, can bring an additional diversity, complementary to A(S)DI, in order to unmix the signal of very faint objects from the nuisance component. However, \cite{males2015orbital} also emphasized that ODI requires a dedicated (statistical) framework since the application of detection metrics commonly used in the direct imaging community for single-epoch analysis leads to a false alarm rate that is significantly higher than expected -- and even more so when multiplying the number of (possibly quite similar) tested orbits. This precludes the application of their approach to blind searches, namely, those conducted without prior information about the orbit and/or about tight distributions of the orbital elements (e.g., obtained with other observational techniques), similarly to the Proxima Centauri b search from \cite{2020Gratton}. The K-Stacker algorithm \citep{coroller2015k, nowak2018k, coroller2020k, coroller2022k}, also used in this latter work, is the first method addressing multi-epoch staking for blind search of exoplanets in direct imaging at high contrast. It combines (i) a brute-force step testing a large amount of orbits pre-defined on a grid, with (ii) a local refinement of the best orbits from the first step by gradient-descent optimization. K-Stacker is able to detect point-like sources that remain undetectable in each individual epoch, without any knowledge or strong priors on the source's orbits. The algorithm also delivers, as a byproduct, an estimate of the orbital elements of the detected sources. Very recently, \cite{thompson2022deep} proposed an alternative to K-Stacker integrating: (i) a dynamic orbit sampler based on Markov chain Monte-Carlo (MCMC) and (ii) a joint probabilistic model of the orbit and of a common photometry consistent across epochs. Integrating this model in a Bayesian framework and marginalizing over all the orbital elements allows to derive metrics to evaluate the significance of a detection. Besides, an estimate of the orbit uncertainties can be derived from the underlying posterior distributions. The idea of combining multi-epoch direct imaging data with other methods was also studied. For $\epsilon$ Eridani b, \cite{Mawet2019} made use of direct imaging data and radial velocity (RV) measurements to constrain the orbital elements of the planet historically confirmed via astrometry. This idea was later carried and improved by \cite{LlopSayson2021}, who added astrometric measurements to the two other methods. They showed that even if not direct imaging detection were made alone, combining it with astrometry and RVs consequently helped constraining the orbit of $\epsilon$ Eridani b.

For all these multi-epoch combination algorithms, the detection sensitivity and the astrophysical interpretability of the estimates are limited. A major drawback is related to the use (as input) of individual residual images (i.e., those constructed from the subtraction of an estimation of the on-axis PSF) produced by standard single-epoch algorithms (e.g., based on a principal component analysis, \citep{soummer2012detection, amara2012pynpoint}) that are known to reach moderate detection sensitivities and that are prone in some cases to a large number of false alarms especially near the star (e.g., see \cite{Flasseur2018, Flasseur2020, Cantalloube2020, Chomez2023}). These issues are even more problematic since the inner star environment corresponds to the area where the room for improvement with respect to a single-epoch analysis is the most substantial. These side effects are the direct consequence of the absence of explicit modeling of the non-stationarity and of the multiple correlations (spatial, temporal, or spectral) of A(S)DI observations. In addition, most existing multi-epoch combination algorithms lack a dedicated statistical framework to properly propagate the single-epoch uncertainties. As a result, combined multi-epoch metrics (i.e., detection confidence, achievable contrast, orbital elements, etc.) cannot be fully interpreted as strict measures. Another important source of limitation is related to the photometric calibration of the datasets. Most of the algorithms assume that the companion's relative photometry (i.e., contrast) is constant across epochs, while it is well known that this is a critical issue in direct imaging \citep{biller2021high}. The sources of relative photometric variability are multiple: evolution of the companion's angular separation and its associated biases, instrumental calibration issues, variability of the atmospheric conditions, and (only marginally) the intrinsic variability of the companion brightness. This strong assumption leads to some loss of sensitivity and can even lead to an increase of the false alarm rate in the multi-epoch combination. Most advanced algorithms cope with these different issues only partially by resorting to empirical, and possibly time consuming, correction steps: for instance, by filtering, massive injections of synthetic exoplanets, annular corrections of the estimated detection confidence by the variance of the flux maps, and via small-sample statistics \citep{mawet2014fundamental}. Once again, these additional processing steps ignore the complex non-stationarity and the multiple correlations of the data. In that context, the analysis of multi-epoch results lacks of automaticity and often requires the close inspection of the combined detection maps by an expert, whose judgement remains subject to interpretation. The numerous candidate companions (e.g., around $\beta$ Pictoris, \cite{coroller2020k}; HD 95086, \cite{coroller2020k, desgrange2022depth}; $\alpha$ Centauri A, \cite{coroller2022k}; HR 8799 \cite{thompson2022deep}) targeted by multi-epoch algorithms  
and that currently remain unconfirmed is a crude illustration of this lack of statistically grounded measures in the field. 

Based on the analysis of the current limitations of existing multi-epoch combination algorithms, we list some requirements for a new approach, namely: (i) achieving an improved detection sensitivity; (ii) deriving interpretable combined metrics (i.e., detection scores interpretable in terms of probability of detection and of probability of false alarm, as well as reliable estimates of the orbits and of the associated error bars) by a end-to-end propagation of the uncertainties; (iii) remaining robust to the highly variable quality of each individual epoch and to the lack of absolute calibration of the relative photometric variability across epochs; and (iv) integrating some prior domain knowledge in the form either of additional parameters to be optimized (e.g., common stellar mass) or of additional constraints (e.g., known system resonances, when available). The method we propose in this paper\footnote{A preliminary version of this work was presented in the form of a conference contribution in \cite{dallant2022optimal}.}, dubbed \PACOME (for \PACO Multi-Epoch)\footnote{The actual code of the algorithm is available here:~\url{https://github.com/JulesDallant/PACOME}.}, specifically addresses these points by capitalizing on the statistical framework of the \PACO algorithm dedicated to the post-processing of single-epoch datasets. Rather than directly using the outputs of \PACO as inputs of \PACOME, we extend the statistical formalism of \PACO for optimal multi-epoch combination of A(S)DI observations by accounting for the Keplerian orbital motion of the sought-for objects. This new framework maximizes the likelihood of the observations given its underlying model. 

This paper is organized as follows. Section \ref{sec:me_comb_formalism} formalizes the statistical framework we derived for multi-epoch combination. Section \ref{sec:pacome_algorithm} presents the practical algorithm we propose by combining the multi-epoch statistical framework with an exploration of the possible orbits of the sought-for objects. Section \ref{sec:results} assesses the performance of \PACOME on A(S)DI observations from the VLT/SPHERE instrument, both by resorting to injections of synthetic exoplanets and by considering a study-case example on the HR 8799 system. Finally, Sect. \ref{sec:conclusion} concludes this paper and suggests a number of future prospects. 

\section{Multi-epoch combination formalism}
\label{sec:me_comb_formalism}
 
Throughout this section (and when possible throughout the paper), the algorithm is described in a general fashion considering ASDI observations, without any explicit differentiation between ADI and ASDI. The formalism remains valid for ADI observations by downgrading the model to a single spectral channel without any loss of generality.

\subsection{Direct model of the observations}
\label{subsec:direct_model_obs}

\noindent Classical ASDI datasets are temporal sequences of coronagraphic images, decoupled into spectral channels and recorded at different epochs. The apparent position on the sky $\bm\theta_t(\bm\mu)$ of a (point-like) celestial body following a Keplerian motion depends on a set of orbital elements $\bm\mu$ and on the epoch of observation $t$. We model a $N$-pixel observed intensity image, $\bm r_{t, \ell, k} \in \mathbb{R}^N$, taken at epoch, $t$, spectral channel, $\ell$, and temporal frame, $k,$ via:
\begin{equation}
\label{eq:directModel}
    \bm r_{t, \ell, k} = \alpha_{t, \ell} \, \bm h_{t, \ell}(\bm\theta_t(\bm\mu)) + \bm f_{t, \ell, k}\,,
\end{equation}
with $\bm\theta_t(\bm\mu)$ as the 2D angular location of the companion at epoch, $t,$ for a given set of orbital elements, $\bm\mu$, $\bm h_{t, \ell}(\bm\theta)  \in \mathbb{R}^N$, the off-axis PSF centered at angular location, $\bm \theta,$ and identical for all frames, $k,$ of the same epoch, $\alpha_{t, \ell}$ as the contrast of the companion\footnote{\samepage Due to data calibration issues and to  varying angular separation of the companion among epochs, the companion contrast is epoch-dependent. We discuss this aspect in Sect. \ref{subsec:multi_epoch_snr}. Given the typical photometric uncertainties currently encountered in ground-based direct imaging at high-contrast, we neglect the intrinsic variability of the companion intensity during a single observing epoch.} at epoch, $t,$ and spectral channel, $\ell$, and $\bm f_{t, \ell, k}$ as a nuisance term representing all other contributions other than the mean signal due to the companion. The nuisance term notably accounts for the stellar leakages, as well as for the thermal background for the sources of noise such as detector readout noise and photon noise. The main notations used in this paper are summarized in Table \ref{tab:notations}.

Given the number of photons collected by any pixel of the detector is sufficient, a weighted multi-variate Gaussian distribution has been shown to be a good approximation for the nuisance term \citep{Flasseur2018, Flasseur2020,Flasseur2020robust}. However, when the scale of the spatial (or spectral) correlations is larger than a few tens of pixels (as may occur near the coronagraph or with broad band observations), this approximation does not  hold perfectly (see Sect.~\ref{sec:search5thplanet} for a discussion of this effect). Images in different spectral channels and/or epochs are mutually independent and nearly identically distributed for a given epoch and spectral channel:
\begin{equation}
  \label{eq:nuisance-distrib-law}
  \bm f_{t, \ell, k} \sim \mathcal{N}(\bar{\bm f}_{t,\ell},\, \sigma^2_{t,\ell,k}\,\bm \C_{t, \ell})\,,
\end{equation}
with $\bar{\bm f}_{t,\ell}$ and $\bm \C_{t, \ell}$ as the expectation and the typical covariance matrix of the nuisance term, $\bm f_{t, \ell, k}$, and where $\sigma^2_{t,\ell,k}$ are the temporo-spectral correction factors accounting for the uneven quality of the frames \citep{Flasseur2020}.

\begin{table}[t]
    \centering
    \caption{Summary of the main notations used in this paper.}
    \begin{tabular}{@{}c@{}c@{\hspace*{1ex}}l@{}}
        \toprule     
        \textbf{Not.}\; & \textbf{Range}\; & \textbf{Definition} \\
        \midrule
        \multicolumn{3}{c}{$\blacktriangleright$ Constants}\\ \midrule
        $T$ & $\mathbb{N}$ & number of epochs\\
        $K_t$ & $\mathbb{N}$ & number of frames per epoch $t$\\
        $L$ & $\mathbb{N}$ & number of spectral  channels$^\text{(a)}$\\
        $N$ & $\mathbb{N}$ & number of pixels per frame$^\text{(a)}$\\
        
        \midrule
        \multicolumn{3}{c}{$\blacktriangleright$ Indexes}\\ \midrule
        $t$ & $\llbracket 1, T \rrbracket$ & epoch index\\
        $k$ & $\llbracket 1, K_t \rrbracket$ & temporal frame index\\
        $\ell$ & $\llbracket 1, L \rrbracket$ & spectral channel index\\
        
        \midrule
        \multicolumn{3}{c}{$\blacktriangleright$ Physical quantities}\\ \midrule
        $\bm \alpha$ & $\mathbb{R}_+^{T L}$ & source's spectral energy distribution \vspace{0.75mm}\\
        $\bm \alpha^\text{int}$ & $\mathbb{R}_+^{T}$ & source's spectrally integrated energy distribution \vspace{0.75mm}\\
        $\bm f$ & $\mathbb{R}^{N T K_t L}$ & nuisance (stellar leakages and noise)\\
        $\bm h$ & $\mathbb{R}^{N T L}$ & off-axis PSF\\
        $\bm r$ & $\mathbb{R}^{N T K_t L}$ & observed intensity\\
        $\bm \gamma$ & $\mathbb{R}^{T L}$ & source's prior spectrum\\
      
        \midrule
        \multicolumn{3}{c}{$\blacktriangleright$ Positions}\\ \midrule
        $\bm{\mu}$ & $\mathbb{R}^{7}$ & orbital elements$^\text{(b)}$\\
        $\bm\theta_t(\bm\mu)$ & $\mathbb{R}^{2}$ & 2D position on the detector\\
        
        \midrule
        \multicolumn{3}{c}{$\blacktriangleright$ Other quantities and metrics}\\ \midrule
        $\bm \C$ & $\mathbb{R}^{N \times N}$ & spatial covariance matrix of $\bm f$$^\text{(c)}$ \\
        $\bm \Sigma$ & $\mathbb{R}^{L \times L}$ & spectral covariance matrix of $\bm f$ \\
        ${\SNR}_t$ & $\mathbb{R}_+$ & mono-epoch S/N at epoch $t$\vspace{0.5mm}\\
        $\SNR$ & $\mathbb{R}_+$ & multi-epoch S/N\\
        $\CostFunc$ & $\mathbb{R}_+$ & multi-epoch cost function \\
        \bottomrule
    \end{tabular}
    \label{tab:notations}
    \tablefoot{$^\text{(a)}$The number of spectral channels and the number of pixels per frame are constant for a given instrument so that $L$ and $N$ are independent of $t$. $^\text{(b)}$See Sect. \ref{sec:pacome_algorithm} and Table \ref{tab:tableOrbitalElements} for the allowable ranges of each orbital element. $^\text{(c)}$As discussed in the text, with \PACO, covariance matrices are evaluated locally at a scale of small patches so that the full covariance, $\M C,$ is never evaluated explicitly.}
\end{table}

\subsubsection{Maximum likelihood estimation}
\label{subsec:maxLikelihoodEstimation}

Extending the \PACO formalism \citep{Flasseur2018, Flasseur2020,Flasseur2020robust} to multi-epoch ASDI observations, the total log-likelihood of the data results from the statistical model assumed in Eq.~\eqref{eq:nuisance-distrib-law} for the nuisance term:
\begin{align}
         \LogLike(\bm \alpha, \bm \mu, \bar{\bm f}, \bm \sigma, \bm \C)
         &= \sum_{t,\ell,k} \log p \left(\bm r_{t, \ell, k} | \, \alpha_{t, \ell}, \bm\mu \right) \notag \\
         &= \text{c}_1 - \frac12 \sum_{t,\ell,k} \log \, \text{det}\left( \sigma^2_{t,\ell,k} \bm \C_{t,\ell}\right)\notag \\
         &\hspace{-10mm}\quad - \frac12 \sum_{t,\ell,k} \norm{ \bm r_{t, \ell, k} - \alpha_{t, \ell} \, \bm h_{t, \ell}(\bm\theta_t(\bm\mu)) - \bar{\bm f}_{t,\ell} }^2_{\sigma^{-2}_{t,\ell,k} \bm \C_{t, \ell}^{-1}}\,,
\end{align}
where $\text{c}_1$ is an irrelevant constant and $\norm{\bm x}_{\M A}^2 = \V x \transp \, \M A \, \V x$ denotes the squared Mahalanobis norm of $\bm x$. 

The maximum likelihood estimation of $\bar{\bm f}_{t,\ell}$, $\bm \C_{t,\ell}$, and $\sigma^2_{t,\ell,k}$ is performed locally by the \PACO algorithm in small patches at the scale of a few tens of pixels. To simplify  the equations, we introduce the precision matrix $\bm \W_{t,\ell,k} = \sigma^{-2}_{t,\ell,k} \bm \C_{t, \ell}^{-1}$. The unknowns are now just $\bm\alpha$ and $\bm\mu$, so that the multi-epoch log-likelihood can be re-expressed as:
\begin{align}
         \LogLike(\bm \alpha, \bm \mu)
         &= \text{c}_2 - \frac12 \sum_{t,\ell,k} \norm{ \bm r_{t, \ell, k} - \alpha_{t, \ell} \, \bm h_{t, \ell}(\bm\theta_t(\bm\mu)) - \bar{\bm f}_{t,\ell} }^2_{\bm\W_{t,\ell,k}} \notag \\
         &= \text{c}_3 + \sum\limits_{t,\ell} \bigg( \alpha_{t,\ell} \,  b_{t,\ell}(\bm\theta_t(\bm\mu)) - \frac12 \alpha_{t,\ell}^2 \,  a_{t,\ell}(\bm\theta_t(\bm\mu))  \bigg)\,, \label{eq:TotLogLikelihood_alpha_mu}
\end{align}
where $\text{c}_2$ and $\text{c}_3$ are irrelevant constants and $a_{t, \ell}$ and $b_{t, \ell}$ are defined as:
\begin{equation}
\label{eq:ab_terms}
     \begin{cases}
          a_{t,\ell}(\bm\theta) &= \sum_k  \bm h_{t,\ell}(\bm\theta)\transp \, \bm \W_{t,\ell,k} \, \bm h_{t,\ell}(\bm\theta) \,, \vspace{1mm}\\
          b_{t,\ell}(\bm\theta) &= \sum_{k}\bm h_{t,\ell}(\bm\theta)\transp \, \bm \W_{t,\ell,k} \, \big( \bm r_{t,\ell,k}-\bar{ \bm f}_{t,\ell} \big)\,.
    \end{cases}
\end{equation}
The term $b_{t, \ell}(\bm\theta)$ represents the correlation between the centered plus whitened\footnote{We refer to a \textit{whitened} quantity when its correlations are removed (in a certain direction) by application of the Cholesky's factorization $\mathbb{W}\transp$ of the precision matrix $\M W$ such that $\mathbb{W} \, \mathbb{W}\transp = \M W$.} data and the whitened off-axis PSF centered at $\bm\theta$. The term $a_{t, \ell}(\bm\theta)$ acts as a normalization factor and represents the auto-correlation of the whitened off-axis PSF centered at $\bm\theta$.
In practice, terms $a_{t,\ell}(\bm\theta)$ and $b_{t,\ell}(\bm\theta)$ can be approximated by interpolating at angular position, $\bm\theta,$ the 2D maps, $\bm a_{t,\ell}$ and $ \bm b_{t,\ell}$. Conveniently, these maps are already pre-calculated by \PACO for a grid of angular positions (usually corresponding to the grid of pixels). This shows that the previous formalism of \PACO can be extended to multi-epoch observation combination when these byproducts are properly added together.

Assuming that the flux of the companion may change from one epoch to another, the maximum likelihood estimator of $\alpha_{t,\ell}$ can be expressed by:
\begin{equation}
\label{eq:SEDestimator_unconstrained}
         \widehat{\alpha}_{t,\ell}(\bm \mu) = \argmax_{\alpha_{t,\ell}} \, \LogLike(\bm\alpha, \bm\mu) = \frac{  b_{t,\ell}(\bm\theta_t(\bm\mu))}{ a_{t,\ell}(\bm\theta_t(\bm\mu))} \,.
\end{equation}
As the source flux is necessarily non-negative, we can impose a positivity constraint while maximizing the multi-epoch log-likelihood in Eq.~\eqref{eq:TotLogLikelihood_alpha_mu}. This constrained problem has a simple closed-form solution which amounts to thresholding the unconstrained maximum likelihood estimator:
\begin{equation}
         \widehat{\alpha}_{t,\ell}^{+}(\bm \mu) = \argmax_{\alpha_{t,\ell} \geq 0} \, \LogLike(\bm\alpha, \bm\mu) = \frac{\big[  b_{t,\ell}(\bm\theta_t(\bm\mu)) \big]_+}{ a_{t,\ell}(\bm\theta_t(\bm\mu))} \,,
    \label{eq:SEDestimator}
\end{equation}
where $[x]_+ = \max(x,0)$ denotes the nonnegative part of $x$.
Substituting the estimator given by Eq.~\eqref{eq:SEDestimator} of the intensity of the companion in the multi-epoch log-likelihood in Eq.~\eqref{eq:TotLogLikelihood_alpha_mu} yields:
\begin{align}
         \LogLike(\bm \mu) &= 
         \LogLike(\widehat{\bm \alpha}^{+}(\bm \mu), \bm\mu) = \text{c}_3 + \frac12 \sum_{t,\ell} \dfrac{\big( \big[  b_{t,\ell}(\bm\theta_t(\bm\mu)) \big]_+ \big)^2}{ a_{t,\ell}(\bm\theta_t(\bm\mu))}\,.
         \label{eq:TotLogLikelihood_mu}
\end{align}
Hence, the maximum likelihood estimator of the orbital elements, $\bm\mu,$ is expressed as:
\begin{equation}
         \widehat{\bm \mu} =  \argmax_{\bm{\mu}}  \sum_{t, \ell} \frac{\big( \big[ b_{t,\ell}(\bm\theta_t(\bm\mu))\big]_+\big)^2}{ a_{t,\ell}(\bm\theta_t(\bm\mu))}\,.
         \label{eq:mu_opt}
\end{equation}

Having no closed-form expression, the estimator $\widehat{\bm \mu}$ of the orbital elements can only be found via numerical methods of global optimization. Finally, our problem comes down to maximizing the following multi-epoch objective function:
\begin{equation}
         \CostFunc(\bm\mu) = \sum_{t,\ell} \frac{\big(\big[ b_{t,\ell}(\bm\theta_t(\bm\mu))\big]_+\big)^2}{ a_{t,\ell}(\bm\theta_t(\bm\mu))} \,.
         \label{eq:costfuncC}
\end{equation}
The criterion $\CostFunc$ of Eq.~\eqref{eq:costfuncC} combines optimally the information provided by the data and should enable the detection of sources yet undetectable in individual epochs and simultaneously provide an estimation of some plausible orbital elements. It can be noted that similar expressions of Eqs.~\eqref{eq:TotLogLikelihood_mu},~\eqref{eq:mu_opt}, and~\eqref{eq:costfuncC} hold without positivity constraint  using $\widehat{\alpha}_{t,\ell}$ instead of $\widehat{\alpha}_{t,\ell}^+$ in the multi-epoch log-likelihood (Eq.~\ref{eq:TotLogLikelihood_alpha_mu}). However, as shown by \cite{thiebaut2005maximum,smith2008detection,mugnier2009optimal}, it is beneficial to enforce a positivity constraint on the flux $\alpha_{t,\ell}$ when deriving the detection criterion, that is, to use (as we do) the estimate $\widehat{\alpha}_{t,\ell}^+$ instead of $\widehat{\alpha}_{t,\ell}$ in the multi-epoch log-likelihood (Eq.~\ref{eq:TotLogLikelihood_alpha_mu}).

\subsubsection{Multi-epoch signal-to-noise ratio}
\label{subsec:maxLikelihoodEstimation}

We show in Appendix \ref{subsec:multi_epoch_snr} that using the flux maximum likelihood estimator $\widehat{\alpha}_{t,\ell}^+$ and a matched filter approach \citep{kay1998fundamentals,kay1998fundamentals2} allows us to establish a link between the multi-epoch criterion of Eq.~\eqref{eq:costfuncC} and the best possible multi-epoch S/N of a linear combination of the data:
\begin{equation}
        \SNR(\bm\mu) =\sqrt{  \sum_{t,\ell} \frac{\big(\big[ b_{t,\ell}(\bm\theta_t(\bm\mu))\big]_+\big)^2}{ a_{t,\ell}(\bm\theta_t(\bm\mu))} } = \sqrt{\CostFunc(\bm \mu)}\,.
\end{equation}
We note that this quantity is exactly the square root of the criterion derived from our direct model in Eq.~\eqref{eq:costfuncC}. Hence, maximizing $\SNR(\bm\mu)$ or $\CostFunc(\bm\mu)$ in $\bm \mu$ yields the exact same estimator for $\bm \mu$.

This analysis shows that searching for the maximum likelihood estimator of $\bm \mu$ given our direct model is equivalent to searching for the orbital elements for which the best possible S/N is reached among all possible linear combinations of the reduced data collected along the apparent trajectory of the companion. The derived combination criterion is therefore optimal both in the maximum likelihood sense and in terms of S/N. For the rest of the paper, we refer to the objective function, $\CostFunc(\bm \mu),$ of Eq.~\eqref{eq:costfuncC} to find the best estimator and assess the potential detection relevance.

\subsubsection{Multi-epoch noise distribution}
\label{sec:StatisticalGuarantees}

In the expression of the estimator, $\widehat{\alpha}_{t,\ell}^{+}$, of the companion intensity, the $a_{t,\ell}$ term is supposed  deterministic\footnote{This approximation neglects the dependency of the term $a_{t,\ell}$ with respect to the estimated precision matrix $\M W_{t, \ell, k}$ (which is not completely deterministic since it depends explicitly on the observed intensity $\V r_{t, \ell, k}$). This approximation is reasonable since the variance of $\M W_{t, \ell, k}$ is negligible compared to the variance of  $\V r_{t, \ell, k}$; $\M W_{t, \ell, k}$ being estimated from a relatively large number of samples.} so that only the $b_{t,\ell}$ term fluctuates. Given our statistical model, $b_{t,\ell}$ is Gaussian-distributed and of variance:
\begin{align}
\Var \{b_{t,\ell}(\bm \theta) \} &= \sum_k \Var \left\{ \bm h_{t,\ell}(\bm \theta)\transp \, \bm \W_{t,\ell,k} \, (\bm r_{t,\ell,k}-\bar{\bm f}_{t,\ell}) \right\} \notag\\
&= \sum_k \bm h_{t,\ell}(\bm \theta) \transp \, \bm \W_{t,\ell,k} \,  \Cov\left\{ \bm r_{t,\ell,k}-\bar{\bm f}_{t,\ell} \right\}  \, \bm \W_{t,\ell,k} \, \bm h_{t,\ell}(\bm \theta) \notag \\
&= \sum_k \bm h_{t,\ell}(\bm \theta) \transp \, \bm \W_{t,\ell,k} \, \bm h_{t,\ell}(\bm \theta) \notag \\
&=  a_{t,\ell}(\bm \theta) \,,
\end{align}
as the frames are mutually independent. Remarkably, $b_{t,\ell}(\bm\theta_t(\bm\mu))$ and $a_{t,\ell}(\bm\theta_t(\bm\mu))$ for any spectral channel, $\ell,$ and epoch, $t,$ provide "sufficient statistics" to study a potential source with orbital elements $\bm\mu$. In addition, these terms account for dominant instrumental effects (e.g., transmission of the coronagraph \citep{Flasseur2021}).

We assume that the variance of $\widehat{\alpha}_{t,\ell}^{+}$ given in Eq.~\eqref{eq:SEDestimator} can be approximated by the variance of the unconstrained estimator:
\begin{align}
\Var \{ \widehat{\alpha}_{t,\ell}^{+}(\bm\mu) \} \approx \Var\{ \widehat{\alpha}_{t,\ell}(\bm\mu) \} &= \Var\left\{\frac{b_{t,\ell}(\bm\theta_t(\bm\mu))}{ a_{t,\ell}(\bm\theta_t(\bm\mu))}\right\} = \frac{ \Var \{  b_{t,\ell}(\bm\theta_t(\bm\mu)) \} }{ a^2_{t,\ell}(\bm\theta_t(\bm\mu))}\notag\\
&= \dfrac{1}{a_{t,\ell}(\bm\theta_t(\bm\mu))}.
\end{align}

In \PACOs formalism \citep{Flasseur2018,Flasseur2020,Flasseur2020robust}, the mono-epoch temporo-spectral S/N at position $\bm\theta$ is computed independently for each epoch in a statistical sense as:
\begin{align}
\SNR_{t,\ell}(\bm\theta) = \dfrac{\E \big\{ \widehat{\alpha}_{t,\ell}(\bm \theta) \} }{\sqrt{\Var \{ \widehat{\alpha}_{t,\ell}^{+}(\bm \theta) \big\} }} \approx \dfrac{b_{t,\ell}(\bm \theta)}{\sqrt{a_{t,\ell}(\bm \theta)}}\,,
\label{eq:single_epoch_detection_criterion}
\end{align}
which follows a normal law in the absence of companion. Thanks to this key property, the single-epoch detection criterion defined in Eq.~\eqref{eq:single_epoch_detection_criterion} is directly interpretable in terms of probability of detection and of probability of false alarm for any given spectral channel. 

\medskip

\begin{figure}[t]
    \centering
    \includegraphics[width=0.4\textwidth]{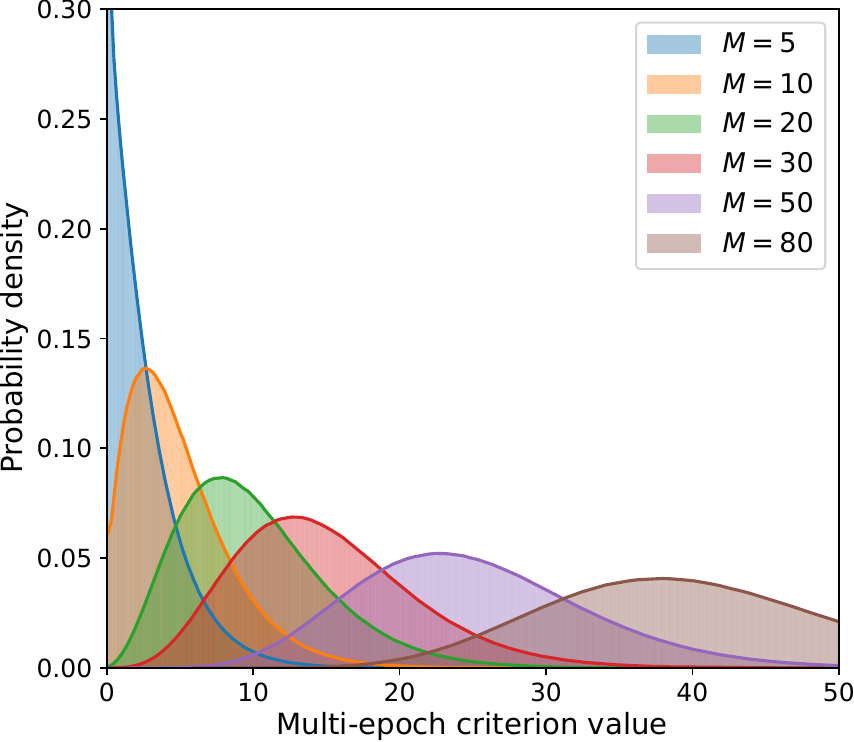}
    \caption{Theoretical probability density functions of the multi-epoch detection criterion of Eq.~\eqref{eq:costfuncC} for different degrees of freedom, $M$.}
    \label{fig:distribcriteriontheoret}
\end{figure}

\begin{figure}[t]
    \centering
    \includegraphics[width=0.4\textwidth]{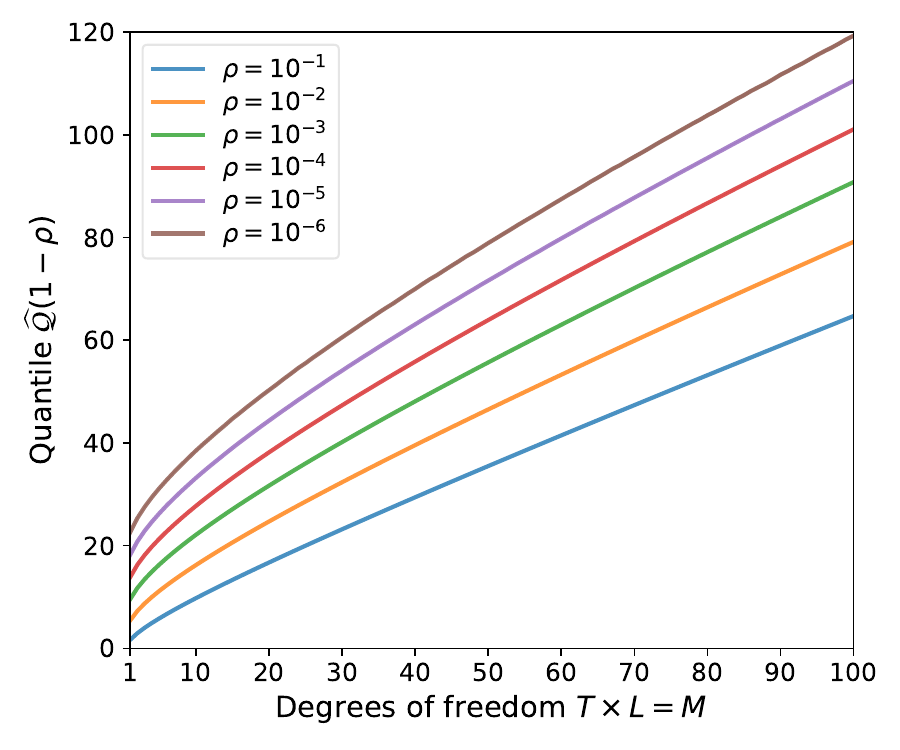}
    \caption{Sample quantile function $\widehat{Q}_M(1-\rho)$ of the multi-epoch criterion for different degrees of freedom, $M,$ and at different confidence levels, $\rho$.}
    \label{fig:qquantileabacus}
\end{figure}

Noting that the criterion derived in Eq.~\eqref{eq:costfuncC} is the sum of all squared non-negative temporo-spectral S/N, it can be used to assert the relevance of multi-epoch detections. Indeed, the statistical distribution of the criterion corresponds to the sum of $T \times L = M$ independent squared so-called "rectified Gaussian" distributions, where $T$ and $L$ represent respectively the total number of epochs and spectral channels. To the best of our knowledge, this distribution has no analytical expression and computing its probability density function numerically would imply integrating $M - 1$ intertwined convolution products which is difficult to perform and very time consuming. Instead, we resort to a Monte-Carlo approach to estimate the upper bound of the multi-epoch confidence interval associated to the confidence level $1-\rho \in [0,1]$ of a detection, with $\rho$ a small number representing the targeted probability of false alarm\footnote{For example, a classical significance at ``$5\sigma$'' (for a Gaussian distributed criterion with unit variance) corresponds to $\rho \simeq 2.9 \times 10^{-7}.$}. To do so, we build a collection of $N_s$ (pseudo)-random numbers $\big\{\sum_{m=1}^{M} ([x_{n,m}]_+)^2 \big\}_{n=1:N_s}$, where the $x_{n,m}$ are independently drawn from a Gaussian normal distribution, and we compute the value of the sample quantile function $\widehat{Q}_M(1-\rho)$ (threshold value below which random draws from the given distribution would fall $100 \times (1-\rho)$ percent of the time) as defined in \cite{Hyndman1996}. The probability density functions of the criterion computed by this Monte Carlo procedure are shown in Fig.~\ref{fig:distribcriteriontheoret} and some values of the sample quantile function evaluated at different thresholds and for several degrees of freedom are given in Fig.~\ref{fig:qquantileabacus}. Besides, we empirically found that the distribution of the sample quantile function $\widehat{Q}_M$ with respect to the confidence level $\rho$ is well approximated by a law of the form:
\begin{equation}
    \widehat{Q}_M(1-\rho) \approx \kappa_1^{-\log{\rho}} \kappa_2 + \kappa_3
\label{eq:power_law_criterion_quantile}
,\end{equation}
where $\kappa_1$, $\kappa_2$, and $\kappa_3$ are constants for a given number of degrees of freedom, $M$. This law is useful to roughly estimate the sample quantile function at very high confidence levels (typically beyond $\rho= 10^{-7}$), which are otherwise very time-consuming to compute empirically.

\begin{figure}[t]
    \centering
    \includegraphics[width=0.4\textwidth]{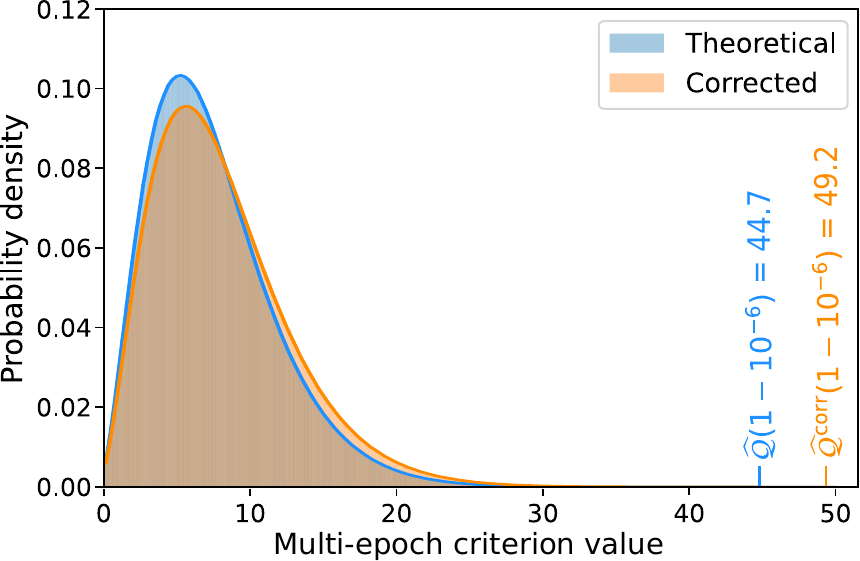}
    \caption{Theoretical and corrected probability density functions of the multi-epoch detection criterion for $M=15$ degrees of freedom. The corrected distribution is built out of Gaussian distributions of variable means and variances ranging in $[-0.036, 0.048]$ and $[0.92, 1.14]$ respectively.}
    \label{fig:distribcriteriontheoretVScorrected}
\end{figure}

This empirical distribution is however optimistic as is it built out of a perfect "normal"\ distribution. With real data, we often observe slightly different variances in the statistical distribution of the mono-epochs S/N (from $-10\%$ up to $+10\%$) which tends to change the distribution of the multi-epoch criterion and such even more at higher degrees of freedom. In these cases, our optimistic sample quantile function $\widehat{Q}_M$ can either under- or overestimate the prescribed confidence threshold if the individual variances (resp.,\ the means) of the data tend to be smaller or larger than $1$ (resp.\ smaller or larger than 0). To cope with this tendency, we introduced a corrected sample quantile function, $\widehat{Q}^{\text{corr}}_M$, which is made up of Gaussian distributions whose variances and means are those of the data itself. Known sources are masked and robust estimators (median absolute deviation and median) are used to estimate respectively the variance and mean values of each mono-epoch distribution. A comparison between the theoretical and corrected criterion distribution is shown in Fig.~\ref{fig:distribcriteriontheoretVScorrected} for $M=15$ degrees of freedom and considering variable mono-epoch means and variances ranging in $[-0.036, 0.048]$ and $[0.92, 1.14],$ respectively.

For a given orbit, comparing the value of the criterion of Eq.~\eqref{eq:costfuncC} to the sample quantile at confidence level $\rho$ is the only way to interpret the "goodness" of the multi-epoch combination. We note that the desired confidence level $\rho$ should be chosen according to the number of points drawn  (i.e., the number of multi-epoch combinations).

\subsection{Accounting for spectral correlations in the model}
\label{subsec:asdi_formalism}

We show in Sect.~\ref{sec:StatisticalGuarantees} that the mono-epoch signal-to-noise ratios, $\SNR_{t,\ell}$, follow a normal distribution. However, since the stellar leakages are caused by the diffraction of the light, the spectral channels of a given data frame are not mutually independent. Therefore, the corresponding spectral correlations need to be learned and whitened before all multi-epoch S/N may be combined. 
Taking into account the spatial and spectral correlations jointly in the model, based on the multi-epoch observed intensity data, would be too computationally demanding and very difficult to perform (if not altogether impossible). In addition, these spectral correlations are difficult to capture at the patches scale. An efficient alternative, proposed in \cite{Flasseur2020}, is first to account for the spatial correlations only to derive the individual signal-to-noise ratio, $\SNR_{t,\ell}$, (as done in Sect.~\ref{subsec:direct_model_obs}) and then account for the spectral correlations with these byproducts in the second stage.

In that context, the reasoning behind deriving the multi-epoch detection criterion is quite similar to the one where the spectral correlations are not taken into account, except that we now base the model on the collection, $\bm x_t$, of all temporo-spectral S/N at epoch, $t$. In addition, we assume a prior spectrum to improve the detection of sources having a similar spectral energy distribution. Hence, at a given epoch $t$, the model is expressed as:
\begin{equation}
    \bm x_t = \alpha_t^{\text{int}} \, \bm\gamma_t \, \bm\beta_t(\bm \theta_t(\bm\mu)) + \bm\epsilon_t\,,
\end{equation}
where $\alpha^{\text{int}}_{t}$ is the spectrally integrated flux, $\bm\gamma_t$ is the assumed spectrum of the point source (normalized between 0 and 1), $\bm\beta_{t}(\bm\theta_t(\bm\mu))$ the vector whose $\ell$-th element is equal to $\sqrt{a_{t,\ell}(\bm\theta_t(\bm\mu))}$ (with $a_{t,\ell}$ described in Eq.~\eqref{eq:ab_terms}) and $\bm\epsilon_t$ is a random vector accounting for the fluctuations of the temporo-spectral S/N values that have a Gaussian distribution with zero mean and spectral covariance, $\bm\Sigma_t$ (i.e., $\bm\epsilon_t \sim \mathcal{N}(\bm 0, \bm\Sigma_t)$). 

Considering all epochs to be independent of one another, the multi-epoch log-likelihood of the spectrally correlated data under the assumption of a prior spectrum, $\bm\gamma = \{  \bm\gamma_t \}_{t=1:T}$, is expressed as:
\begin{align}
    \LogLike_{\bm\gamma}(\bm\alpha^{\text{int}}, \bm\mu)
    &= \sum_{t} \LogLike_{t,\bm\gamma_t}(\alpha^{\text{int}}_{t}, \bm\theta_t(\bm\mu)) \notag \\[-0.1cm]
    &= c_4 - \frac12 \sum_{t} \norm{  \bm x_t - \alpha^{\text{int}}_{t} \bm\gamma_t \, \bm\beta_t(\bm\theta_t(\bm\mu))}_{\bm\Sigma_t^{-1}}^2 \notag \\[-0.1cm]
    = c_5 + &\sum_t \left[ \alpha^{\text{int}}_{t} B_{t,\bm\gamma_t}(\bm\theta_t(\bm\mu)) - \frac12 \left(\alpha^{\text{int}}_{t} \right)^2 A_{t,\bm\gamma_t}(\bm\theta_t(\bm\mu)) \right]\,,
\end{align}
where the $A_{t,\bm\gamma_t}$ and $B_{t,\bm\gamma_t}$ quantities are still pre-calculated by \PACO and correspond to a normalization term and to the data whitened spatially and spectrally and filtered by the shape of PSF. These expressions are:
\begin{equation}
     \begin{cases}
          A_{t,\bm\gamma_t}(\bm\theta) = \big( \bm\gamma_t \, \bm\beta_t(\bm\theta) \big)\transp \, \bm\Sigma_t^{-1} \, \big( \bm\gamma_t \, \bm\beta_t(\bm\theta) \big), \vspace{1mm}\\
          B_{t,\bm\gamma_t}(\bm\theta) = \big( \bm\gamma_t \, \bm\beta_t(\bm\theta) \big) \transp \, \bm\Sigma_t^{-1} \, \bm x_t.
    \end{cases}
\end{equation}
The source flux estimate, in the maximum likelihood sense, has an analytical expression and is expressed with the pre-calculated terms, $A_{t,\bm\gamma_t}$ and $B_{t,\bm\gamma_t}$, by:
\begin{equation}
    \widehat{\alpha}_{t,\bm\gamma_t}^{\text{int}^+}(\bm\mu) = \argmax_{\alpha^{\text{int}}_{t} \geq 0} \, \LogLike_{t,\bm\gamma_t}(\alpha^{\text{int}}_{t}, \bm\theta_t(\bm\mu)) = \frac{\big[B_{t,\bm\gamma_t}(\bm\theta_t(\bm{\mu}))\big]_+}{A_{t,\bm\gamma_t}(\bm\theta_t(\bm{\mu}))}.
\end{equation}
By injecting the spectrally integrated fluxes estimators into $\LogLike_{\bm\gamma_t}(\bm\alpha^{\text{int}},\bm\mu)$, the multi-epoch log-likelihood now only depends on the assumed prior spectrum, $\bm\gamma,$ and the orbital elements, $\bm\mu,$ such that:
\begin{align}
    \LogLike_{\bm\gamma}(\bm\mu) &= c_5 + \frac12 \sum_{t} \frac{\big(\big[B_{t,\bm\gamma_t}(\bm\theta_t(\bm{\mu}))\big]_+\big)^2}{A_{t,\bm\gamma_t}(\bm\theta_t(\bm{\mu}))}.
\end{align}
Echoing what is described in Sect.~\ref{subsec:maxLikelihoodEstimation}, the optimal set of orbital elements given the data and the assumed prior spectrum, $\bm\gamma,$ is written as: 
\begin{align}
    \widehat{\bm\mu}_{\bm\gamma} &=  \argmax_{\bm{\mu}} \sum_{t} \frac{\big(\big[B_{t,\bm\gamma_t}(\bm\theta_t(\bm\mu))\big]_+\big)^2}{A_{t,\bm\gamma_t}(\bm\theta_t(\bm\mu))}\,.
\end{align}
Therefore, finding the optimal orbital elements amounts to maximizing the following multi-epoch objective function:
\begin{align}
    \CostFunc_{\bm\gamma}(\bm\mu) = \sum_{t} \frac{\big(\big[B_{t,\bm\gamma_t}(\bm\theta_t(\bm\mu))\big]_+\big)^2}{A_{t,\bm\gamma_t}(\bm\theta_t(\bm\mu))}. \label{eq:costfuncC_specCorr}
\end{align}
which is equivalent to maximizing the multi-epoch S/N (see Sect.~\ref{subsec:maxLikelihoodEstimation}):
\begin{align}
    \SNR_{\bm\gamma}(\bm\mu) = \sqrt{ \sum_{t} \frac{\big( \big[B_{t,\bm\gamma_t}(\bm\theta_t(\bm\mu))\big]_+\big)^2}{A_{t,\bm\gamma_t}(\bm\theta_t(\bm\mu))} }. \label{eq:SNR_spec_corr}
\end{align}

\section{The PACOME algorithm}
\label{sec:pacome_algorithm}

\subsection{Reduced orbital elements}

The orbits of \PACOME were parametrized using a mix of the conventions described in \cite{Murray2010} and \cite{Blunt2020}. We used the semi-major axis ($a$), eccentricity ($e$), inclination ($i$), epoch of periapsis passage ($\tau$), argument of periapsis ($\omega$) and longitude of ascending node ($\Omega$). To these six conventional orbital elements, we added a seventh and final parameter, namely, Kepler's constant ($K$), to link the orbital period $P$ to the semi-major axis while accounting for the total mass of the system without making any strict assumption on it nor on the distance of the object from the Earth.

\begin{figure}[t!]
    \centering
    \includegraphics[width=0.48\textwidth]{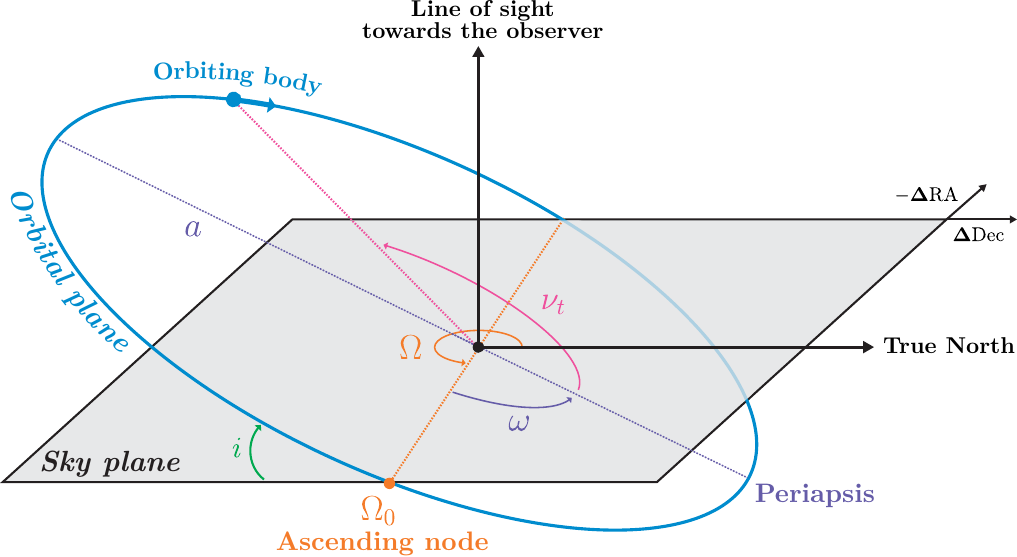}
    \caption{Diagram of the main orbital elements of a celestial body moving along its orbit (blue) intersecting the sky plane (grey) at the ascending node $\Omega_0$.}
    \label{fig:orbitalElements}
\end{figure}

The semi-major axis is expressed in milliarcseconds and all angular quantities ($i$, $\omega$, $\Omega$) in degrees. The longitude of ascending node $\Omega$ is oriented with respect to the true north direction and increases counterclockwise. 
The orientation of the inclination angle is such that $i=\qty{0}{\degree}$ denotes a prograde orbit seen face-on whereas $i=\qty{90}{\degree}$ denotes an edge-on orbit. The epoch of periapsis passage, $\tau,$ is expressed as a fraction of the orbital period with $\tau = t_0/P \pmod 1$, where $t_0$ is the traditional epoch of periapsis passage in years. Finally, Kepler's constant $K=a^3/P^2$ is expressed in milliarcseconds cubed per year squared. The derived period $P$ is therefore expressed in years. 
The 2D reference frame coincides with the sky plane. Its origin is the central body (i.e., the host star), the $x$-axis is in the positive declination direction ($\Delta\text{Dec}$, north is up or true north) and the $y$-axis is in the negative right ascension direction ($-\Delta\text{RA}$, east is left). Most of these orbital elements are represented in Fig. \ref{fig:orbitalElements} and the chosen parameters are summarized in Table \ref{tab:tableOrbitalElements}. 

\begin{table}[t!]
    \centering
    \caption{Summary of the orbital elements used in this work.}
    \begin{tabular}{c c c c}
         \toprule
         \textbf{Not.} & \textbf{Definition} & \textbf{Units} & \textbf{Range} \\
         \midrule
         $a$ & semi-major axis & mas & $\mathbb{R}_+$ \\ 
         $e$ & eccentricity & - & $[0,1[$ \\
         $i$ & inclination & deg & $[0,180]$ \\
         $\tau$ & epoch of periapsis passage & - & $[0,1]$ \\
         $\omega$ & argument of periapsis & deg & $[0,360]$ \\
         $\Omega$ & longitude of ascending node & deg & $[0,360]$ \\
         $K$ & Kepler's constant & mas$^3$/yr$^2$ & $\mathbb{R}_+$ \\ \bottomrule
    \end{tabular}
    \label{tab:tableOrbitalElements}
\end{table}

Orbits are usually degenerated as several combinations of orbital elements give the same projection on the 2D plane. A typical example is the pairing $(\omega,\Omega),$ which is perfectly equivalent to $(\omega + \pi,\Omega +  \pi)$. Orbits also become very degenerated for nearly face-on orbits ($i \sim 0$) and/or small eccentricities. While it is not possible to circumvent these degeneracies without additional constraints (e.g., system resonances), they do not affect the detection sensitivity of the proposed method (see Sects.~\ref{subsec:ResultsSearchInjections} and~\ref{subsec:resultsRealData_HR8799bcde}).

To solve Kepler's equation and project the 2D positions of a celestial body on the sky plane, we used Brent's \texttt{fzero} method \citep{Brent1973}. The equations for the Keplerian motion, the 2D projection, and the solver are described in more detail in Appendix \ref{sec:keplerianMotionSolver}.

\subsection{Optimization strategy}

The cost function derived in Sect.~\ref{sec:me_comb_formalism} has several local maxima mostly due to the high dimensionality of the problem, the presence of noise in the data, and the speckles being similar to the PSF. Hence, finding the optimal solution cannot be performed by a local optimization method. To search for the global maximum of the criterion $\CostFunc(\bm\mu)$ of Eq.~\eqref{eq:costfuncC}, we proceeded to sample $\CostFunc(\bm\mu)$ on a large 7D regularly spaced grid followed by a local optimization to refine the parameters.

\subsubsection{Search space sampling}

We used a regularly spaced 7D grid to exhaustively explore the parameter space but all orbital elements are not necessarily sampled at the same precision. This strategy has proven to be effective for the tests on the semi-synthetic data and on the HR~8799 system, detailed in Sects.~\ref{subsec:results_ss_data} and \ref{subsec:results_real_data}. We stress that the 7D grid sampling should be thinner for far-out exoplanets with highly covered orbits.

Given the fairly small computation times required by the algorithm to explore a large number of orbits ($\simeq 20$ min for $10$ epochs, two spectral channels, and $10^9$ orbits on $12$ CPUs), we did not find any urgent need for more complex methods though more advanced sampling and on-search grid refinement strategies will be explored in future works to decrease redundancies, while continuing to maintain an exhaustive approach, to derive the full orbital elements posterior distributions and ensure that the global maximum is found.

\subsubsection{Local optimization refinement}
\label{sec:LocalOptiRefinement}

After establishing the best on-grid orbit, a local refinement of the orbital elements is performed around the $N_\text{opt}$ best solutions (typically the first 100 or 1000). To maximize the objective function, we use the VMLM-B optimization method \citep{Thiebaut2002}, which is a memory-limited quasi-Newton method with bound constraints. It is quite similar to L-BFGS-B \citep{byrd1995limited} but has less overheads per iteration. Algorithms of this kind are particularly suitable for large-scale nonlinear problems compared to, for instance, constrained conjugate gradients. VMLM-B requires the objective function to optimize as well as its gradient. Given the reasonable number of iterations needed for the algorithm to converge, we set the tolerance thresholds for deciding of the convergence of the optimization so that the precision on the sought-for variables is close to the machine precision.

The required gradient of the multi-epoch criterion (Eq.~\eqref{eq:costfuncC}) is expressed as:
\begin{align}
    \frac{\partial\CostFunc(\bm\mu)}{\partial \bm\mu} = \sum_{t,\ell} & \Bigg[ 2 \, \frac{[ b_{t,\ell}(\bm\theta_t(\bm\mu))]_+}{ a_{t,\ell}(\bm\theta_t(\bm\mu))} \frac{\partial  b_{t,\ell}(\bm\theta_t(\bm\mu))}{\partial \bm\mu} \notag \\
    &- \left( \frac{[b_{t,\ell}(\bm\theta_t(\bm\mu))]_+}{ a_{t,\ell}(\bm\theta_t(\bm\mu))} \right)^2 \frac{\partial  a_{t,\ell}(\bm\theta_t(\bm\mu))}{\partial \bm\mu} \Bigg]\,,
\end{align}
and can be computed semi-analytically using the chain rule:
\begin{equation}
    \begin{cases}
        \dfrac{\partial b_{t,\ell}(\bm\theta_t(\bm\mu))}{\partial \bm\mu} &= \dfrac{\partial b_{t,\ell}(\bm\theta)}{\partial \bm\theta}\bigg|_{\bm \theta = \bm\theta_t(\bm\mu)} \cdot \dfrac{\partial \bm\theta_t(\bm\mu)}{\partial \bm\mu}\,, \vspace{1mm}\\
         \dfrac{\partial a_{t,\ell}(\bm\theta_t(\bm\mu))}{\partial \bm\mu} &= \dfrac{\partial a_{t,\ell}(\bm\theta)}{\partial \bm\theta}\bigg|_{\bm \theta = \bm\theta_t(\bm\mu)} \cdot \dfrac{\partial \bm\theta_t(\bm\mu)}{\partial \bm\mu}\,,
    \end{cases}
\end{equation}
where the derivatives of the $a_{t, \ell}$ and $b_{t, \ell}$ terms with respect to the projected positions $\partial a_{t, \ell}/\partial \bm \theta$ and $\partial b_{t, \ell}/\partial \bm \theta$ are approximated using the derivatives of the interpolation function and where the derivatives $\partial \bm\theta_t/\partial \bm \mu$ of the projected positions with respect to the orbital elements are computed analytically. The details of the latter derivatives can be found in Appendix~\ref{sec:projPosDerivWRTOrbElem}.

Two other classical constrained optimization methods\footnote{Implementation of the optimization functions can be found at \href{https://github.com/emmt/OptimPackNextGen.jl}{\texttt{https://github.com/emmt/OptimPackNextGen.jl}}.} without derivatives were also tested: NEWUOA \citep{Powell2006NEWUOA} and BOBYQA \citep{Powell2009BOBYQA}. However, VMLM-B showed better performance by refining the solution deeper, thus giving higher multi-epoch detection scores. Optimizing the solution with VMLM-B is also fast. It takes about \qty{50}{\micro\second} per evaluation of the cost function and its gradient with a cubic spline interpolator and a dataset consisting of $T=10$ epochs, $L=2$ spectral channels (see Sect.~\ref{sec:interpolationStrat} for more details).

\subsection{Inference of the uncertainties}

 We developed two methods to infer the local uncertainties associated with the optimal orbital solutions. The first is a perturbation method in which random realizations of Gaussian noise of the same order of magnitude as the variance of the signal are injected in the data and where the solution is re-optimized locally with the perturbed data to calculate confidence intervals on the orbital elements. The second method exploits the Cramér-Rao lower bounds (CRLBs) which are good estimates of the covariance of maximum likelihood estimators when the number of samples is large enough \citep{kendall1948advanced}.

\subsubsection{Numerical method based on perturbations}
\label{sec:NumericalMethodErrors}

For each epoch and spectral channel, we perturb the data by adding to the $b_{t, \ell}$ terms random noise realizations drawn in a centered Gaussian distribution with variance $a_{t, \ell}$ (see Sect. \ref{sec:StatisticalGuarantees} for the derivation of the distribution of $b_{t, \ell}$). 
For these new perturbed data, we re-optimize the criterion around the optimal solution, $\widehat{\bm\mu,}$ found for unperturbed data. We repeat this process a large number of times, $N_p$, to form a collection of optimal orbital parameters obtained with perturbed data, $\{\widetilde{ \, \bm\mu}_{(n)} \}_{n=1:N_p}$. Then, we compute the lower and upper bounds of the confidence intervals (CI) associated to the obtained orbital elements distribution at a specified confidence level (e.g., $95\%$). As the samples are drawn independently, the number of draws $N_p$ does not need to be very large ($N_p \simeq 10^5$ is usually enough).

\subsubsection{Analytical method based on CRLBs}
\label{sec:CramerRaoMethodError}

From the expression of the multi-epoch log-likelihood of Eq.~\eqref{eq:TotLogLikelihood_mu}, we derived the Fisher information matrix of the orbital elements, $\bm\mu,$ at epoch, $t$. It is expressed as a $7 \times 7$ matrix, $\mathcal{I}_{t}(\bm\mu),$ whose elements $\big[ \mathcal{I}_{t}(\bm\mu) \big]_{i,j}$ are given by:
\begin{equation}
\label{eq:FisherInfoMatrixCoeffs}
    \big[ \mathcal{I}_{t}(\bm\mu) \big]_{i,j} = \E \left\{ \,  \frac{\partial \LogLike_t(\bm\mu)}{\partial \bm \mu_i}   \, \frac{\partial \LogLike_t(\bm\mu)}{\partial \bm \mu_j} \, \right\},
\end{equation}
where $\LogLike_t(\bm\mu)$ is the log-likelihood at epoch, $t$.
Distributing the derivative over the sum gives:
\begin{align}
    \frac{\partial\LogLike_t(\bm\mu)}{\partial \bm \mu_i} = \sum_{\ell} \Bigg( & \frac{\big[  b_{t,\ell}(\bm\theta_t(\bm{\mu})) \big]_+}{ a_{t,\ell}(\bm\theta_t(\bm{\mu}))} \frac{\partial  b_{t,\ell}(\bm\theta_t(\bm{\mu}))}{\partial \bm\mu_i}
    \notag \\[0cm]
    &  - \frac12 \Bigg(\frac{\big[  b_{t,\ell}(\bm\theta_t(\bm{\mu})) \big]_+}{ a_{t,\ell}(\bm\theta_t(\bm{\mu}))}\Bigg)^2 \frac{\partial  a_{t,\ell}(\bm\theta_t(\bm{\mu}))}{\partial \bm\mu_i} \Bigg)\,.
\end{align}
The derivatives of the $b_{t,\ell}$ and $a_{t,\ell}$ terms with respect to the orbital elements, $\bm\mu,$ are computed via the chain rule mentioned in Sect. \ref{sec:LocalOptiRefinement}. 
By approximating the expectation of Eq.~\eqref{eq:FisherInfoMatrixCoeffs} to simply the product of derivatives, the Fisher information matrices of all independent epochs can be computed individually and gathered to encompass all multi-epoch information by simply summing over time:
\begin{equation}
\mathcal{I}(\bm\mu) = \sum_t \mathcal{I}_t(\bm\mu).
\end{equation}
Thus, the multi-epoch covariance matrix of the orbital elements satisfies the CRLBs:
\begin{equation}
    \Cov^{\text{(CRLB)}}\{ \bm\mu \} \geq \mathcal{I}(\bm\mu) ^{-1}\,,
    \label{eq:crlb_orbit}
\end{equation}
where the square roots of the diagonal coefficients yield the lower bounds, ${\bm\sigma}_{\bm\mu}^{(\textup{CRLB})}$, of the standard deviations associated to the orbital elements. As epochs provide independent samples, the inequality of Eq.~\eqref{eq:crlb_orbit} tends to the equality as the number of epoch grows.

\subsubsection{Comparison between both approaches}
\label{subsec:ComparisonCRLBsPerturbations}

The analytical method based on CRLBs is faster than the numerical one but it can be sensitive to possible orbital elements degeneracies since it requires inverting an estimate of the Fisher's information matrix. It also supposes that the errors are symmetric and Gaussian, however, they were shown not to be \citep{Konopacky2016hr8799, wertz2017vlt, Wang2018}. This approximation can only quantify the local error around the solution and fails totally to provides information on the global orbital elements distribution. The perturbation method explores more of the parameter space but still stays in the vicinity of the solution which is not enough. It is also biased \citep{Ford2005} as the multiple re-optimization processes always start from the optimal solution identified by PACOME.

Furthermore, both approaches account for the "local data fitting error" only, namely, the local uncertainty induced by the model given the data. In other words, the global distribution of the orbital elements on the full parameter space or additional sources of systematics errors related to the instrument itself, to calibration, and/or to pre-reduction issues are not accounted for. As both methods act locally on the orbital elements they do not deal with the degeneracies (i.e., multi-modal peaks in their distribution). This generally results in uncertainties different than those obtained with dedicated sampling algorithm such as MCMC \citep{Ford2005} or nested sampling \citep{Skilling2004} that will be explored in future studies.
Even though the main focus of this work is the detection aspect, we use the perturbation method in the following (Sect.~\ref{sec:results}) to assess the local uncertainties of the orbital solutions with full awareness of the limits of our approach.

\subsection{Implementation details}

\subsubsection{Interpolation strategy}
\label{sec:interpolationStrat}

\begin{table}[t!]
    \centering
    \caption{Scores of different interpolation functions.}
    \begin{tabular}{c c c c}
        \toprule
        \textbf{Interpolator} & \textbf{Class} & \textbf{AMPRS (\%)} & \textbf{RMSE} \\
        \midrule
        Nearest & - & $77.34 \pm 5.76$ & $1.67$ \\
        Bilinear & $\mathcal{C}^0$ & $95.47 \pm 2.30$ & $0.51$ \\
        Cubic B-spline & $\mathcal{C}^2$ & $92.61 \pm 3.28$ & $0.87$ \\
        Lanczos kernel & $\mathcal{C}^1$ & $96.99 \pm 1.06$ & $0.22$ \\
        Mitchel \& Netravali & $\mathcal{C}^1$ & $96.66 \pm 1.94$ & $0.37$ \\
        Catmull \& Rom & $\mathcal{C}^1$ & $98.29 \pm 1.55$ & $0.16$ \\
        BCCS & $\mathcal{C}^1$ & $98.49 \pm 1.21$ & $0.11$ \\
        \bottomrule
    \end{tabular}
    \tablefoot{The two metrics are defined in Eqs.~\eqref{eq:AMPRS} and~\eqref{eq:RMSE}.}
    \label{tab:interpStudy}
\end{table}

\begin{figure}[t!]
    \centering
    \includegraphics[width=0.45\textwidth]{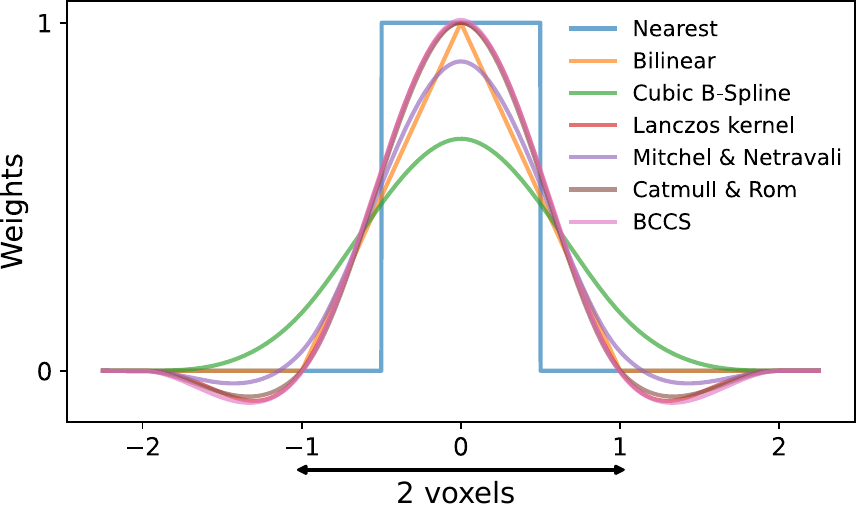}
    \caption{Kernel functions of the tested interpolation methods.}
    \label{fig:interpolationKernelsPlots}
\end{figure}

\usetikzlibrary{positioning, calc, fit}

\tikzstyle{RectangleGris} = [rectangle, draw, fill=gray!10, node distance=1.4cm, text width=3.75cm, text centered, rounded corners, minimum height=1cm, thick]
\tikzstyle{RectangleGris2} = [rectangle, draw, fill=gray!10, node distance=4.75cm, text width=3.75cm, text centered, rounded corners, minimum height=1cm, thick]
\tikzstyle{gs} = [rectangle, draw, fill=green!40, node distance=3.167cm, text width=1.7cm, text centered, rounded corners, minimum height=1cm, thick]
\tikzstyle{rs} = [rectangle, draw, fill=red!40, node distance=3.167cm, text width=1.7cm, text centered, rounded corners, minimum height=1cm, thick]
\tikzstyle{RectangleGris3} = [rectangle, draw, fill=gray!10, node distance=2.8cm, text width=3.75cm, text centered, rounded corners, minimum height=1cm, thick]
\tikzstyle{RectangleGris3Dashed} = [rectangle, draw, fill=white, node distance=2.8cm, text width=3.75cm, text centered, rounded corners, minimum height=1cm, dashed, thick]
\tikzstyle{RectangleOrange} = [rectangle, draw, color=orange!95!black, fill=white, node distance=3.167cm, text width=1.75cm, text centered, rounded corners, minimum height=1cm, dashed, line width=0.4mm]

\tikzstyle{line} = [draw, -, orange, dashed, line width=0.5mm]
\tikzstyle{line2} = [draw, -, black, dashed, line width=0.4mm]
\tikzstyle{arrow} = [draw, -to, line width=0.5mm]

\tikzstyle{RectangleVert} = [rectangle, draw, minimum height=5cm, minimum width=2.67cm, green!65!black, line width=0.6mm]
\tikzstyle{lineVert} = [draw, -to, green!65!black, line width=0.6mm]
\tikzstyle{RectangleBleu} = [rectangle, draw, minimum height=5cm, minimum width=3cm, blue!60, line width=0.6mm]

\begin{figure*}[t!]
    \centering
    \small
    
    \begin{tikzpicture}[scale=1, transform shape]
        \node [RectangleGris] (selection) {Selection of orbit $\bm\mu_{(n)}$};
        \node [RectangleGris, below of=selection] (projection) {Computation of the 2D projected positions $\bm\theta_t(\bm\mu_{(n)})$};
        \node [RectangleGris, below of=projection] (interpolation) {Interpolation of the $\V a_{t,\ell}$ and $\V b_{t,\ell}$ maps at $\bm\theta_t(\bm\mu_{(n)})$};
        \node [RectangleGris, below of=interpolation] (evaluation) {Evaluation of the cost function $\CostFunc(\bm\mu_{(n)})$};
        \node [RectangleGris, below of=evaluation] (optimization) {Local optimization of the $N_{\text{opt}}$ best orbits};
        
        \node [RectangleGris2, right of=optimization] (identification) {Identification of the optimal solution $\bm{\widehat{\mu}}$};
        
        \node [gs, above right=0.36cm and -1.92cm of identification] (detection) {\normalsize{Detection}};
        \node [rs, above left=0.36cm and -1.92cm of identification] (nodetection) {\small{No detection}};

        \node [RectangleGris3, above of=identification] (errors) {Error estimation of the optimal solution};
        
        \node [RectangleOrange, above left=0.36cm and -1.92cm of errors] (perturbation) {\small{Perturbation\\ method}};
        \node [RectangleOrange, above right=0.36cm and -1.92cm of errors] (carmerrao) {\small{Cramér-Rao\\ lower bounds}};
    
        \node [RectangleGris3Dashed, above of=errors] (kepler) {\small{Solving Kepler's equation \\ $\forall t$, $M_t = E_t - e \sin E_t$}};
        
        \node [left of=selection, yshift=-0.785cm, node distance=4.3cm, text width=2.5cm, text centered, label=below:$\{\bm b_{t,\ell} \}_{t=1:T}$ maps] (Bpic) {\includegraphics[width=1\linewidth]{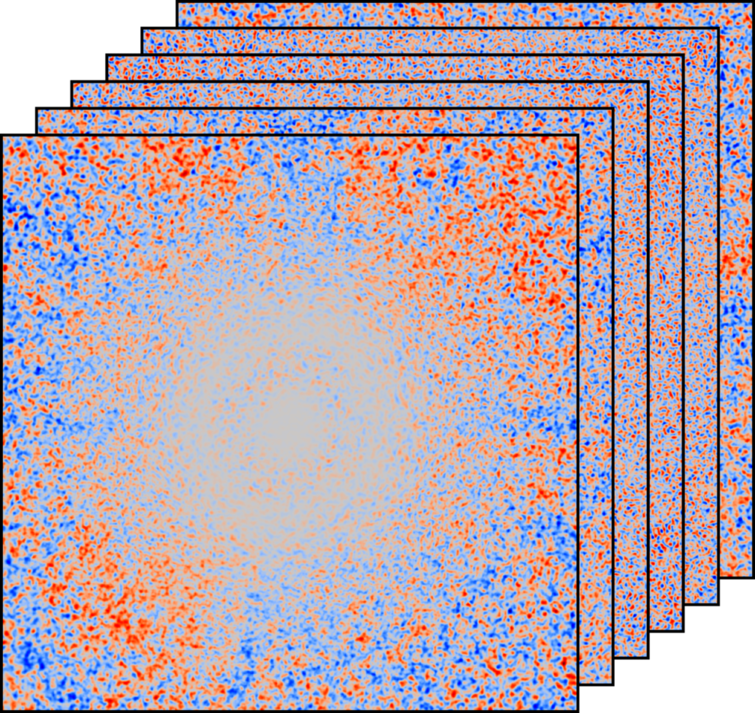}};
        \node [left of=Bpic, node distance=2.75cm, text width=2.5cm, text centered, label=below:$\{\bm a_{t,\ell} \}_{t=1:T}$ maps] (Apic) {\includegraphics[width=1\linewidth]{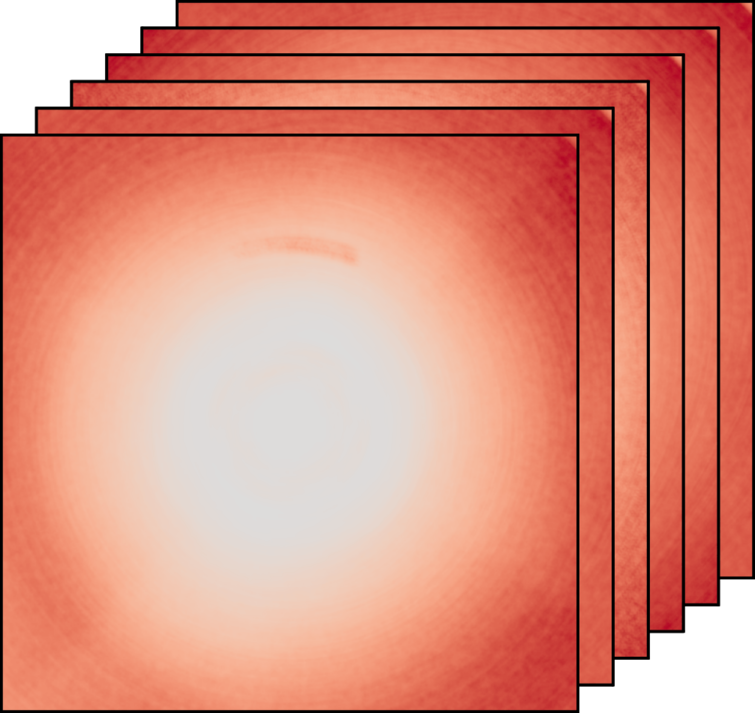}};
        \node [below of=Apic, node distance=2.5cm, text width=2.5cm, text centered, label={[align=center]below:Search space\\(7D grid)}] (searchSpace) {\includegraphics[width=0.5\linewidth]{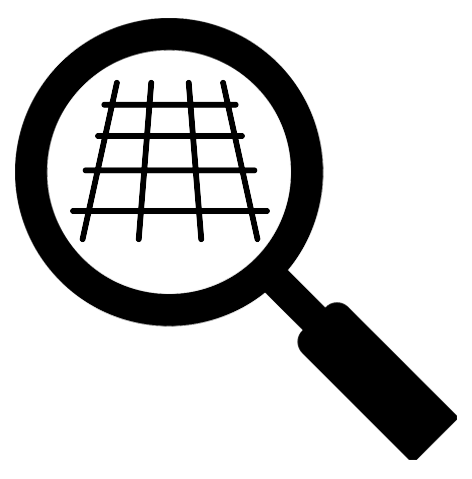}};
        \node [below of=Bpic, node distance=2.5cm, text width=2.5cm, text centered, label={[align=center, name=lsettings]below:Additional settings\\ \scriptsize{(interpolator, number of }\\ \scriptsize{threads, saving options, etc.)}}] (settings) {\includegraphics[width=0.5\linewidth]{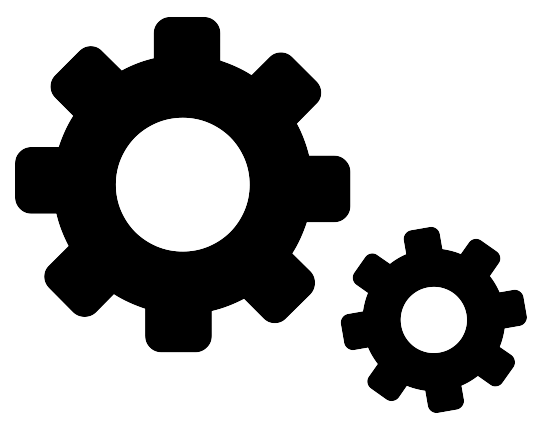}};
        
        \node[RectangleVert, label=above:\textcolor{green!65!black}{Inputs}, fit=(Apic) (lsettings)] (inputs) {};
        \node[RectangleBleu, label=above:\textcolor{blue!60}{Parallelized ($n \in [1,N_{\text{orb}}]$)}, fit=(selection) (evaluation)] (parallelized) {};
        \path [lineVert] (inputs) -| ($(inputs)!0.57!(selection)$) |- ([xshift=-0.125cm]selection.west);
        
        \path [arrow] (selection) -- (projection);
        \path [arrow] (projection) -- (interpolation);
        \path [arrow] (interpolation) -- (evaluation);
        \path [arrow] (evaluation) -- (optimization);
        \path [arrow] (optimization) -- (identification);
        \path [arrow] (identification) -- (nodetection);
        \path [arrow] (identification) -- (detection);
        \path [arrow] (detection) -- (errors);
        \path [line] (errors) -- (perturbation);
        \path [line] (errors) -- (carmerrao);
        
        \path [line2] (projection) -| ($(projection)!0.51!(kepler)$) |- (kepler);

    \end{tikzpicture}
    
    \normalsize
    \caption{Schematic diagram of the \PACOME algorithm.}
    \label{fig:PACOMEscheme}
\end{figure*}
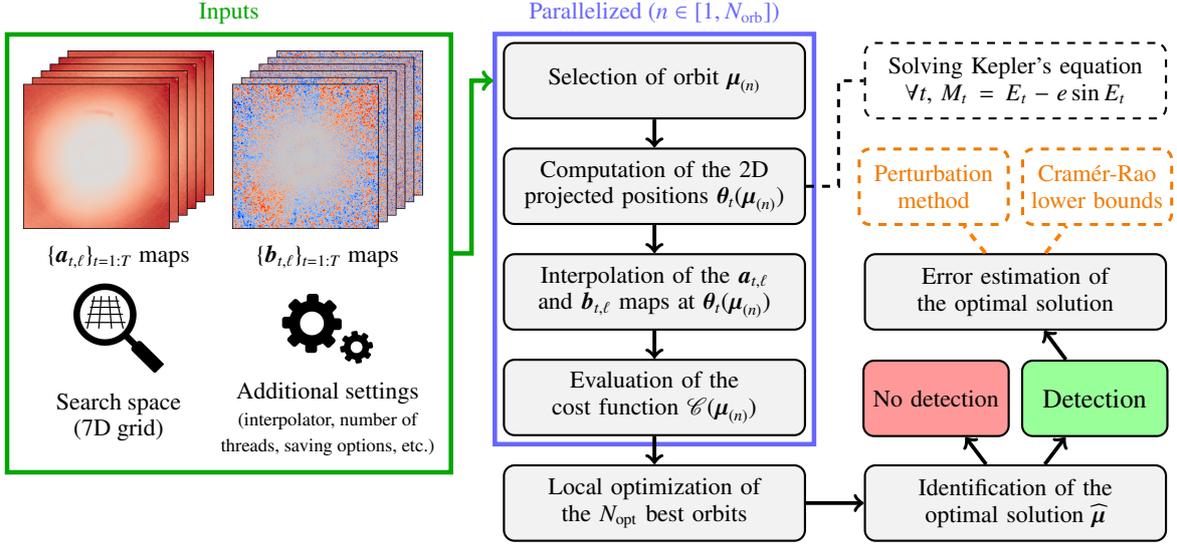

To cope with the fact that the projected positions, $\bm\theta_{t}(\bm\mu),$ do not necessarily coincide with the sampling positions of the maps, $\V a_{t, \ell}$ and $\V b_{t, \ell}$, computed by \PACO, we interpolate these maps to evaluate the criterion. The choice of the interpolation method is quite critical, as discussed hereafter.

To assess which interpolator is most suitable for our problem, we need a ground truth to quantify the goodness of the interpolation and, therefore, an assumption with respect to the shape of the signal of interest must be made. The off-axis PSF can be approximated by a 2D isotropic Gaussian. 
To identify the best possible interpolator, we generated $S=10^4$ 2D Gaussian PSFs, whose positions and amplitudes are known. The amplitudes of the synthetic PSFs are drawn uniformly between $5$ and $30$, and their full widths at half maximum were chosen to match the experimental conditions of the VLT/SPHERE-IRDIS and IFS instruments (wavelengths $\lambda \in [\qty{0.95}{\micro\meter},\qty{2.32}{\micro\meter}]$ and a telescope diameter $D=\qty{8.2}{\meter}$). Then, we interpolated each synthetic PSF in a $M$-pixel circular region of diameter $d=4\,\sigma$ with a sampling step six times smaller than the pixel size.
We computed the absolute mean percentage of the recovered signal (AMPRS), as well as the root mean square error (RMSE), between the interpolated and ground truth amplitude values:
\begin{equation}
\label{eq:AMPRS}
    \text{AMPRS (\%)} = 100 - \frac{100}{S M}  \sum\limits_{s=1}^S \sum\limits_{m=1}^M \left| \frac{\V g_{s,m}^{\text{int}} - \V g_{s,m}^{\text{gt}}}{\V g_{s,m}^{\text{gt}}} \right| \,,
\end{equation}
\begin{equation}
\label{eq:RMSE}
    \text{RMSE} = \sqrt{\frac{1}{S M} \sum_{s=1}^S \sum\limits_{m=1}^M \left(\V g_{s,m}^{\text{int}} - \V g_{s,m}^{\text{gt}}\right)^2}\,,
\end{equation}
where $\V g^{\text{int}}$ and  $\V g^{\text{gt}}$ are the interpolated and ground truth synthetic PSFs, respectively.
Several interpolation functions\footnote{Implementation of the interpolation functions can be found at \href{https://github.com/emmt/InterpolationKernels.jl}{\texttt{https://github.com/emmt/InterpolationKernels.jl}}.} were tested: nearest neighbor, bilinear, cubic B-spline, and four-neighbor Lanczos function, as well as two cubic cardinal splines: Mitchel \& Netravali ($\psi=-1/2$, $\chi = 1/18$) and Catmul \& Rom  ($\psi=-1/2$, $\chi = 0$), where $\psi$ and $\chi$ represent, respectively, the derivative and the value of the kernel function at position $x=1$.
We also performed a quick optimization with VMLM-B to find the best (in the RMSE sense) cubic cardinal spline (BCCS) coefficients for our problem. It yielded $\psi \approx  -0.600$ and $\chi \approx -0.004$. 

The detailed results of the study are given in Table \ref{tab:interpStudy} and a graphical representation of the tested kernel functions are represented in Fig.~\ref{fig:interpolationKernelsPlots}.
The BCCS and the Catmull \& Rom spline give very similar results and are more efficient than the other comparative methods. 
We chose the BCCS interpolation for the rest of this work.

\subsubsection{Storing of the useful data}
\label{sec:StoringTheData}

As the algorithm explores a very large number of orbits ($10^9$ typically), the amount of data to be saved is substantial. Indeed, an orbit being encoded by seven parameters, storing all explored orbits as well as their multi-epoch scores would take about $500$ Gb of space for double-precision floating-point numbers ($64$ bits $\times \, 8 \times 10^9$), which is excessive. 

Our parameter space is sampled regularly so the values of the orbital elements we explore are deterministic. We use this to our advantage and only store seven lists of variable sizes containing the discrete values of the orbital elements to explore, rather than storing all the orbital elements combinations. We assign a unique integer index to each orbital combination and use that instead. Doing so reduces the space taken to save the data by a factor of 4  (instead of 8), as we only need to store an integer and a double precision floating-point number.

In addition, most of the explored orbits do not fall on the projected position of the putative exoplanet and therefore contain no interesting information as they only combine noise. Hence, we only save the orbits whose multi-epoch cost function scores are greater than a given value that we typically set to the quantile function value associated to the confidence level of detection $\rho=0.01$ (i.e., a $99\%$ confidence level). Doing so reduces the number of saved orbits by a factor 100 which, combined with what is described above, brings the total space taken in memory by the \PACOME products down to $\simeq 1$ Gb, whatever the number of epochs.

Finally, the indexes of the retained orbits and their multi-epoch scores are stored dynamically in memory-mapped files to avoid overcharging the random-access memory by loading a very large matrix.

\subsubsection{Massive code parallelization and execution time}

The entire algorithm is written in Julia \citep{Julia2017}, an open-source and strongly typed language designed for high performance programming.
Coupling the intrinsic speed of Julia with massive code  parallelization enables the algorithm to process very large numbers of orbits in a reasonable computation time.
Even though \PACOME can be run on a laptop, we used a local machine equipped with an Intel\texttrademark~Xeon E5-2620 v3 CPU running at \qty{2.40}{\giga\hertz} with $12$ cores. The computational power achievable by this CPU is estimated at a maximum of $325$ Gflops by the Intel\texttrademark's Math Kernel Library Benchmarks for Linux. Our implementation of \PACOME does not use any GPUs.

Testing an orbit is done by selecting the orbit, computing its projected positions by solving Kepler's equations, interpolating the $a_{t,\ell}$ and $b_{t,\ell}$ terms\footnote{The 2D maps $\V a_{t,\ell}$ and $\V b_{t,\ell}$ are pre-computed only once per epoch with \PACO, so that \PACOME can be launched multiple times, e.g., with various bound constrains on the orbital parameters, without the need to relaunch the data reduction step with \PACO. Similarly, including new available epochs is performed efficiently by reducing the new epochs with \PACO and by relaunching \PACOME.}, computing the cost function, assessing its value, and saving it in a memory-mapped file. This is done in approximately $12.6\times10^3$ operations on average, which takes a typical execution time of \qty{9}{\micro\second} for $T=10$ epochs and $L=2$ spectral channels.
Altogether, the \PACOME algorithm takes about $20$ minutes to assess $10^9$ orbits for the considered case (i.e., 12 cores, $T=10$, and $L=2$).

The $a_{t,\ell}$ and $b_{t,\ell}$ maps are pre-calculated via \PACOs reduction step upstream of \PACOME and, hence, they do not need to be recomputed each time we run the algorithm. The execution time of this phase is variable but it usually takes several minutes up to several hours. For example, the ADI version of the \texttt{PACO} algorithm takes an hour to process (with ten threads on our machine) a rather large VLT/SPHERE-IRDIS hypercube of 500 frames, $1024 \times 1024$ pixels per frame, and two spectral channels with a patch radius of five pixels.

\subsection{Summary of PACOME's workflow}

The \PACOME algorithm starts by sampling the criterion of Eq.~\eqref{eq:costfuncC} on a 7D grid mapping the orbital elements search space. For that purpose, in a parallel loop, the subpixelic 2D projected positions $\bm \theta_t$ are computed for each sample of the parameter space and for all epochs. The values of $a_{t,\ell}$ and $b_{t,\ell}$ terms at these positions are interpolated thus allowing the evaluation of the cost function. Then, the algorithm proceeds with the local optimization of the few best $N_\text{opt}$ orbital solutions to find the one that maximizes the objective function. The score of the optimal solution is finally compared to the multi-epoch detection limit (i.e., at a confidence level of $1 - \rho$) estimated numerically via the empirical statistical distribution of the criterion to decide  on a potential detection (or otherwise). If a detection is claimed, the local uncertainties associated to the solution are computed via the perturbation method and/or with the CRLBs. A schematic diagram of the \PACOME algorithm is represented in Fig. \ref{fig:PACOMEscheme} and a pseudo-code is given in Algorithm \ref{alg:pseudoCodePacome}.

\section{Results}
\label{sec:results}

\subsection{Datasets description}
\label{subsec:datasets}

\begin{table*}[h!]
                \centering 
        \caption{
                        Observing conditions of A(S)DI sequences of HR\,8799 from the VLT/SPHERE instrument considered in this paper.}
                \begin{scriptsize} \begin{tabular}{cccccccccccccc}
                \toprule 
                ESO ID & Obs. date & Obs. mode & Spec. band & $N_\textup{frames}$ & $\Delta_{\text{par}}$ (\si{\degree}) & DIT (s) & TDIT (s) &$\tau_0$ (ms) & Seeing ('') & Related paper\\ \midrule
        
        60.A-9249(B) & 2014-07-13 & IRDIS & DB H23 & 639 & 16.75 & 4 & 2556 & 5.64 & 0.569 &  (1)\\ 
        60.A-9249(B) & 2014-07-16 & IRDIS/IFS & DB K12/YH & 91/48 & 17.38/17.22 & 30/60 & 2730/2880 & 3.35 & 1.186 &  (1)\\ 
        60.A-9249(C) & 2014-08-12 & IRDIS/IFS & DB K12/YH & 102/51 & 18.43 & 30/60 & 3060 & 2.88 & 0.867 &  (1)\\ 
        60.A-9249(C) & 2014-08-14 & IRDIS & BB J & 100 & 27.47 & 16 & 1600 & 2.78 & 0.590 &  (1)\\ 
        60.A-9352(A) & 2014-12-05 & IRDIS & BB H & 218 & 8.70 & 8 & 1744 & 2.97 & 1.005 &  (1)\\ 
        60.A-9352(A) & 2014-12-06 & IRDIS & BB H & 203 & 8.55 & 8 & 1624 & 2.56 & 1.170 &  (1)\\ 
        60.A-9352(A) & 2014-12-07 & IRDIS & BB H & 211 & 8.45 & 8 & 1688 & 2.69 & 1.252 &  (1)\\ 
        60.A-9352(A) & 2014-12-09 & IRDIS & BB H & 209 & 7.84 & 8 & 1672 & 2.68 & 0.863 &  (1)\\ 
        095.C-0298(C) & 2015-07-04 & IRDIS/IFS & DB K12/YH & 112/58 & 17.91/24.09 & 32/64 & 3584/3712 & 3.10 & 0.654 &  (2)\\ 
        095.C-0689(A) & 2015-07-30 & IRDIS & DB J23 & 952 & 83.40 & 16 & 15232 & 6.15 & 0.560 &  (3)\\ 
        095.C-0689(A) & 2015-07-31 & IRDIS & DB K12 & 951 & 90.70 & 16 & 15216 & 30.52 & 0.659 &  (3)\\ 
        095.C-0298(D) & 2015-09-28 & IRDIS/IFS & DB K12/YH & 499/63 & 23.94/23.61 & 8/16 & 3992/1008 & 24.02 & 0.556 &  (2)\\ 
        \textbf{198.C-0209(B)} & \textbf{2016-11-18} & \textbf{IRDIS}/IFS & \textbf{DB H23}/YJ & \textbf{64}/64 & \textbf{16.84}/18.77 & \textbf{32}/64 & \textbf{2048}/4096 & \textbf{4.24} & \textbf{0.930} &  \textbf{(2)}  \\ 
        198.C-0209(J) & 2017-06-15 & IRDIS/IFS & DB H23/YJ & 136/107 & 19.28/19.35 & 24/32 & 3264/3424 & 7.11 & 0.652 &  (2)\\ 
        099.C-0588(A) & 2017-10-08 & IRDIS/IFS & BB H/YJ & 344/242 & 76.08/78.63 & 16/64 & 5504/15488 & 19.07 & 0.446 &  (3)\\ 
        099.C-0588(A) & 2017-10-12 & IRDIS/IFS & BB H/YJ & 743/245 & 74.92/79.47 & 16/64 & 11888/15680 & 11.08 & 0.435 &  (3)\\ 
        099.C-0588(A) & 2017-10-13 & IRDIS/IFS & BB H/YJ & 1162/231 & 79.45/79.64 & 8/64 & 9296/14784 & 10.0 & 0.434 &  (3)\\ 
        1100.C-0481(H) & 2018-06-19 & IRDIS/IFS & DB H23/YJ & 62/64 & 34.40/34.38 & 96 & 5952/6144 & 7.77 & 0.666 &  (2)\\ 
        0101.C-0315(A) & 2018-08-18 & IRDIS/IFS & BB H/YJ & 895/224 & 73.94/73.76 & 16/64 & 14320/14336 & 28.78 & 0.452 &  (3)\\ 
        0101.C-0315(A) & 2018-08-20 & IRDIS/IFS & BB H/YJ & 894/224 & 72.59/73.65 & 4/64 & 3576/14336 & 26.47 & 0.267 &  (3)\\ 
        2103.C-5076(A) & 2019-11-01 & IRDIS/IFS & DB K12/YH & 215/44 & 23.76/14.14 & 8/32 & 1720/1408 & 15.48 & 0.591 &  (4)\\ 
        2103.C-5076(A) & 2019-11-02 & IRDIS/IFS & DB K12/YH & 444/93 & 49.34/43.27 & 8/32 & 3552/2976 & 10.21 & 0.556 &  (4)\\ 
        1104.C-0416(E) & 2021-08-21 & IRDIS/IFS & DB H23/YJ & 40/40 & 20.48/20.53 & 96 & 3840 & 23.26 & 0.418 &  (5)\\ 
                \bottomrule
                \end{tabular}
                \end{scriptsize} 
                \label{tab:dataset_logs}
  \tablefoot{Columns are: ESO survey ID, observation date, considered observation mode (IRDIS, IFS or both simultaneously), spectral band: broad band (BB) or dual band (DB), number $N_\textup{frames}$ of selected temporal frames, total amount of field rotation $\Delta_{\text{par}}$ of the field of view, individual exposure time DIT, total exposure time TDIT, DIMM coherence time $\tau_0$, average seeing, and the first paper reporting analysis of the same data. All the observations are performed with the apodized Lyot coronagraph (APLC; \cite{carbillet2011apodized}) of the VLT/SPHERE instrument. Both seeing and coherence time come from SPARTA estimations except for the observations of 2014 and 2016 where ESO's Astronomical Site Monitor estimations were used.
   The central wavelengths of the J, H, H2, H3, K1 and K2 filters are $\lambda = 1.245$, 1.625, 1.593, 1.667, 2.110, and \qty{2.251}{\micro\meter} respectively. The IRDIS epoch in boldface is used in Sect.~\ref{subsec:results_ss_data} as a reference dataset to build nine semi-synthetic epochs.}
   \tablebib{(1)~\cite{zurlo2016first}; (2)~\cite{langlois2021sphere}; (3)~\cite{biller2021high}; (4)~\cite{wahhaj2021search}; (5)~\cite{zurlo2022orbital}.}
\end{table*}

\noindent In this section, we evaluate the performance of the proposed algorithm on 23 datasets from both the InfraRed Dual Imaging Spectrograph (IRDIS; \cite{dohlen2008infra, dohlen2008prototyping}) and from the Integral Field Spectrograph (IFS; \cite{claudi2008sphere, claudi2010sphere}) of the VLT/SPHERE instrument \citep{beuzit2019sphere}. The observations were conducted with the A(S)DI technique using the pupil tracking mode of the instrument (see Sect. \ref{sec:introduction}). 
The observations were scheduled so that the star was observed during meridian passage to take benefit, as best as possible, of the apparent rotation of the sought-for objects. All datasets correspond to observations of HR\,8799 (HIP\,114189), obtained under highly variable observing conditions between 2014 and 2021 (see Table \ref{tab:dataset_logs} summarizing the main logs parameters).

HR\,8799 is a A5V type star of the Pegasus constellation of mass $M_\star = 1.47^{+0.12}_{-0.17} \, M_\odot$ \citep{sepulveda2022dynamical} and parallax of $\pi = 24.462 \pm 0.046$\,mas \citep{GaiaEDR3}. Until now, it has remained the sole star hosting four massive exoplanets discovered by direct imaging \citep{marois2008direct, marois2010images}. These exoplanets orbit nearly coplanarly between 15 and 80\,au, with a low eccentricity, and with a quasi-resonance of type 1:2:4:8. Three of them (HR\,8799 c, d, and e) are within the IRDIS and IFS field of view, while the fourth one (HR\,8799 b) lies only in the larger field of view of the IRDIS instrument. This system has been widely studied, in particular to characterize the orbits (see e.g., \cite{maire2015leech,Zurlo2016,wertz2017vlt,Wang2018,lacour2019first}), the atmospheres composition (see e.g., \cite{currie2014deep,ruffio2021deep,wang2021detection}), and the masses (see e.g., \cite{marley2012masses,brandt2021first,sepulveda2022dynamical}) of the known exoplanets. The presence of a debris disk in the circumstellar environment and orbital stability studies showed that there is room for additional exoplanets, either interior or exterior of the known four exoplanets \citep{currie2014deep, gozdziewski2014multiple, wahhaj2021search, thompson2022deep}. Despite the substantial efforts put into the analysis and combination of the available data, no candidate companion has been robustly identified. The current best results show a possible candidate of $4-7\,M_{\text{Jup}}$ at $4-5$\,au, at a detection confidence of about $3\sigma$ in the L spectral band with KECK/NIRC2 \citep{thompson2022deep}. However, this candidate companion is not detected with on-par or deeper observations from the VLT/SPHERE instrument at shorter wavelengths \citep{wahhaj2021search}. As a byproduct of the evaluation of the performance of the proposed algorithm, we also aim to put tight constrains (in terms of contrast) on an additional candidate companion in the HR\,8799 system, as described in Sect. \ref{subsec:results_real_data}.

\subsection{Pre-reduction}
\label{subsec:pre_reduction}

\begin{algorithm}[t!]
\DontPrintSemicolon
    \KwIn{\PACO reduced data $\{ \bm a_{t,\ell}, \bm b_{t,\ell} \}_{t=1:T, \ell=1:L} $.}
    \KwIn{Number $\{N_{\textup{orb}}$, $N_\textup{opt}\}$ of tested \& optimized orbits.}
    \KwIn{List of all discrete orbital elements to explore $\M \Gamma$.}
    \KwIn{Targeted confidence level $\rho$.}
    \KwIn{Interpolator function.}
    
    \vspace{1mm}
    
    \KwOut{Optimal orbit $\widehat{\bm\mu}$.} \vspace{0.5mm}
    \KwOut{Score $\CostFunc(\widehat{\bm\mu})$ of the optimal orbit.}
    \KwOut{Error bars $\widehat{\V \sigma}_{\V \mu}^{\text{CRLB}}$ or confidence intervals $\widehat{\V \sigma}_{\V \mu}^{\text{Pert}}$.}
    \KwOut{Memory-mapped file $s$ of best detection scores.}
    
    \hrule
    \medskip
    
    $\blacktriangleright$ \textbf{Step 1.} Test all orbits and save the \textit{best} ones. \vspace{0.5mm}
    
    \For{$i = 1$ \KwTo $N_{\textup{orb}}$}{
        $\CostFunc \leftarrow 0 $ \tcp*[r]{(cost initialization)} 
        Get $\V \mu_i$ from $\M \Gamma$ \tcp*[r]{(orbit selection)} 
        \For{$t = 1$ \KwTo $T$}{
        $\bm\theta \leftarrow \bm\theta_t(\bm\mu)$ \tcp*[r]{(sky to sensor projection)} 
            \For{$\ell = 1$ \KwTo $L$}{
            $a \leftarrow a_{t,\ell}(\bm\theta)$ \tcp*[r]{(interpolation, Sect. \ref{sec:interpolationStrat})}
            $b \leftarrow b_{t,\ell}(\bm\theta)$ \tcp*[r]{(interpolation, Sect. \ref{sec:interpolationStrat})}
            $\CostFunc \leftarrow \CostFunc + \max(b,0)^2/a$ \tcp*[r]{(cost, Eq.~\eqref{eq:costfuncC})} 
            } 
        }
        \If{$\CostFunc > \widehat{\mathcal{Q}}_{T \times L}(1-\rho)$}{\vspace{0.5mm}
        Append $[\CostFunc, n]$ to $s$ \tcp*[r]{(storing, Sect. \ref{sec:StoringTheData})}
        }
    }
    Sort $s$ on $\CostFunc$ by descending order
    
    \medskip
    
    $\blacktriangleright$ \textbf{Step 2.} Optimization of the $N_\textup{opt}$ \textit{best} on-grid orbits. \vspace{0.5mm}
    
    Get $j$ from line $1$ of $s$\; 
    Get $\V \mu_j$ from $\M \Gamma$ \tcp*[r]{(refined orbit initialization)} 
    $\widehat{\bm\mu}, \widehat{\CostFunc} \leftarrow$ VMLM-B($\bm\mu, \{ \bm a_{t,\ell}, \bm b_{t,\ell} \}_{t=1:T\,, \ell=1:L}$) \tcp*[r]{(Sect. \ref{sec:LocalOptiRefinement})}
            
    \For{$j = 2$ \KwTo $N_\textup{opt}$}{
            Get $j$ from line $j$ of $s$ \tcp*[r]{(index selection)} 
            Get $\V \mu_j$ from $\M \Gamma$ \tcp*[r]{(orbit selection)} \vspace{0.5mm}
            $\widehat{\bm\mu}_j, \widehat{\CostFunc}_j \leftarrow$ VMLM-B($\bm\mu_j, \{ \bm a_{t,\ell}, \bm b_{t,\ell} \}_{t=1:T\,, \ell=1:L}$) \tcp*[r]{(\ref{sec:LocalOptiRefinement})}
            \If{$\widehat{\CostFunc}_j > \widehat{\CostFunc}$}{
                $\widehat{\CostFunc} \leftarrow \widehat{\CostFunc}_j$ \tcp*[r]{(cost update)}
                $\widehat{\bm\mu} \leftarrow \widehat{\bm\mu}_j$ \tcp*[r]{(orbit update)}
            }
    } 
        
    \medskip
    
    $\blacktriangleright$ \textbf{Step 3.} Error estimation of the optimal solution. \vspace{0.5mm}
    
    \If{$\widehat{\CostFunc} > \widehat{\mathcal{Q}}_{T \times L}(1-\rho)$}{

    $\widehat{\bm\sigma}_{\bm\mu}^{(\textup{Pert})} \leftarrow $ CI on $\{\widetilde{ \, \bm\mu}_{(n)} \}_{n=1:N_p}$  \tcp*[r]{(Sect. \ref{sec:NumericalMethodErrors})}
    
    $\widehat{\bm\sigma}_{\bm\mu}^{(\textup{CRLB})} \leftarrow \sqrt{\textup{diag}(\Cov^{\text{(CRLB)}}\{ \widehat{\bm\mu} \})}$  \tcp*[r]{(Sect. \ref{sec:CramerRaoMethodError})}
        
    }
\caption{Pseudo-code of the \PACOME algorithm.} \label{alg:pseudoCodePacome}
\end{algorithm}

Raw observations were pre-processed with the pre-reduction and handling pipeline of the SPHERE consortium \citep{pavlov2008advanced}. Background, flat-field, bad pixels, registration, true north, wavelength, and astrometric calibrations are performed during this step. Additional custom steps implemented at the SPHERE Data Center \citep{delorme2017sphere} were also applied to refine the wavelength calibration, reduce the crosstalk, and improve the identification of bad pixels. We also designed a new strategy to refine the conventional frame centering implemented in the SPHERE Data Center, as described in the following.

All the datasets of this work were acquired in pupil-tracking mode, namely, at a given epoch the field rotates around the star from a frame to another (see Sect. \ref{sec:introduction}). In that context, having a precise knowledge of the rotation center is critical to maximizing the detection confidence and deriving accurate astro-photometric estimates of the putative sources in single-epoch datasets. In our case, the sought-for sources also orbit around the same rotation center defined by the star itself across epochs, so it is even more crucial to ensure that the rotation center is precisely known for all epochs. Unfortunately, it is not possible to use the central star to measure directly the rotation center because it is masked by the coronagraph. Conveniently, applying a waffle pattern to the deformable mirror while acquiring the observations creates (by diffraction) four replicas of the PSF so-called "satellite spots",  located around the rotation center at about $14\,\lambda/D$, where $\lambda$ is the wavelength and $D$ the diameter of the telescope \citep{langlois2012infrared}. The rotation center of all the frames of an A(S)DI sequence can thus be measured using the satellite spots, if their positions are precisely estimated. We re-addressed the previous numerical centering strategy that was used on the SPHERE Data Center by developing an algorithm that is faster and less sensitive to outliers (e.g., bad pixels and large stellar leakages). This leads to a better recentering procedure which improves the S/N of real sources (see Table~\ref{tab:gainSNRcenteringProcedure} in appendix for more details). Similarly to the work of \cite{Wang2014} for GEMINI/GPI, we implemented a new routine that fits 2D Gaussian functions to the satellite spots via a nonlinear least mean square constrained optimization method to estimate their subpixel locations. Based on these measurements, all the frames are shifted accordingly to match the measured rotation center to an arbitrary position satisfying the SPHERE Data Center and the \PACO pipeline requirements. Our centering method has been delivered and integrated in the SPHERE Data Center (see  Appendix \ref{sec:dataCent_satSpots} for more details and results). 

After this pre-reduction step, raw observations were assembled in calibrated A(S)DI datasets. Each pre-reduced IRDIS observation is composed of $L=2$ spectral channels that are processed both independently (i.e., akin to ADI sequences) and jointly (i.e., akin to ASDI sequences). Each pre-reduced IFS observation is composed of $L=39$ spectral channels that are  processed jointly. All pre-reduced datasets are first processed by the \PACO pipeline (see Sect. \ref{subsec:PACOreduction}), whose byproducts are then used by the proposed \PACOME algorithm, see Sects. \ref{subsec:results_ss_data} and \ref{subsec:results_real_data}.

\subsection{Reduction with \PACO}
\label{subsec:PACOreduction}

The calibrated A(S)DI datasets described in Sect. \ref{subsec:pre_reduction} were processed with the \PACO algorithm to produce, for each epoch, the $\bm a_{t,\ell}$ and $\bm b_{t,\ell}$ maps. These outputs serve as inputs of the \PACOME algorithm (see Sect.~\ref{subsec:direct_model_obs}). As discussed in Sect. \ref{subsec:asdi_formalism}, for the reduction of ASDI datasets, it is possible to enforce one or multiple spectral prior(s) $\V \gamma \in \mathbb{R}^{T L}$ to boost the S/N of detection of point-like sources having a similar spectra\footnote{It has been shown in \cite{Flasseur2020} that the S/N of detection is only marginally degraded in case of the prior spectra differ significantly from the true spectrum of the sources. In any case, the probability of false alarms remains controlled.}. For ASDI processing of IRDIS observations, we chose seven spectral priors for the source corresponding to slopes of $0.25$, $0.5$, $0.75$ and their opposites, namely, priors $(1,1)$, $(1, 0.25)$, $(0.25, 1)$, $(1, 0.5)$, $(0.5, 1)$, $(1, 0.75)$, $(0.75, 1)$. For the IFS observations, we chose 23 spectral priors that are representative of the variety of potential exoplanets spectra, as described in \cite{Chomez2023}, for the mode of selection of these spectral priors. The data are processed independently for each spectral prior (but jointly on the whole spectral range). We give equal importance to all the spectral priors used in this work. The only criterion to decide in favor of the presence of a planet is the combined multi-epoch score, whatever the underlying spectral prior, since no other information is known a priori.

\subsection{Analysis of semi-synthetic data}
\label{subsec:results_ss_data}

\subsubsection{Description of the datasets and of the injected orbits}

\begin{table*}[t!]
                \centering 
        \caption{Summary of the ground truth fluxes, S/N and separations of the four injected sources (1), (2), (3), and (4) at all fake epochs.}
                \begin{tiny} \begin{tabular}{c||cccc|cccc||cccc|cccc||cccc}
                \toprule
                \multirow{3}{*}{\textbf{Epoch}} & \multicolumn{8}{c||}{$\bm{\alpha^{\textup{gt}}_{t,\ell}}$ $\left(\times 10^{-6}\right)$} & \multicolumn{8}{c||}{$\bm{\mathcal{S}/\mathcal{N}^\textup{gt}_{t,\ell}}$} & \multicolumn{4}{c}{\multirow{2}{*}{\makecell{\textbf{Separation} \textbf{('')}}}}\\
                & \multicolumn{4}{c}{$\ell=1$} & \multicolumn{4}{c||}{$\ell=2$} & \multicolumn{4}{c}{$\ell=1$} & \multicolumn{4}{c||}{$\ell=2$} & & & & \\
                & (1) & (2) & (3) & (4) & (1) & (2) & (3) & (4) & (1) & (2) & (3) & (4) & (1) & (2) & (3) & (4) & (1) & (2) & (3) & (4) \\ \midrule
                1963.43 & 1.5 & 2.1 & 3.5 & 7.1 & 1.1 & 0.9 & 2.1 & 7.0 & 2.6 & 3.1 & 2.7 & 2.7 & 2.2 & 1.5 & 1.7 & 2.6 & 0.7 & 0.5 & 0.3 & 0.2\\ 
        1970.11 & 1.6 & 1.9 & 4.0 & 9.6 & 0.9 & 1.0 & 4.1 & 9.8 & 2.9 & 2.1 & 2.7 & 2.9 & 1.7 & 1.4 & 2.8 & 3.0 & 0.7 & 0.4 & 0.3 & 0.2\\ 
        1976.65 & 1.6 & 1.7 & 2.6 & 14.2 & 1.9 & 1.5 & 2.8 & 15.1 & 2.8 & 2.8 & 1.8 & 1.8 & 3.3 & 2.7 & 1.9 & 2.0 & 0.6 & 0.5 & 0.2 & 0.2\\ 
        1981.99 & 2.3 & 2.2 & 3.1 & 8.5 & 1.2 & 1.6 & 3.4 & 9.6 & 4.0 & 3.3 & 3.1 & 2.4 & 2.3 & 2.9 & 3.6 & 2.7 & 0.6 & 0.5 & 0.3 & 0.2\\ 
        1986.05 & 1.7 & 2.8 & 1.3 & 6.1 & 1.7 & 1.5 & 4.2 & 7.8 & 3.5 & 2.9 & 0.5 & 1.5 & 3.8 & 1.6 & 1.6 & 1.9 & 0.7 & 0.3 & 0.3 & 0.2\\ 
        1997.76 & 1.7 & 1.2 & 3.3 & 10.1 & 1.0 & 1.3 & 1.7 & 10.8 & 4.0 & 1.9 & 2.6 & 2.5 & 2.2 & 2.4 & 1.6 & 2.8 & 1.0 & 0.6 & 0.3 & 0.2\\ 
        2002.99 & 2.2 & 1.1 & 3.2 & 10.3 & 1.4 & 1.0 & 3.9 & 11.3 & 7.7 & 1.8 & 1.7 & 3.3 & 5.1 & 1.9 & 2.2 & 3.2 & 1.1 & 0.5 & 0.2 & 0.2\\ 
        2007.15 & 2.0 & 2.1 & 3.0 & 6.7 & 1.4 & 1.4 & 2.6 & 8.2 & 7.6 & 2.5 & 3.0 & 1.5 & 5.9 & 1.7 & 3.0 & 1.7 & 1.2 & 0.4 & 0.3 & 0.2\\ 
        2022.15 & 1.8 & 1.0 & 2.8 & 8.5 & 1.2 & 1.7 & 2.4 & 9.9 & 6.1 & 1.1 & 1.6 & 1.2 & 3.8 & 1.9 & 1.4 & 1.4 & 1.1 & 0.4 & 0.3 & 0.2\\ 
        \midrule
        \textbf{Mean value} & 1.8 & 1.8 & 3.0 & 9.0 & 1.3 & 1.3 & 3.0 & 10.0 & 4.6 & 2.4 & 2.2 & 2.2 & 3.4 & 2.0 & 2.2 & 2.4 & 0.9 & 0.5 & 0.3 & 0.2\\
                \bottomrule
                \end{tabular}
                \end{tiny} 
                \label{tab:injectionInformation}
  \tablefoot{Epoch dates are created artificially to match the considered orbits.}
\end{table*}

To quantify the efficiency of \PACOME, we tested it on a semi-synthetic benchmark. We built "fake epochs" by resorting to the injection of synthetic sources on a dataset of HR\,8799 observed with IRDIS in 2016-11-18 (see Table \ref{tab:dataset_logs} for the observations logs). By selecting individual frames from this reference dataset, we built nine fake epochs totalling $5.12$ hours of observation time and spanning over $59$ years. The ADI rotation angles of each fake epoch were chosen in the opposite direction of the original ones, so that signals from real exoplanets were not constructively co-added via the derotation and stacking of the individual frames. The total field of view rotation was set to $\Delta_{\textup{par}} = \qty{67.34}{\degree}$ for each epoch. 

We injected four sources, denoted as (1), (2), (3), and (4) in the following, on known orbits at the corresponding epochs via a \PACO subroutine using the measured off-axis PSF of the reference cube. The injected fluxes of the sources were chosen such that their single-epoch S/N is lower than five on average. A summary of the fluxes, S/Ns, and separations of the injected sources at all epochs is given in Table~\ref{tab:injectionInformation}. 
After the source injections, we reduced the nine semi-synthetic datasets with \PACO. 

\subsubsection{Results of the search}
\label{subsec:ResultsSearchInjections}

To detect multiple sources, we adopted a simple greedy strategy: finding the brightest of the four planets with a first search, estimating its orbit, masking its footprint in \PACOs byproducts, $\bm b_{t,\ell}$, restarting the search for next brighter source, and so on. A summary of the explored search space for the four sought-for sources is given in Table~\ref{tab:searchSpaceSynthetic}. We used the best cubic cardinal spline derived in Sect.~\ref{sec:interpolationStrat} for the interpolation process. In total, $N_{\text{orb}} \approx 17 \times 10^{10}$ orbits were tested, taking $44$ hours of CPU time on our local computer using $12$ threads. For each search, the first $N_\textup{opt}=10^3$ best on-grid orbits were optimized locally and the best of all was retained as the most plausible orbital candidates. The volume of the data is fixed and totals about $10^6$ pixels per epoch for IRDIS. Hence, the total number of possible multi-epoch combinations of all pixels in our data is $10^{6\times T \times L} = 10^{54}$ compared to which the number of explored orbits is completely negligible. For this reason, we put the detection confidence in perspective with the number of explored orbits $N_{\text{orb}}$ and set it at a high conservative value ($\rho = 0.1 / N_{\text{orb}} = 2.33 \times 10^{-12}$) so that it is expected to experience one combined detection score above the multi-epoch detection threshold in the background $10\%$ of the time with different \PACOME runs. This threshold is arbitrary and it is left to the user to choose; in any case the empirical false alarm rate is controlled at the prescribed value.
Given the $T=9$ epochs and $L=2$ spectral channels, the corrected empirical multi-epoch detection threshold at this confidence level is $\widehat{\mathcal{Q}}^{\text{corr}}_{18}(1-2.33 \times 10^{-12}) \approx 48.3$. 

\begin{table}[t!]
                \centering 
        \caption{Orbital elements search space for the four injected sources (1), (2), (3), and (4).}
                \begin{tabular}{ccccc}
                \toprule
                \textbf{Elem.} & \textbf{Unit} & \textbf{Min. val.} & \textbf{Max. val.} & \textbf{Length} \\
                \midrule
                $a$ & mas & \makecell{(1): 852\\(2): 300\\(3): 150\\(4): 100} & \makecell{(1): 1100\\(2): 790\\(3): 500\\(4): 350} & 25 \\
                $e$ & - & 0 & 0.5 & 11 \\
                $i$ & deg & 0 & 180 & 50 \\
                $\tau$ & - & 0 & 1 & 50 \\
                $\omega$ & deg & 0 & 360 & 50 \\
                $\Omega$ & deg & 0 & 360 & 50 \\
                $K$ & mas$^3$/yr$^2$ & 107666.7 & 122359.5 & 25 \\
                \bottomrule
                \end{tabular}
                \label{tab:searchSpaceSynthetic}
  \tablefoot{The "field length" represents the number of tested values per orbital element.}
\end{table}

\begin{figure}[t!]
    \centering
    \includegraphics[width=0.5\textwidth]{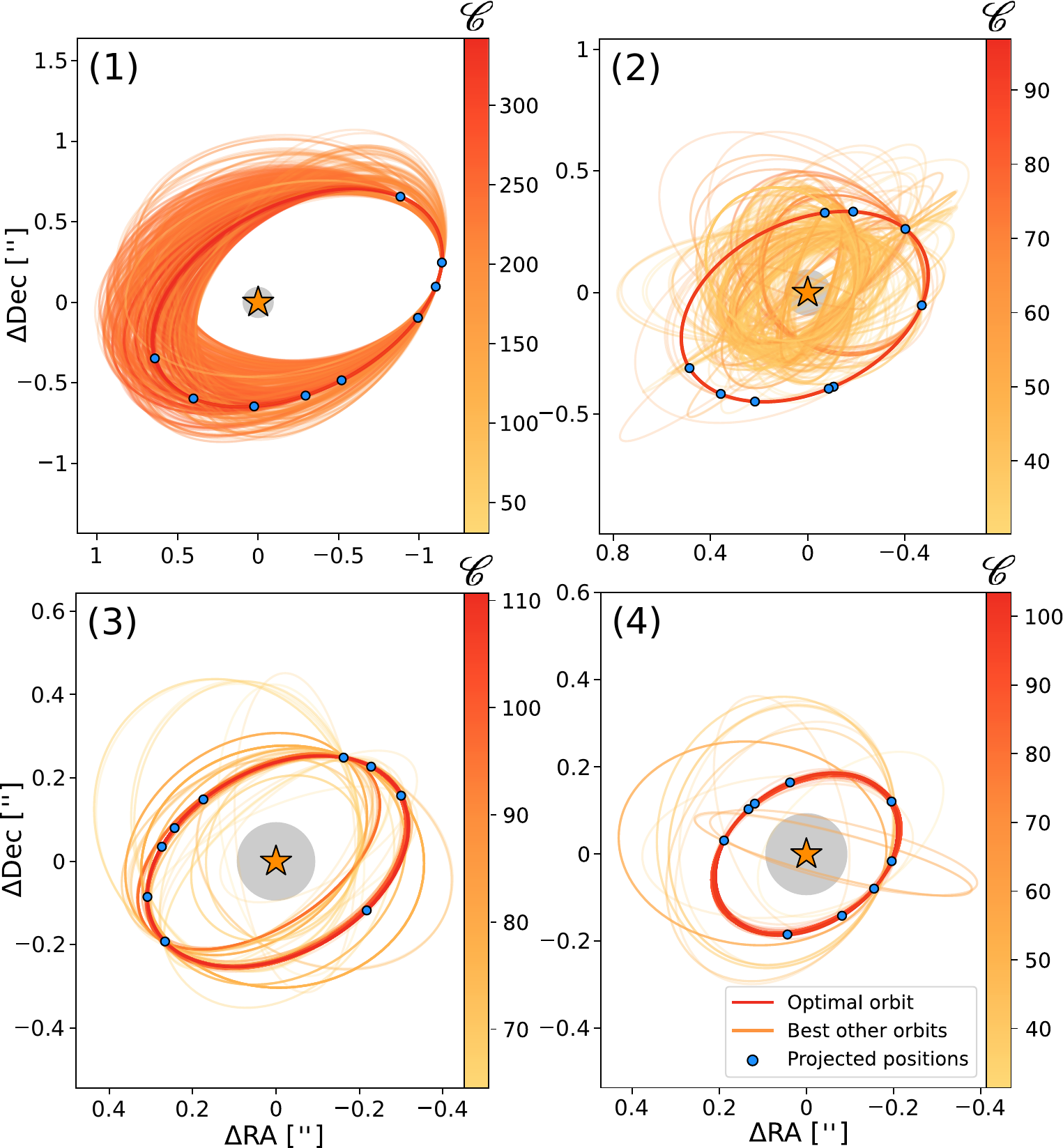}
    \caption{Optimal orbits found for each of the four injected sources along with the best $10^3$ other re-optimized orbits. The red thick line shows the retained optimal orbit, see Table \ref{tab:optimalOrbitalElementsAndInjections}. The blue dots are the projected positions of the signal along the optimal orbit at all epochs and the grey circular area represents the largest coronagraphic mask.}
    \label{fig:projOrbitInjection}
\end{figure}

\begin{figure}[t!]
    \centering
    \includegraphics[width=0.5\textwidth]{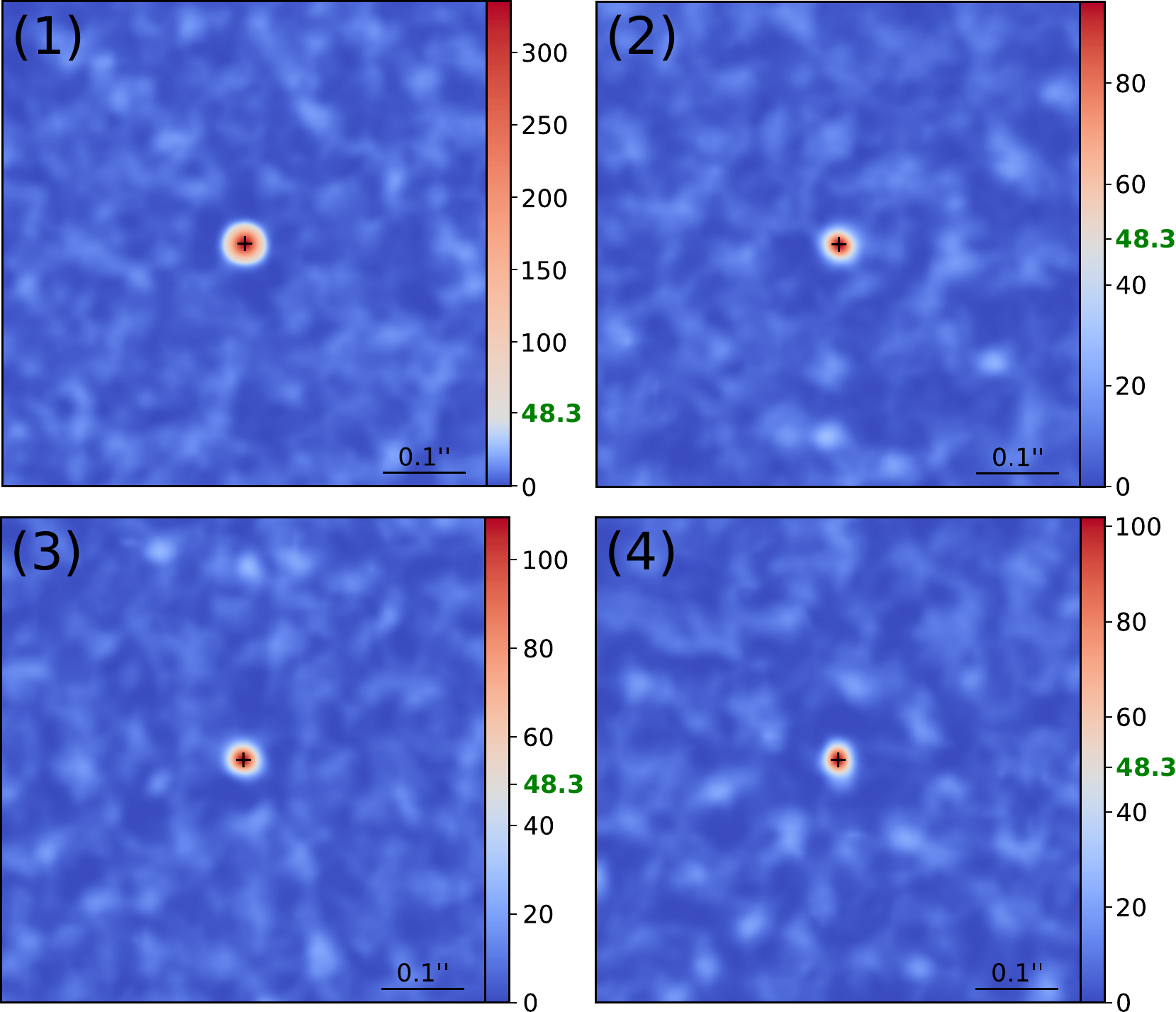}
    \caption{Cost function maps around the optimal solution estimated by \PACOME for each injected source in a ROI of $50$ pixels wide sampled with four nodes per pixel. The value of the empirical multi-epoch detection threshold $\widehat{\mathcal{Q}}_{18}(1-2.33 \times 10^{-12}) \approx 48.3$ is highlighted in green in the color bar.}
    \label{fig:costFuncMapInjection}
\end{figure}

The orbital elements of the best source candidates are given in Table~\ref{tab:optimalOrbitalElementsAndInjections} and compared to the injected ones. The upper and lower confidence intervals for the orbital elements shown in this table were computed at a confidence level of $95\%$ via the perturbation method with $N_p=10^4$. To evaluate the similarity between the 2D projected positions of the retrieved orbits $\widehat{\bm\mu}$ and the (ground truth) injected orbits $\bm\mu^\textup{gt}$, we compute their root mean square distance (RMSD) as:
\begin{equation}
    \textup{RMSD}(\widehat{\bm\mu},\bm\mu^\textup{gt}) = \sqrt{ \frac{1}{T} \sum^{T}_{t=1}  \norm{ \bm\theta_t(\widehat{\bm\mu})-\bm\theta_t(\bm\mu^\textup{gt}) }^2 }\,.
    \label{eq:rmsd_gt}
\end{equation}
The RMSD formulation is a convenient tool that can be used to assess how good the algorithm is when it comes to fit the original injected orbit. The projections of the optimal retrieved orbits of Table~\ref{tab:optimalOrbitalElementsAndInjections} along with the best other $10^3$ optimized orbits are shown in Fig.~\ref{fig:projOrbitInjection}. In addition, we plot the cost function maps of the criterion centered on each of these optimal solutions in regions of interest (ROIs) of 60 pixels wide sampled with four nodes per pixel in Fig.~\ref{fig:costFuncMapInjection}. The color maps are specifically centered on the corrected multi-epoch detection threshold $\widehat{\mathcal{Q}}^{\text{corr}}_{18}(1-\rho)$ such that any signal above the limit (in dark red) can be interpreted as statistically significant regarding the set detection confidence $\rho = 2.33 \times 10^{-12}$.

\begin{table*}[t!]
                \centering 
        \caption{Orbital elements and multi-epoch scores of the four injected orbits and the optimal ones found with \PACOME.} 
                \begin{tiny} \begin{tabular}{cc||cc|cc|cc|cc}
                \toprule
                \multirow{2}{*}{\textbf{Elem.}} & \multirow{2}{*}{\textbf{Unit}} & \multicolumn{2}{c|}{\textbf{Source (1)}} & \multicolumn{2}{c|}{\textbf{Source (2)}} & \multicolumn{2}{c|}{\textbf{Source (3)}} & \multicolumn{2}{c}{\textbf{Source (4)}} \\
                 & & \textit{Estimated} & \textit{Injected} & \textit{Estimated} & \textit{Injected} & \textit{Estimated} & \textit{Injected} & \textit{Estimated} & \textit{Injected} \\ \midrule
                    $a$ & mas & $976.26_{-4.09}^{+5.22} $ & $ 976.23 $ & $ 542.71_{-6.24}^{+5.86} $ & $ 543.75 $ & $ 334.88_{-3.89}^{+3.62} $ & $ 331.66 $ & $ 229.68_{-5.04}^{+3.86} $ & $ 229.01 $  \\[0.1cm]
                $e$ & - & $ 0.28_{-0.01}^{+0.01} $ & $ 0.28 $ & $ 0.15_{-0.01}^{+0.01} $ & $ 0.15 $ & $ 0.01_{-0.01}^{+0.02} $ & $ 0.01 $ & $ 0.00_{-0.00}^{+0.02} $ & $ 0.0 $  \\[0.1cm]
                $i$ & deg & $ 52.19_{-0.65}^{+0.63} $ & $ 52.19 $ & $ 49.86_{-1.57}^{+1.51} $ & $ 50.45 $ & $ 48.19_{-2.38}^{+2.51} $ & $ 48.13 $ & $ 44.62_{-3.56}^{+2.96} $ & $ 45.33 $  \\[0.1cm]
                $t_0$ & yr & $ 1958.88_{-0.42}^{+0.42} $ & $ 1958.90 $ & $ 1986.05_{-0.30}^{+0.34} $ & $ 1986.02 $ & $ 2004.99_{-1.87}^{+3.68} $ & $ 2004.52 $ & $ 2011.71_{-1.29}^{+7.43}  $ & $ 2014.27 $  \\[0.1cm]
                $\omega$ & deg & $ 146.51_{-1.86}^{+2.41} $ & $ 146.12 $ & $  63.93_{-2.96}^{+2.59} $ & $ 63.39 $ & $ 145.79_{-43.61}^{+61.60} $ & $ 134.21 $ & $ 259.95_{-112.67}^{+46.78} $ & $ 350.32 $  \\[0.1cm]
                $\Omega$ & deg & $ 298.92_{-0.96}^{+0.79} $ & $ 298.99 $ & $ 295.17_{-1.90}^{+1.96} $ & $ 295.76 $ & $ 298.62_{-2.77}^{+2.94} $ & $ 299.11 $ & $ 302.97_{-5.91}^{+5.93} $ & $ 302.02 $  \\[0.1cm]
                $K \, (\times 10^4)$ & mas$^3$/yr$^2$ & $ 11.44_{-0.27}^{+0.25} $ & $ 11.50 $ & $ 11.48_{-0.39}^{+0.38} $ & $ 11.50 $ & $ 11.86_{-0.41}^{+0.38} $ & $ 11.50 $ & $ 11.62_{-0.73}^{+0.62} $ & $ 11.50 $  \\[0.1cm]
                \midrule
                \multicolumn{2}{c||}{\textbf{RMSD} [pix]} & \multicolumn{2}{c|}{0.12}  & \multicolumn{2}{c|}{0.34}  & \multicolumn{2}{c|}{0.61}  & \multicolumn{2}{c}{0.20} \\
                \multicolumn{2}{c||}{\textbf{min} $ \SNR_{t,\ell}$} & 1.76 & 1.74  & 1.05 & 1.12  & 0.74 & 0.47 & 1.05 & 1.16  \\ 
                \multicolumn{2}{c||}{\textbf{mean} $\SNR_{t,\ell}$} & 3.98 & 3.97 & 2.22 & 2.18 & 2.37 & 2.19 & 2.28 &  2.28\\ 
                \multicolumn{2}{c||}{\textbf{max} $\SNR_{t,\ell}$} & 7.67 & 7.69 & 3.50 & 3.29 & 3.65 & 3.64 & 3.28 & 3.30  \\ 
                \multicolumn{2}{c||}{\textbf{Multi-epoch} $\SNR$} & 18.50 & 18.46  & 9.84 & 9.66  & 10.52 & 9.86  & 10.17 & 10.12  \\ 
                \multicolumn{2}{c||}{\textbf{Criterion} $ \CostFunc $}  & 342.10 & 340.58  & 96.90 & 93.39  & 110.71 & 97.12  & 103.46 & 102.39  \\      
                \multicolumn{2}{c||}{\textbf{Threshold} $ \widehat{\mathcal{Q}}^{\text{corr}}_{18}(1-\rho) $}  & \multicolumn{2}{c|}{48.3}  & \multicolumn{2}{c|}{48.3}  & \multicolumn{2}{c|}{48.3}  & \multicolumn{2}{c}{48.3}  \\ 
                \bottomrule
                \end{tabular}
                \end{tiny} 
                \label{tab:optimalOrbitalElementsAndInjections}
        \tablefoot{Because of their degeneracy, any pair, $(\omega, \Omega),$ is perfectly equivalent to $(\omega+\pi, \Omega+\pi)$. For better interpretability, the epoch of periapsis passage $t_0$ is shown instead of $\tau$. The RMSD between the projected positions of the injected sources and the ones retrieved are computed with Eq.~\eqref{eq:rmsd_gt}. The uncertainties were estimated via the perturbation method described in Sect. \ref{sec:NumericalMethodErrors} with $N_p=10^4$. The central values are the optimal orbital elements found by PACOME and the lower and upper bounds correspond to the bounds of the $95\%$ confidence interval. The corrected multi-epoch detection threshold $\widehat{\mathcal{Q}}^{\text{corr}}_{18}(1-\rho) $ associated to a confidence level of $\rho = 2.33 \times 10^{-12}$ is to compare to the criterion score $\CostFunc$.}
\end{table*}

The retrieved orbits agree very well with the ground truth (Table~\ref{tab:optimalOrbitalElementsAndInjections}) except for the argument of periapsis, $\omega,$ of  source (4), which is significantly different and is not captured by our uncertainties (underestimated because they were performed locally; see Sect.~\ref{subsec:ComparisonCRLBsPerturbations}). However, the projected solution found by \PACOME is still very close to the ground truth as it gives almost the same positions on the detector at the given epochs ($\textup{RMSD} = 0.2$ pixels). Remarkably, for each of the four sources, the RMSD of the projected positions is always lower than 3/4 of a pixel, which indicates the closeness between the solutions retrieved by \PACOME and the projections of the ground truth orbits.
We also notice in Fig.~\ref{fig:projOrbitInjection} a very different variety of orbits among the $10^3$ best optimized orbits for sources (2), (3), and (4). Most of the atypical orbits would correspond to putative sources with a multi-epoch $\SNR$ lower than $3.3$. In practice, they correspond to the recombination of residual noise, which is statistically expected in this low S/N regime.

\begin{figure}[ht!]
    \centering
    \includegraphics[width=0.9\linewidth]{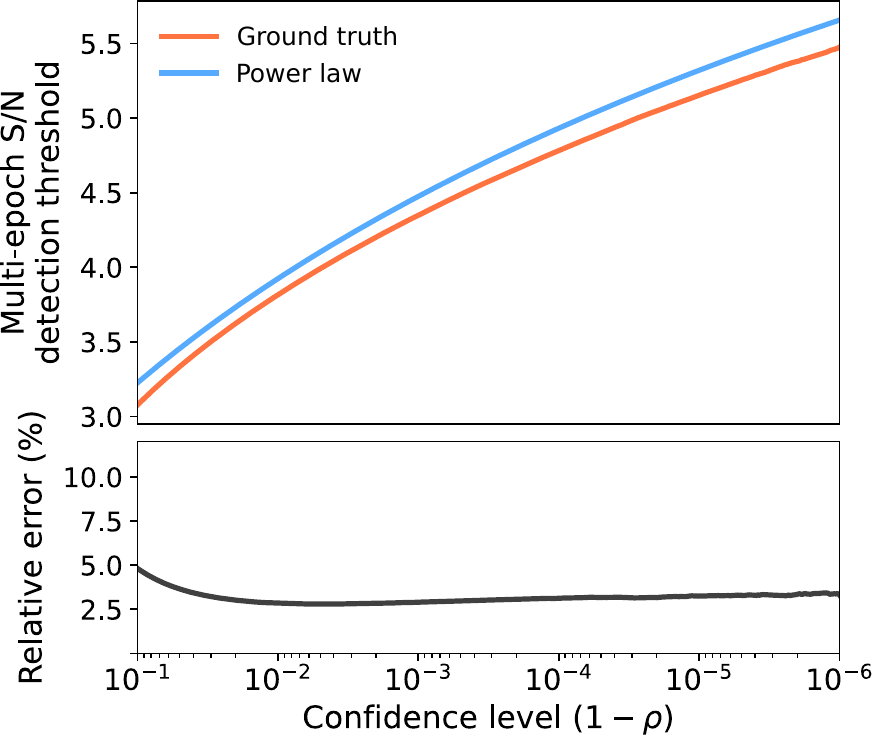}
    \caption{Comparison between the multi-epoch detection threshold estimated on the distribution out of the data in the absence of source and the one approximated by the power law of Eq.~\eqref{eq:power_law_criterion_quantile}.}
    \label{fig:comparison_injections_distribTheo_VS_distribEmpirical}
\end{figure}

Apart from the four injected sources effectively retrieved with \PACOME, no other signals were found above the chosen multi-epoch detection threshold, as statistically expected by our statistical model. To validate the reliability of our detection metric, we masked the four retrieved sources and re-explored the orbital parameters space on the masked data to build the effective distribution of the multi-epoch S/N in the absence of source. In total, $2 \times 10^{8}$  orbits were drawn uniformly in the parameter space, with Table~\ref{tab:searchSpaceSynthetic} displaying the allowed ranges. In Fig.~\ref{fig:comparison_injections_distribTheo_VS_distribEmpirical}, the multi-epoch S/N detection thresholds associated to this distribution (i.e., the ground truth) are compared to the actually used detection thresholds estimated with the power law of Eq.~\eqref{eq:power_law_criterion_quantile}. The absolute relative error between the two estimations is $3.2\%$ on average and always below $5\%$, over the whole range $\rho \in [10^{-6},10^{-1}],$  while the corresponding RMSE is $0.14$ in terms of the S/N. Given these errors are relatively low, this indicates that the detection metric is well controlled even though the threshold estimated via the power law (used in this paper) is systematically (mildly) over-estimated, which is conservative. As a result, we experience slightly fewer false alarms than theoretically expected for a given threshold.

As expected (and more precisely discussed in Sect. \ref{subsec:ComparisonCRLBsPerturbations}), the uncertainties on the orbital elements are estimated locally. As a consequence, they are optimistic and smaller than what would be obtained using more classical orbit-fitting methods in the literature that effectively explore the full parameter space \citep{Ford2005}.

The criterion values for the optimal orbital solutions found by \PACOME (Table~\ref{tab:optimalOrbitalElementsAndInjections}) are all significantly higher than the empirical multi-epoch detection limit at $\rho = 2.33 \times 10^{-12}$ (by a factor of at least 2), while the vast majority of the point-like sources remain undetected in each individual epoch ($\SNR_{t,\ell} < 5$ by applying only the \PACO algorithm; see Table~\ref{tab:injectionInformation} and Appendix \ref{sec:details_ind_snr} for a local view of $\SNR_{t,\ell}$ maps). In particular, the mean mono-epoch $\SNR_{t,\ell}$ of the optimal solutions (averaged over the $T=9$ epochs and $L=2$ spectral channels) are $4.0$, $2.2$, $2.4$, $2.3$ and their multi-epoch $\SNR$ are $18.5$, $9.9$, $10.5$, $10.2,$ from source (1) to (4), respectively. This means that switching from mono-epoch to multi-epoch yields a mean S/N gain of about $4.5,$ which behaves (as theoretically expected) as the square root of the number of degrees of freedom ($\sqrt{T \times L} = \sqrt{18} \approx 4.2$). This illustrates a key property of \PACOME: it is indeed able to combine optimally the source signals of single-epoch datasets. 
For completeness, we notice in practice that the values of the criterion associated with the optimal orbits found by \PACOME are very slightly larger than the theoretically expected ones for the injected orbits. This is due to the presence of noise in the data. Indeed, the algorithm maximizes the signal around the optimal positions found and, thus, it captures the random noise realizations that co-add positively with the signal. As a result, the solutions have systematically slightly stronger criterion values than the injections and the RMSD between them are not perfectly $0$.  
Finally, Fig.~\ref{fig:costFuncMapInjection} shows the cost function maps around the optimal solution estimated by \PACOME. All signals in the field that are not at the source positions combine weakly (from blue to white colors, below the set confidence level at $\rho = 2.33 \times 10^{-12}$).
However, because the sought-for signals are very faint and that the hypothesis of independent spectral channels is not perfectly verified, we sometimes recombine noise or speckles in the background. We show in Sect.~\ref{subsection:takingSpectralCorrelIntoAccount} that explicitly accounting for the spectral correlations reduces this side effect. In any case, the optimal solution found by \PACOME still matches its best with the ground truth. It is worth noticing that the method is very efficient at recovering injected sources that would have remained undetected otherwise. In particular, those are (2), (3), and (4), whose S/N values are below $5$ for all individual epochs.

\subsection{Analysis of HR\,8799 archival data}
\label{subsec:results_real_data}

\subsubsection{Retrieving the four known exoplanets b, c, d, and e}
\label{subsec:resultsRealData_HR8799bcde}

\begin{table}[t!]
                \centering 
        \caption{Orbital elements search space for the four known planets of HR\,8799 b, c, d, and e.}
                \begin{tabular}{ccccc}
                \toprule
                \textbf{Elem.} & \textbf{Unit} & \textbf{Min. val.} & \textbf{Max. val.} & \textbf{Length} \\
                \midrule
                $a$ & mas & 240 & 1850 & 129  \\
                $e$ & - & 0 & 0.4 & 9 \\
                $i$ & deg & 0 & 180 & 15 \\
                $\tau$ & - & 0 & 1 & 35 \\
                $\omega$ & deg & 0 & 360 & 31 \\
                $\Omega$ & deg & 0 & 360 & 31 \\
                $K$ & mas$^3$/yr$^2$ & 18920 & 23403 & 35 \\
                \bottomrule
                \end{tabular}
                \label{tab:searchSpaceHR8799bcde}
  \tablefoot{The "field length" represents the number of tested values per orbital element. Kepler's constant is constrained assuming a stellar mass of $M_\star = 1.47^{+0.12}_{-0.17} \, M_\odot$ \citep{sepulveda2022dynamical} and parallax of $\pi = 24.462 \pm 0.046$\,mas \citep{GaiaEDR3}.}
\end{table}

In this section, we demonstrate the efficiency of the method on real A(S)DI data of HR\,8799 acquired with SPHERE-IRDIS. The four exoplanets have already been detected in each of the individual epochs and spectral channels, hence, the following results show the ability of the algorithm to find matching orbits, rather than to (re)detect the sources.

The orbital elements search space is summarized in Table~\ref{tab:searchSpaceHR8799bcde}. It was deliberately chosen to cover a wide field of view encompassing all four known exoplanets. The cost function was evaluated for $20.5 \times 10^9$ orbits in total. With this setting, the computation took about $22$ hours on 12 cores.

\begin{figure}[t!]
    \centering
    \includegraphics[width=0.48\textwidth]{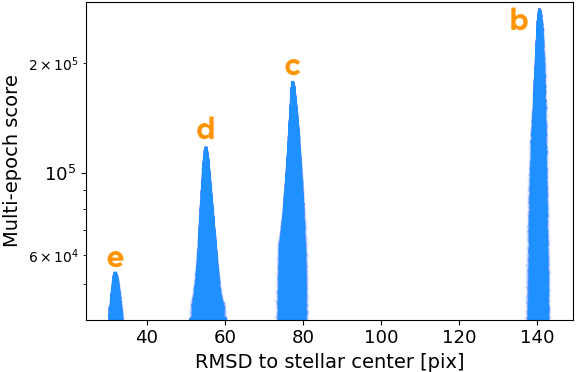}
    \caption{Distribution of the explored orbits with respect to their cost function scores and their RMSD to the center of the images. Each blue dot represents an orbit.}
    \label{fig:CvsRMSDHR8799}
\end{figure}

Considering the distance of the four exoplanets to the central star, the temporal coverage of the data allows to cover only a small portion of their orbit. The respective orbits of the four exoplanets are thus very poorly constrained. In order to disentangle the orbits from each other, we computed their RMSD to the center of the image (thus its pixelic distance to the star), with $\bm \theta_0$ as the position of the center of the images:
\begin{equation}
    \textup{RMSD}(\widehat{\bm\mu}) = \sqrt{ \frac{1}{T} \sum^{T}_{t=1}  \norm{ \bm\theta_t(\widehat{\bm\mu})-\bm \theta_0} }\,.
    \label{eq:rmsd_no_gt}
\end{equation}
This metric is a re-definition of the distance given in Eq.~(\ref{eq:rmsd_gt}), by replacing the (unknown) ground truth $\bm\theta_t(\bm\mu^\textup{gt})$ with $\bm \theta_0$. We show in Fig.~\ref{fig:CvsRMSDHR8799} the distribution of the explored orbits with respect to their cost function and RMSD distance to the star. We distinguish four very sharp peaks at $140$, $77$, $56$ and $32$ pixels from the center. These peaks correspond to the orbits whose 2D projections at all times fall on (or very close to) the effective positions of the exoplanets HR\,8799 b, c, d, and e (from right to left, respectively).

\begin{figure}[t!]
    \centering
    \includegraphics[width=0.45\textwidth]{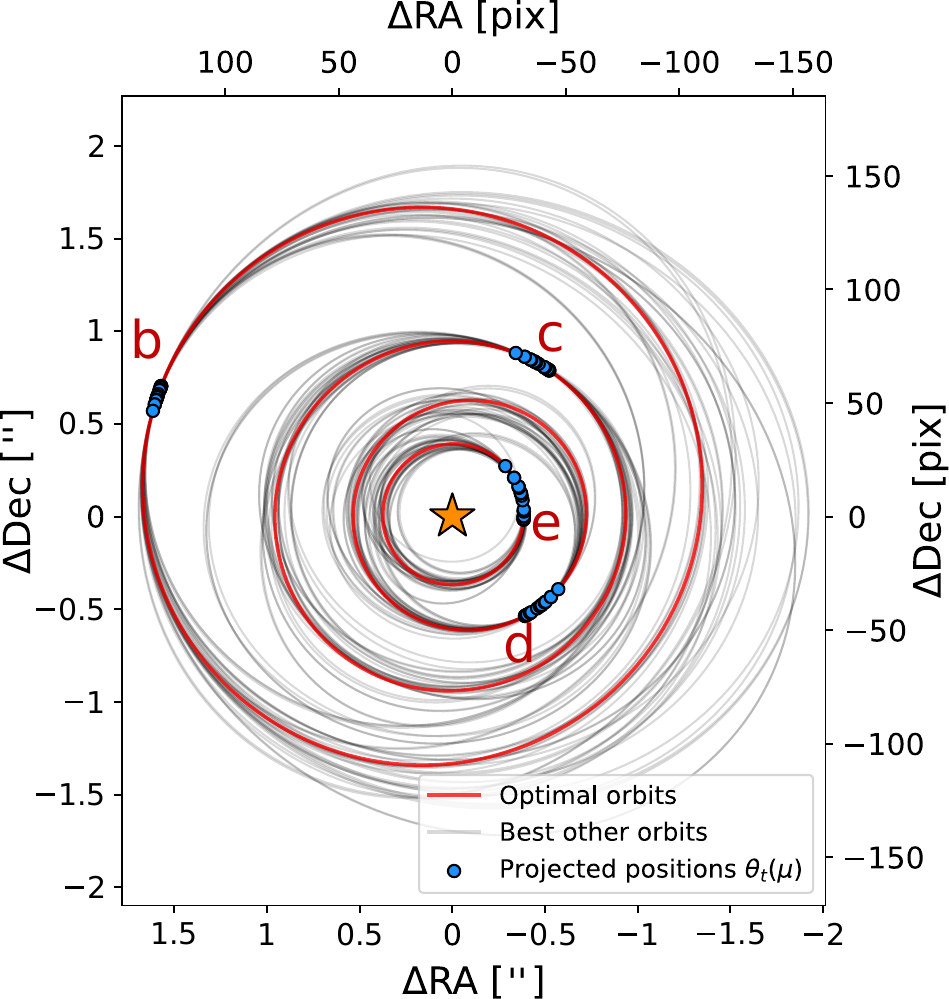}
    \caption{All 37 optimized orbit combinations of the search on the HR\,8799 system satisfying the constraints of coplanarity, near resonance, and identical stellar mass. The best combination is shown in red whereas the 36 others are plotted in grey. Blue dots represent the 2D projected positions of each source at all epochs.}
    \label{fig:all_36_optimal_orbits_coplanar_and_resonance}
\end{figure}

Concerning the orbits of the four planets, \cite{Wang2018} and \cite{Gozdziewski2020} found that a coplanar configuration near the 1$:$2$:$4$:$8 resonance produces orders of magnitude more stable orbits than any other scenario (i.e., orbiting with the same inclination $i$ and longitude of the ascending node $\Omega$). This can be used to constrain the optimization process of each orbit. We considered $\omega$, $\Omega,$ and $K$ to be fixed for each of the four exoplanets (i.e., coplanarity and orbiting the same central star) and therefore we carried out a joint search in the four peaks for the combination of four orbits that satisfy strict coplanarity (i.e., same $i$ and $\Omega$), common host star (i.e., same $K$), and near orbital resonance (i.e., 1:2:4:8$\pm0.25$). A total of 37 orbit quadruplets were found and are shown in Fig.~\ref{fig:all_36_optimal_orbits_coplanar_and_resonance}. For each, the quadratic sum of the individual cost scores was optimized under the constrains described above and the best one was retained (highlighted in red).

\begin{table*}[t!]
                \centering 
        \caption{Orbital elements and multi-epoch scores of HR\,8799 b, c, d, and e found with \PACOME.}
            \begin{tabular}{cc||c|c|c|c}
                \toprule
                \textbf{Elem.} & \textbf{Unit} & \textbf{HR\,8799 b} & \textbf{HR\,8799 c} & \textbf{HR\,8799 d} & \textbf{HR\,8799 e} \\ \midrule
                $a$ & mas  & $ 1525.46_{-134.51}^{+31.34} $ & $ 950.96_{-5.42}^{+24.19} $ & $ 632.6_{-82.29}^{-1.01} $ & $ 382.64_{-1.61}^{+193.82} $ \\[0.1cm]
        $e$ & -  & $  0.15_{-0.04}^{+0.09} $ & $ 0.01_{-0.01}^{+0.04} $ & $ 0.15_{+0.02}^{+0.12} $ & $ 0.04_{-0.03}^{+0.30} $ \\[0.1cm]
        $i$ & deg  & $ 8.87_{-8.87}^{+1.79} $ & $ 8.87_{+3.52}^{+17.39} $ & $ 8.87_{-8.87}^{+4.23} $ & $ 8.87_{+2.58}^{+25.56} $ \\[0.1cm]
        $t_0$ & yr  & $ 1772.63_{-176.49}^{+242.18} $ & $ 1972.65_{-115.54}^{+33.25} $ & $ 1981.31_{-365.69}^{+32.96} $ & $ 1999.35_{-28.43}^{+12.00} $ \\[0.1cm]
        $\omega$ & deg  & $ 278.68_{-115.36}^{+46.41} $ & $ 306.26_{-224.46}^{+27.79} $ & $ 147.97_{-103.54}^{+95.13} $ & $ 213.46_{-164.76}^{+138.63} $ \\[0.1cm]
        $\Omega$ & deg  & $ 305.92_{-231.81}^{+16.55} $ & $ 305.92_{+0.89}^{+25.42} $ & $ 305.92_{-267.3}^{+1.97} $ & $ 305.92_{-302.49}^{+48.70} $ \\[0.1cm]
        $K \, (\times 10^4)$ & mas$^3$/yr$^2$  & $ 2.02_{-0.00}^{+0.32} $ & $ 2.02_{-0.00}^{+0.32} $ & $ 2.02_{-0.00}^{+0.32} $ & $ 2.02_{-0.13}^{-0.00} $ \\[0.1cm]
        $ P $ & yr  & $ 418.81_{-79.70}^{+12.97} $ & $ 206.14_{-16.08}^{+7.95} $ & $ 111.84_{-27.46}^{-0.26} $ & $ 52.61_{-0.33}^{+48.01} $ \\[0.1cm] 
        \midrule
        \multicolumn{2}{c||}{\textbf{min} $\SNR_{t,\ell}$} & 11.22 & 6.19 & 4.37 & 2.22  \\
        \multicolumn{2}{c||}{\textbf{mean} $\SNR_{t,\ell}$} & 56.96 & 44.97 & 35.87 & 23.75 \\
        \multicolumn{2}{c||}{\textbf{max} $\SNR_{t,\ell}$} & 176.08 & 136.66 & 107.39 & 77.84 \\
        \multicolumn{2}{c||}{\textbf{multi-epoch} $\SNR$} & 529.94 & 422.63 & 344.91 & 231.03\\
        \multicolumn{2}{c||}{\textbf{criterion} $ \CostFunc $}  & 280839.68 & 178616.71 & 118962.34 & 53375.24 \\
        \multicolumn{2}{c||}{\textbf{threshold} $ \widehat{\mathcal{Q}}^{\text{corr}}_{46}(1-\rho) $}  & 150.8 & 150.8 & 150.8 & 150.8 \\
                \bottomrule
                \end{tabular}
                \label{tab:optimalOrbitalElementsHR8799}
  \tablefoot{Because of their degeneracy, any pair, $(\omega, \Omega),$ is perfectly equivalent to $(\omega+\pi, \Omega+\pi)$. The uncertainties were estimated via the perturbation method described in Sect. \ref{sec:NumericalMethodErrors} with $N_p=10^4$. The central values are the optimal orbital elements found by PACOME and the lower and upper bounds correspond to the bounds of the $95\%$ confidence interval. The corrected multi-detection detection threshold $ \widehat{\mathcal{Q}}^{\text{corr}}_{46}(1- \rho) $ associated to a confidence level of $\rho = 4.9 \times 10^{-12}$ is to compare to the criterion score $\CostFunc$.}
\end{table*}

\begin{figure}[t!]
    \centering
    \includegraphics[width=0.5\textwidth]{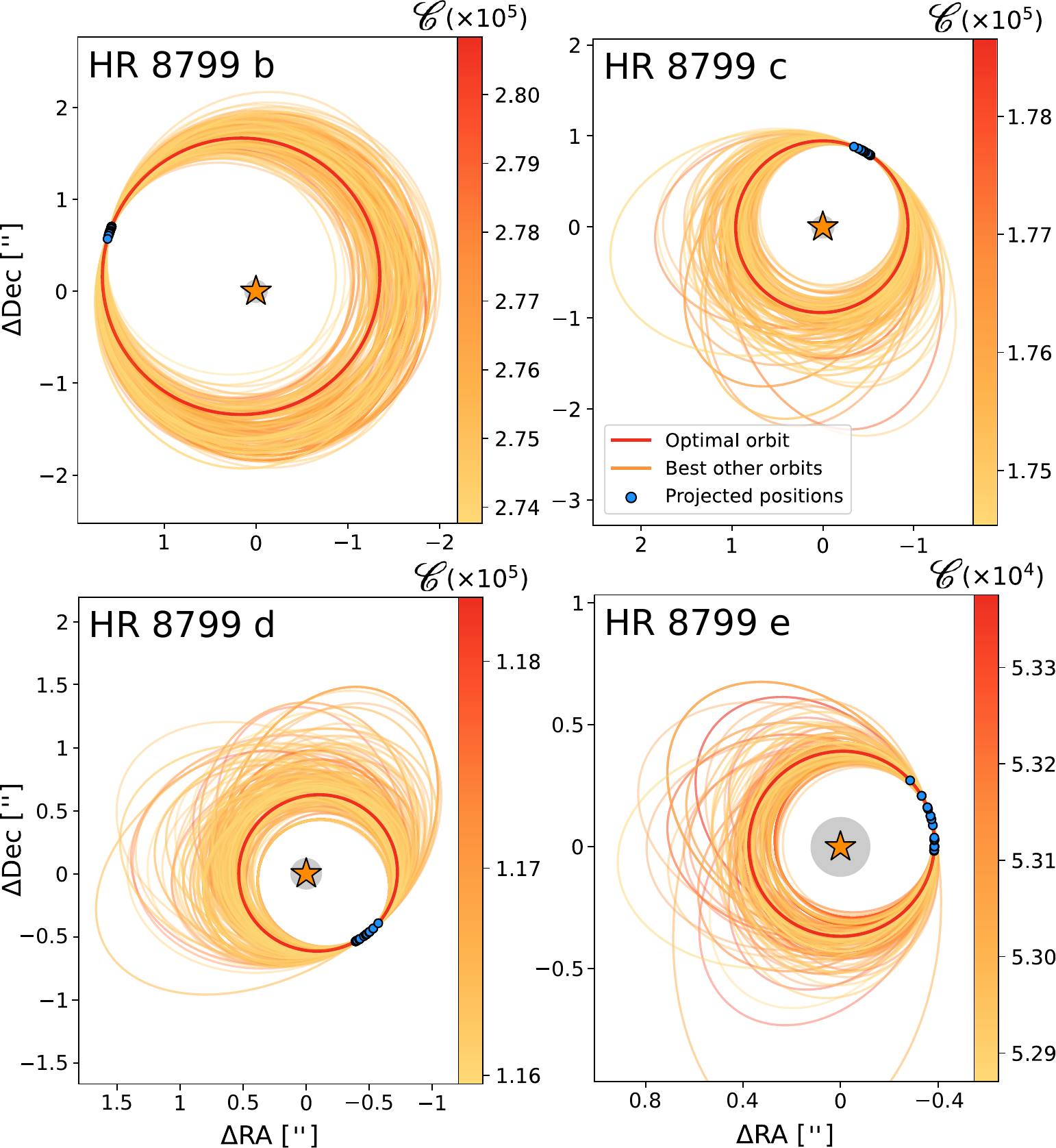}
    \caption{Optimal orbits found for each of the four known exoplanets HR\,8799 b, c, d, and e along with the best $10^3$ other on-grid orbits (in orange to red colors), whose RMSD is less than $1$ pixel away from the optimal solution. The red thick line shows the retained optimal orbit (see Table \ref{tab:optimalOrbitalElementsHR8799}). The blue dots are the projected positions of the signal along the optimal orbit at all epochs and the grey circular area represents the coronagraphic mask.}
    \label{fig:projOrbitHR8799}
\end{figure}

The retained optimal orbital elements are detailed in Table~\ref{tab:optimalOrbitalElementsHR8799}. The semi-major axis, $a$, the eccentricity, $e$, the epoch of periapsis passage, $t_0$, the Kepler constant, $K,$ and the period, $P,$ that we derived are consistent with the literature values \citep{Konopacky2016hr8799,wertz2017vlt,Wang2018,lacour2019first}. However, the inclination, $i$, the argument of periapsis, $\omega,$ and the longitude of the ascending node, $\Omega,$ do not always match. We observed differences in $\Omega$ and $\omega,$ as our uncertainty quantification method does not explore all parameter space and fails to capture the highly degenerate cases of quasi-circular ($e \simeq 0$) and slightly inclined ($i \simeq \qty{0}{\degree}$) orbits (and even more so in cases of poor temporal coverage).
In addition, the retained inclination, $i,$ seems to be slightly underestimated by our method ($i \simeq \qty{8.9}{\degree}$ compared to $\simeq 20-\qty{25}{\degree}$ in the literature). This tends to increase the total number of possible degeneracies on $\Omega$, $\omega$, and $\tau$ and reinforces the difference we observe. On the other hand, 17 orbital combinations out of the 37 re-optimized ones ($\simeq \qty{46}{\percent}$) have inclinations between $20$ and $30$ degrees which means that the commonly accepted range of values for the inclination of the four HR\,8799 exoplanets is found by the algorithm and is very plausible.
The projections of the optimal orbits of the four exoplanets are shown in red in Fig.~\ref{fig:projOrbitHR8799}, whereas the orange trajectories represent the first $10^3$ best other solutions whose RMSD are less than 1 pixel away from the optimal retained solutions (see Eq.~\ref{eq:rmsd_no_gt}). Again, the poor temporal coverage of the data is clearly noticeable and shows that a large number of rather different orbits actually fall on the projected positions of the detected exoplanets.

\begin{figure}[t!]
    \centering
    \includegraphics[width=0.5\textwidth]{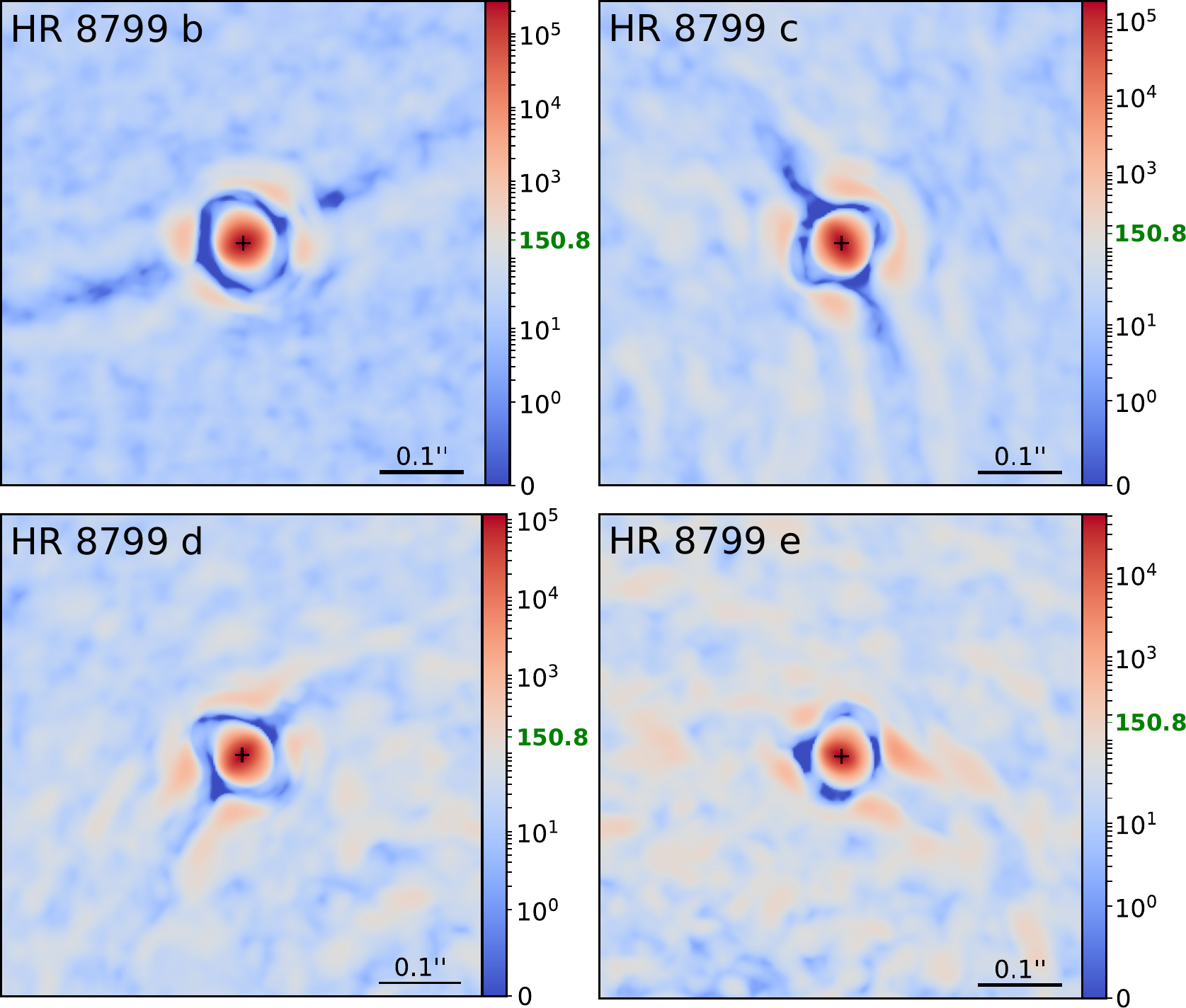} 
    \caption{Cost function maps around the optimal solution estimated by \PACOME for the four known exoplanets of the HR\,8799 system in a ROI of 60 pixels sampled with four nodes per pixel. The value of the corrected empirical multi-epoch detection threshold $\widehat{\mathcal{Q}}^{\text{corr}}_{46}(1-4.9 \times 10^{-12}) \approx 150.8$ is highlighted in green in the color bar.}
    \label{fig:costFuncMapHR8799bcde}
\end{figure}

The associated multi-epoch detection scores and $\SNR$ are extremely high (e.g., HR\,8799 e is detected at $\SNR = 231$), which again highlights the capability of the proposed algorithm to recombine efficiently faint signals from point-like sources, thus increasing the accuracy of the spectral characterisation of exoplanets.
However, the gain in S/N does not scale as the square root of the number of epochs (see Table \ref{tab:optimalOrbitalElementsHR8799}). This behavior was expected as the quality of the data 
is very heterogeneous across the $23$ epochs, resulting in several epochs weighting much more than the others in the multi-epoch combination.
Yet, even the worst epochs bring valuable information and constrain the solution (e.g., by eliminating some possible orbits). Finally, the multi-epoch cost function maps of the retained solutions are displayed in Fig.~\ref{fig:costFuncMapHR8799bcde}.
As for the numerical experiments of Sect. \ref{subsec:results_ss_data}, all signals in the field that are not at the exoplanet locations combine weakly (from blue to white colors, below the set confidence level at $\rho = 4.9 \times 10^{-12}$).

\subsubsection{Benefits of accounting for spectral correlations}
\label{subsection:takingSpectralCorrelIntoAccount}

At the end of Sect.~\ref{subsec:ResultsSearchInjections}, we discuss the fact that \PACOME still managed to find the ground truth even though the hypothesis of independent spectral channels was not perfectly verified. The spectral correlations were in fact negligible because SPHERE-IRDIS data only feature $L=2$ spectral channels. On the other hand, SPHERE-IFS data have $L=39$ different spectral channels; thus, the spectral correlations are much stronger and it becomes essential to take them into account before carrying out the multi-epoch combination.

\begin{figure}[t!]
    \centering
    \includegraphics[width=0.5\textwidth]{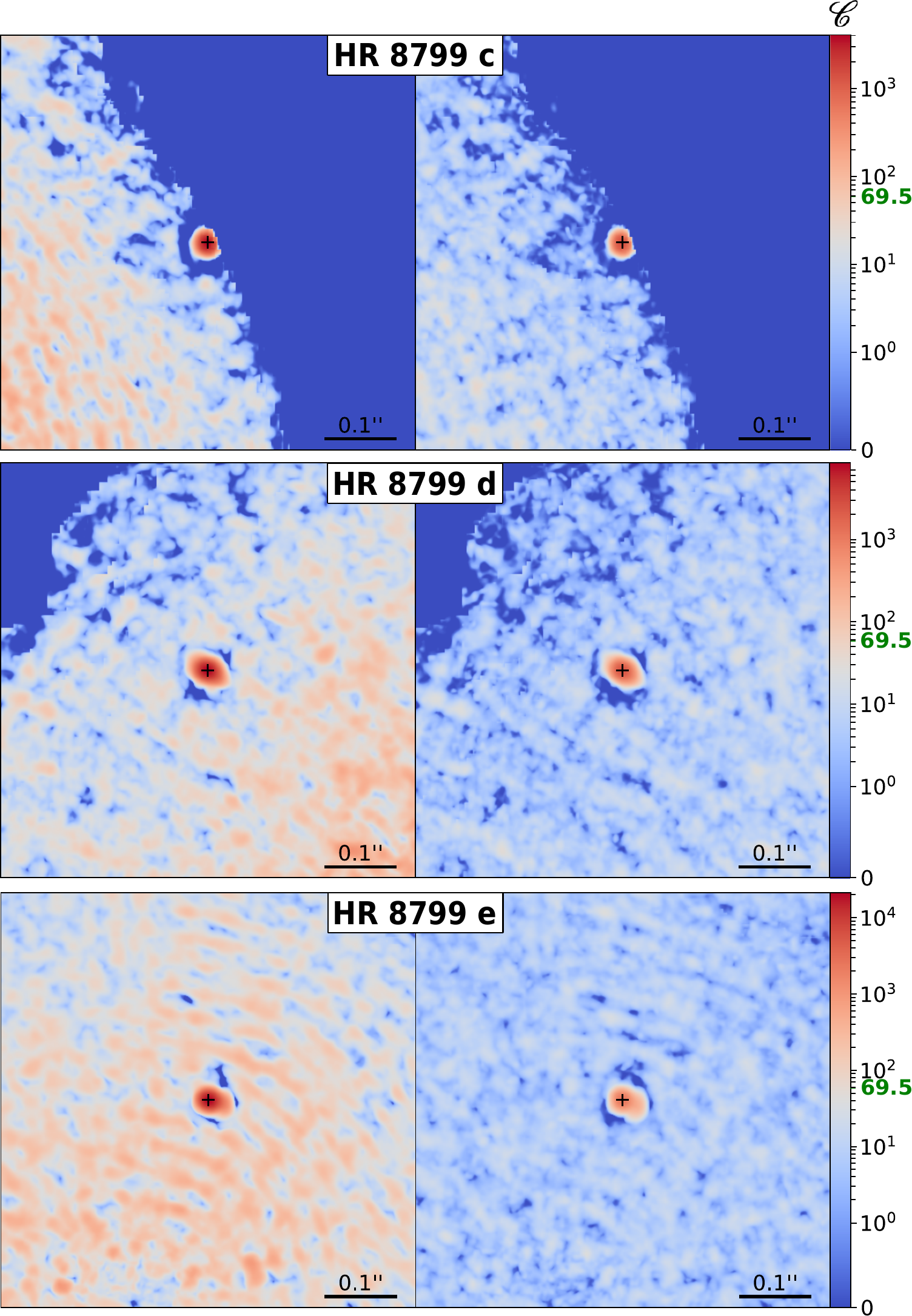}
    \caption{Cost function maps around the optimal solution estimated by \PACOME for three of the four known exoplanets of the HR\,8799 system considering spectrally correlated (left) and spectrally whitened (right) observations from the SPHERE-IFS. A flat spectrum $\bm\gamma$ has been used as spectral prior for the spectral combination by \PACO. The region of interest is 100 pixels wide and sampled with four nodes per pixel. The value of the corrected empirical multi-epoch detection threshold $\widehat{\mathcal{Q}}^{\text{corr}}_{15}(1-1.8 \times 10^{-10}) \approx 69.5$ is highlighted in green in the color bar.}
    \label{fig:costFuncMapHR8799IFS_withoutSpecCorr_VS_withSpecCorr}
\end{figure}

To illustrate the importance of whitening the data spectrally, we show in Fig.~\ref{fig:costFuncMapHR8799IFS_withoutSpecCorr_VS_withSpecCorr} the 2D maps of the multi-epoch criterion centered on the positions of HR\,8799 c, d, and e, described in the section above for IFS data combined with and without the spectral correlation corrections. When we are not taking the spectral correlations into account, a larger number of structured patterns (noise and/or speckles) are combined positively and increase the background signal, thereby going above the prescribed detection level. Explicitly accounting for the spectral correlations reduces this side effect greatly so that solely the real known sources lie above the set detection limit.

\begin{figure}[!ht]
    \centering\includegraphics[width=1\linewidth]{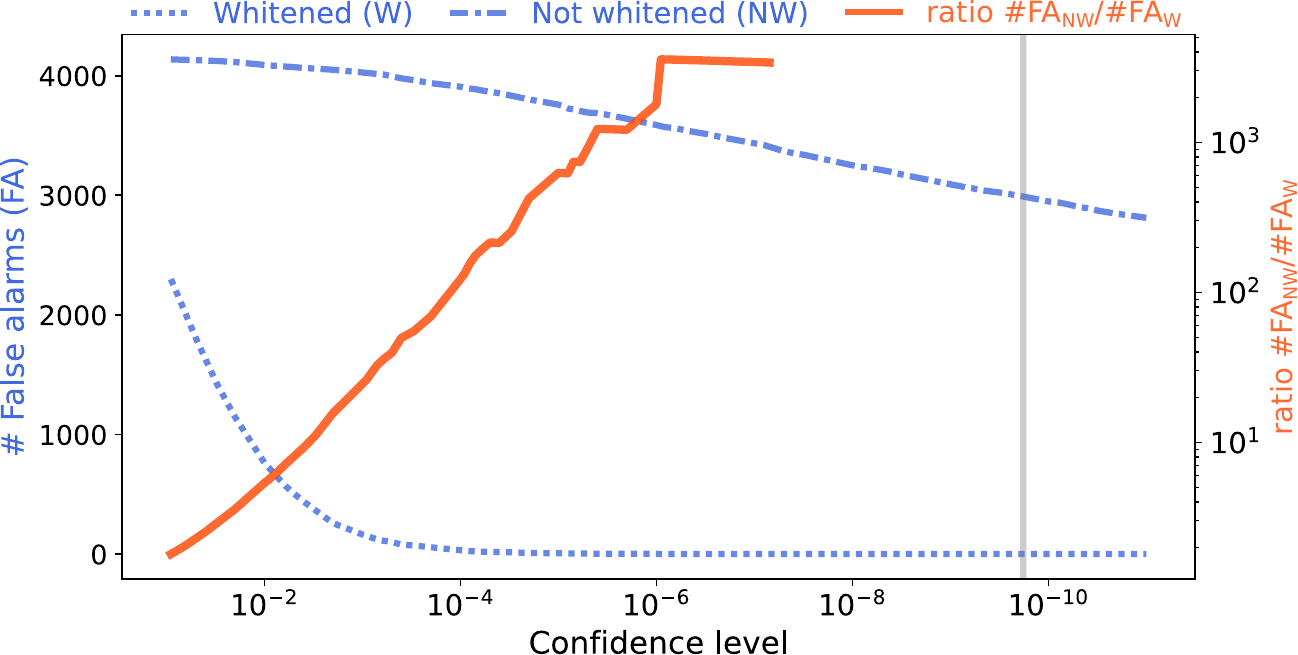}
    \caption{Number of false alarms (FA) for both the spectrally whitened (W) and not-spectrally whitened (NW) multi-epoch SPHERE-IFS datasets of HR 8799. The ratio between the two is shown in red. The number of multi-epoch false alarms was counted considering circular patches of the size of the FWMH. The grey vertical line corresponds to the chosen confidence of detection.}
    \label{fig:numberFA}
\end{figure}

We quantified the importance of accounting for the spectral correlations of the data by counting the number of false alarms (considering circular patches of the size of the FWMH) in the criterion map of the multi-epoch SPHERE-IFS dataset of HR 8799. This study was carried by accounting for the spectral correlations and by ignoring them for different confidence levels and the results are plotted in Fig.~\ref{fig:numberFA}. The spectral whitening effect is undeniably efficient in reducing the number of false alarms (a factor of $600$ at the $\rho = 10^{-5}$ confidence level and a factor of $1800$ at $\rho = 10^{-6}$). At $\rho \approx 6 \times 10^{-8}$ and below, false alarms are no longer found in the spectrally whitened data whereas the opposite case still rates very high number of false alarms ($> 2800$).

\subsubsection{Search for a fifth planet}
\label{sec:search5thplanet}

Several dynamical studies suggested that a fifth companion could orbit HR\,8799 either in the outer region of the system (beyond HR\,8799 b) between $90$ and $110$ au \citep{zurlo2022} or in the innermost part (before HR\,8799 e) between $7.5$ and $9.7$ au \citep{wahhaj2021search}. We probed these two regions with \PACOME to seek a potential candidate or to put constraints on its potential presence.

\paragraph{Derivation of the multi-epoch contrast:}

Building contrast curves is not straightforward when combining multi-epoch direct imaging observations and should be considered with care.
With mono-epoch datasets, the achievable $5\sigma$ contrast is classically  quantified at a given angular separation; whereas in a multi-epoch framework the combined contrast is computed for a given orbit where the angular separation of the source may vary along its trajectory. Hence, to compare the multi-epoch contrast to the more classical mono-epochs ones, we have to restrain the hypothesis and consider only face-on ($i = 0$) circular ($e = 0$) orbits. Given our model, we also need to assume that the flux of the source is constant over the epochs, which is consistent with face-on and circular orbits. These assumptions are restrictive but still more or less hold for the case of HR 8799 \citep{Wang2018, Gozdziewski2020}.

Under these assumptions, the constant source flux equals:
\begin{equation}
         \forall t \in [1,T], \,\, \widehat{\alpha}_{t,\ell}(\bm \mu) = \widehat{\alpha}_{\ell}(\bm \mu) = \frac{\sum\nolimits_t b_{t,\ell}(\bm\theta_t(\bm\mu))}{\sum\nolimits_t a_{t,\ell}(\bm\theta_t(\bm\mu))} \,,
\end{equation}
and, hence, the associated multi-epoch contrast is given by:
\begin{equation}
         \widehat{\sigma}_{\ell}(\bm \mu) = \sqrt{ \Var \{\widehat{\alpha}_{\ell}(\bm \mu) \} } = 1 \big/ \sqrt{\sum\nolimits_t a_{t,\ell}(\bm\theta_t(\bm\mu))}.
\end{equation}
As stated in Sect.~\ref{sec:StatisticalGuarantees}, our direct model implicitly accounts for the attenuation of off-axis PSFs near the coronagraphic mask (encoded in the $a_{t\ell}$ and $b_{t\ell}$ terms) therefore no multiplicative scaling is necessary.

\begin{figure*}[ht!]
    \centering
    \includegraphics[width=1\linewidth]{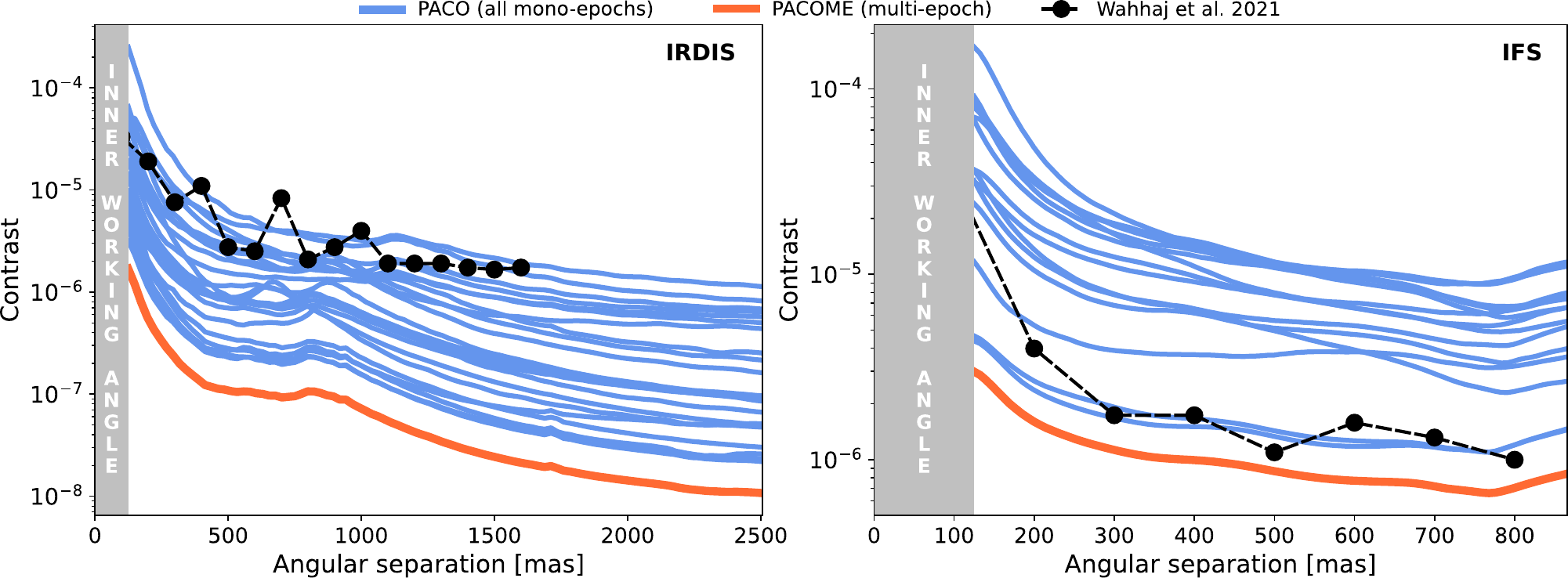}
    \caption{Mono and multi-epoch $5\sigma$ contrast curves of the IRDIS (\textit{left}) and IFS (\textit{right}) datasets of HR 8799 in ASDI mode with flat spectral priors. The multi-epoch contrast is computed as a function of the separation assuming face-on circular orbits and a source flux constant over the epochs. The $5\sigma$ contrasts found in \cite{wahhaj2021search} were extracted and overplotted for comparison.}
    \label{fig:contrastMultiEpoch_HR8799}
\end{figure*}

The computed mono-epoch and multi-epoch $5\sigma$ contrasts curves are given in Fig.~\ref{fig:contrastMultiEpoch_HR8799} and compared to \cite{wahhaj2021search} who presented the deepest contrast limits to date constraining the presence of a potential additional inner planet. Their datasets consist of 4.5 hours of observations in K1-K2 band with IRDIS and Y-H band with IFS. Our work combines more datasets (23 for IRDIS, 15 for IFS) and achieves an even deeper contrast at all separations. 

\paragraph{Outer part:}

The search in the outer part was carried out in ASDI mode with the SPHERE-IRDIS data because the region is not covered by the field of the IFS. We used a combination of (i) the $13$ spectrally whitened dual-band datasets reduced with a flat prior spectrum and (ii) the first spectral channel of the $10$ remaining broad-band datasets (as both channels carry more of less the same information). 

We explored a total of $N_{\text{orb}} \simeq \num{43e9}$ orbits covering the same parameter space (as described in Sect.~\ref{subsec:resultsRealData_HR8799bcde}) for Kepler's constant $K$, semi-major axis values from $2205$ to $2696$\,mas, eccentricities from $0$ to $0.5,$ and all allowed ranges for the remaining orbital elements.

For the same reason described in Sect.~\ref{subsec:ResultsSearchInjections} and given the number of explored orbits, we set the desired confidence level to $\rho= 0.1 / N_{\text{orb}} = 2.3 \times 10^{-12}$ with $\widehat{\mathcal{Q}}^{\text{corr}}_{23}(1-\rho) \approx 76.2$. No signals were found above this threshold. At these separations, the multi-epoch contrast limit is $\widehat{\sigma}(\bm\mu) = 1.2 \times 10^{-8}$ (see Fig.~\ref{fig:contrastMultiEpoch_HR8799}). The family of orbits with the highest multi-epoch criteria was found at projected positions $2582$ mas away from the star and its best optimized orbit scored a $\CostFunc=69.9,$ corresponding to a confidence level of detection $\rho=8.7 \times 10^{-11}$, which is expected to happen around four times given the number of explored orbits. This family of plausible orbits is mostly driven by epochs 2017-10-12 and 2017-10-13, which always give the same projected positions on the detector. Alone, they explain $41\%$ of the cost function score. This could be explained by an instrumental artefact, static on the scale of a day. The assumption of independent epochs on which the algorithm is based is violated in this case and there is no simple way to counter these systematics. In addition, the best optimized orbit (below the detection threshold) is not at all in the same orbital plane as HR\,8799 b, c, d, and e ($i=\qty{77.6}{\degree}$) and its high eccentricity ($e=0.5$) makes it cross the orbits of the other planets of the system which, for stability concerns, weakens considerably the possibility of a false negative candidate.

\paragraph{Inner part:}
For the search in the inner part of the system, we used the $15$ SPHERE-IFS data described in Table~\ref{tab:dataset_logs}. The three known visible sources were masked. Similarly, a part of the region under the coronagraph ($80\%$) was masked since the transmission is very low in this area.
In total, $N_{\text{orb}} = 5.44 \times 10^8$ orbits were explored, probing the semi-major axis parameter space from $122$ to $246$ mas, eccentricities from $0$ to $0.5$ and the same as the outer search for the remaining orbital elements. The confidence level was set to $\rho= 0.1 / N_{\text{orb}} = 1.8 \times 10^{-10}$ yielding a detection threshold of $\widehat{\mathcal{Q}}^{\text{corr}}_{15}(1-\rho) \approx 69.5$. The multi-epoch contrast limit is $\widehat{\sigma}(\bm\mu) = 2.4 \times 10^{-6}$ for these separations (see Fig.~\ref{fig:contrastMultiEpoch_HR8799}).

Among the $23$ spectral priors used for the IFS data reduction step, only $4$ (respectively one flat, two bi-modal, and one tri-modal priors) show optimized orbits with multi-epoch scores $2$ to $11\%$ greater than the prescribed detection threshold. None of the best orbits found for each spectral prior are alike or have the same projected positions on the detector. A closer inspection reveals that for each of the four identified priors the multi-epoch score is only driven by a handful of epochs (two or three). Indeed, the best three epochs of each prior contribute respectively to $64\%$, $76\%$, $72\%,$ and $53\%$ of their associated multi-epoch criterion. 

We analyze these results as follows. First, having several different spectral priors giving a multi-epoch score above the limit is expected. As the same data is used for all priors, they can not be considered independent, so that a detection with one prior could yield an other detection for a different spectral prior. Second, the region that is explored here is very close to the central star, which is typically where slight residual non-stationarities can occur and where the assumed model for the off-axis PSF is not fulfilled (due to high nonlinearities induced by the coronagraph). Thus, the hypothesis on which our model is based (i.e., spatial correlations fully captured by a multi-variate Gaussian, spatially invariant off-axis PSF) are not perfectly verified in this regime and even more so in our multi-epoch case where the nuisance propagates and amplifies when combined temporally. To tackle this problem and reduce these effects, it would be necessary to model the correlations at larger scales and to consider a variable off-axis PSF in the field (e.g., via a grid of physical models) but this is not treated in this paper and is left for future works.
Third, the fact that only a fifth of the epochs contribute to more than half of the criterion reflects an important need to explicitly take into account the consistency of the individual temporal S/N. 
This will avoid giving too much credit to (still rare) quite high detection peaks in some isolated individual epochs. 
Further works will be led to take this aspect into consideration and to quantify its benefits with respect to the current algorithm.

Finally, we ran the same search, but this time with the SPHERE-IRDIS data. No signals were found above the detection threshold considering the $13$ dual-band datasets only. Taking the broad-band ones into account led to a few multi-epoch detections but all were ruled out because only driven by a mono-epoch false alarm detection near the coronagraphic mask in the 2017-10-12 epoch. Again, \PACO suffers from the above-mentioned phenomenon close to the coronagraph, so if there is a false detection in a mono-epoch \PACO map, it will inevitably yield a false alarm in the multi-epoch case.

\section{Conclusion}
\label{sec:conclusion}

In this work, we propose \PACOME, a new algorithm dedicated to the combination of multi-epoch angular and spectral differential observations in high contrast obtained using the pupil tracking mode. As with other algorithms of this category, it integrates a massive search over the possible Keplerian orbits spanned by the putative sources, as well as a novel maximum likelihood multi-epoch detection score. Concerning the exploration of the plausible orbits, we designed an efficient two-steps strategy combining a coarse search on a grid with a local refinement of the best orbits found on this grid. The proposed multi-epoch metric directly derives from an end-to-end statistical framework, which explicitly accounts for the non-stationarity and the multiple and complex correlations of the data (spatial and/or spectral and/or temporal). It conveniently makes use of the byproducts of applying \PACO, a powerful post-processing algorithm of mono-epoch observations at high contrast. Our new metric offers a reliable multi-epoch detection criterion, which is interpretable both in terms of probabilities of detection and false alarms. Jointly, \PACOME also produces a few plausible estimates of the detected sources orbital elements and assesses their uncertainties locally. From an implementation point of view, we also integrated  a new centering routine of the individual frames in our
pipeline to increase the accuracy of the host star centering and, consequently, to improve the detection confidence of real sources. 

We showed from realistic numerical experiments that the proposed approach is able to detect faint point-like sources that remain undetectable by advanced post-processing of each individual epoch. We obtained a gain in terms of detection sensitivity very close to the square root of the number of epochs, which corresponds to an optimal combination of the available information. As a case-study example, we applied \PACOME on 23 VLT/SPHERE observations of the HR\,8799 star. From these observations, we were able to re-detect the four known exoplanets at unprecedented levels of confidence (from S/N $= 231$ for planet e to $530$ for planet b). In addition, we obtained the deepest level of contrast especially at close angular separation ($3.0 \times 10^{-6}$ for IFS and $1.7 \times 10^{-6}$ for IRDIS at 0.125'') by combining all the available VLT/SPHERE datasets for this system. 

We believe that the high detection sensitivity of \PACOME opens the door to a massive re-exploration of the archived high contrast observations, possibly from different instruments.As an illustration for the VLT/SPHERE instrument, this analysis would at least concern HD 95086, HIP 65426, 51 Eridani, GJ 504, and Proxima Centauri that have been observed multiple times representing at the very least $80$ observations (IRDIS and IFS combined) totalling a total exposure time of about $\qty{90}{hours}$. Such dedicated \PACOMEs study will be conducted in a future paper. Besides, in the context of the forthcoming 30-meter class telescopes, our approach would be key to probe the habitable zone of the nearby stars and reach the deepest possible contrast limits. This will require long exposure times of several tens of hours that will only be achieved by combining several observations conducted days, weeks, or months apart. At these timescales and separations, the orbital motion of exoplanets will no longer be negligible and a proper orbital modelling will be crucial to combine multi-epoch observations without drastically degrading the detection confidence and the achievable contrast. \PACOMEs outputs could also allow for the simulation of the experienced combined S/N for a given set of parameters (e.g., orbits, observing conditions, spectral band, parallactic rotation, exposure time, etc.). This will be essential for optimizing the use of the observational facilities by predicting the best observational times and parameters to maximize the scientific return. 
Concerning methodological developments, we are currently working on the refinement of the proposed algorithm by (i) further improving its robustness against the highly variable observation quality, (ii) improving the statistical model of the nuisance to account for correlations at longer spatial scale, and (iii) accounting for the variations of the exoplanetary signature across epochs and during the sequence of observations itself. As a longer term goal and in the context of the next generation of instruments designed to scrutinize the vicinity of nearby stars, we aim to develop a multi-epoch combination algorithm accounting jointly for the presence of point-like sources and of spatially resolved objects such as circumstellar disks.

\begin{acknowledgements}

We thank Frédéric Vachier (IMCCE, Observatoire de Paris, France) for fruitful discussions. We Thank Antoine Chomez and Philippe Delorme for their help with the data preprocessing. This work has made use of the SPHERE Data Center, jointly operated by OSUG/IPAG (Grenoble, France), PYTHEAS/LAM/CESAM (Marseille, France), OCA/Lagrange (Nice, France), Observatoire de Paris/LESIA (Paris, France), and Observatoire de Lyon/CRAL (Lyon, France). This work has been supported by the French National Programs (PNP and PNPS), and by the Action Spécifique Haute Résolution Angulaire (ASHRA) of CNRS/INSU co-funded by CNES. This work used ESO archive data for HR8799 from various observing programms with ESO-IDs listed in Table \ref{tab:dataset_logs}.

\end{acknowledgements}

\bibliographystyle{aa}
\bibliography{ref}

\begin{appendix}

\section{Deriving an optimal multi-epoch S/N}
\label{subsec:multi_epoch_snr}

\subsection{Matched filter formalism}

Here, we consider a general case where some data, $\bm d \in \mathbb{R}^n$, are measured under reproducible conditions parameterized by $\bm \phi \in \mathbb{R}^p$. These parameters account for the experimental conditions, the object of interest, and so on. It is always possible to express the following:
\begin{equation}
        \bm d = \bm m + \bm z\,,
\end{equation}
with $\bm m = \E\{ \bm d \, | \, \bm \phi\}$ as the expectation of the measurements under given conditions, $\bm \phi,$ and $\bm z \in \mathbb{R}^n$ representing a nuisance term accounting for the measurement noise. It directly follows that $\E\{ \bm z \, | \, \bm \phi\} = \M 0$; in other words, the noise term, $\bm z,$ is centered knowing conditions, $\bm \phi$.

We then consider a measurable quantity, say, $\eta \in \mathbb{R}$, which linearly depends on the data:
\begin{equation}
        \eta = \bm q \transp \, \bm d\,,
\end{equation}
with $\bm q \in \mathbb{R}^n$ the vector of coefficients of the linear combination, which may depend on $\bm \phi$. Our goal is to find the coefficients $\bm q$ that yield $\eta$ with the highest possible S/N given the data. The expectation and variance of $\eta$ under given conditions, $\bm \phi,$ are:
\begin{equation}
    \begin{cases}
            \, \E\{ \eta \, | \, \bm \phi\} &= \bm q \transp \, \bm m\,, \\
        \, \Var \{ \eta \, | \, \bm \phi \} &= \bm q \transp \,  \M S \, \bm q\,,
    \end{cases}
\end{equation}
with $\M S = \E\{ \bm z \, \bm z\transp \, | \, \bm \phi\} = \Cov\{\bm d\,\vert\,\bm\phi\}$ the covariance of the nuisance term $\V z$ under given conditions $\bm \phi$, which is also the covariance of the data $\V d$ under the same conditions. The S/N of $\eta$ under the given observing conditions is then:
\begin{align}
\label{eq:SNRmatchedFilter}
        \SNR(\eta \, | \, \bm \phi) = \dfrac{\E\{ \eta \, | \, \bm \phi\}}{\sqrt{\Var \{ \eta \, | \, \bm \phi \}}} = \dfrac{\bm q \transp \, \bm m}{\sqrt{\bm q \transp \,  \M S \, \bm q}}\,.
\end{align}
Noting that for any $\gamma>0$, the S/N of $\eta$ and $\gamma\,\eta$ are equal, it follows that the linear quantity of highest S/N is determined up to a positive factor. We can therefore maximize the S/N of $\eta$ under the constraint that, say, its expectation, $\E\{\eta \, | \, \bm \phi\},$ has a given value. Under that constraint, maximizing the S/N amounts to minimize the denominator in the right hand side of Eq.~\eqref{eq:SNRmatchedFilter}, that is $\bm q \transp \,  \M S \, \bm q$ the variance of $\eta$. The Lagrangian of this constrained problem is expressed as:
\begin{align}
        \LogLike = \frac12 \, \bm q \transp \,  \M S \, \bm q - \xi \, \bm q \transp \, \bm m\,,
\end{align}
with $\xi \in \mathbb{R}$ the Lagrangian multiplier associated to the constraint that $\E\{\eta \, | \, \bm \phi\}$ has a given value (the $1/2$ factor is for convenience). Provided that the covariance $\M S$ is invertible, the optimal solution of the Lagrangian is given by:
\begin{align}
        \bm q = \xi \, \M S^{-1} \, \bm m\,,
\end{align}
which is known as the "matched filter" \citep{kay1998fundamentals,kay1998fundamentals2}. 
If $\bm q$ is the matched filter, the best possible S/N of $\eta$ is then obtained for any $\xi \not= 0$ and is equal to:
\begin{align}
\label{eq:theorBestSNR}
        \max_{\bm q} \SNR(\eta \, | \, \bm \phi) = \sqrt{ \bm m \transp \,  \M S^{-1}  \, \bm m}\,,
\end{align}
which, as expected, does not depend on the multiplier $\xi \not= 0$. Also, as expected, the optimal filter, $\bm q,$ and thus the optimal quantity, $\eta,$ are defined up to a factor of $\xi$.

\subsection{Application to multi-epoch detection}

To apply the matched filter formalism described above to the direct model of A(S)DI observations given in Eq.~\eqref{eq:directModel}, we consider the residual data:
\begin{align}
        \bm d_{t,\ell,k} = \bm r_{t,\ell,k} - \bar{\bm f}_{t,\ell}\,,
\end{align}
where $\bm d_{t,\ell,k}$ is assumed Gaussian distributed with expectation $\alpha_{t,\ell} \, \bm h_{t,\ell}(\bm\theta_t(\bm \mu))$ and variance $\sigma^2_{t,\ell,k} \, \bm \C_{t,\ell}$\,.
Assuming that residual data from different epochs, frames, and/or spectral channels are mutually independent, the quantity of maximal S/N is:
\begin{align}
        \eta &= \xi \, \sum_{t,\ell} \alpha_{t,\ell} \sum_{k} \bm h_{t,\ell}(\bm\theta_t(\bm\mu))\transp \, \bm \W_{t,\ell,k} \, \bm d_{t,\ell,k} \notag \\
        &= \xi \, \sum_{t,\ell} \alpha_{t,\ell} \, b_{t,\ell}(\bm\theta_t(\bm\mu))\,.
\end{align}
Given the data and accounting for the assumed independencies, the best achievable S/N is obtained from Eq.~\eqref{eq:theorBestSNR}:
\begin{align}
        \max \SNR(\bm\mu) &= \sqrt{ \sum_{t,\ell,k} \alpha_{t,\ell}^2 \, \bm h_{t,\ell}(\bm\theta_t(\bm\mu))\transp \, \bm \W_{t,\ell,k} \, \bm h_{t,\ell}(\bm\theta_t(\bm\mu))} \notag \\
         &= \sqrt{ \sum_{t,\ell} \alpha_{t,\ell}^2 \,  a_{t,\ell}(\bm\theta_t(\bm\mu))}\,.
\end{align}
When considering the maximum likelihood estimator of $\alpha_{t,\ell}$, this theoretical value can be approximated by the following (biased) estimator:
\begin{equation}
        \SNR(\bm\mu) =\sqrt{  \sum_{t,\ell} \frac{\big(\big[ b_{t,\ell}(\bm\theta_t(\bm\mu))\big]_+\big)^2}{ a_{t,\ell}(\bm\theta_t(\bm\mu))} } = \sqrt{\CostFunc(\bm \mu)}\,.
\end{equation}
 
\section{Keplerian motion solver}
\label{sec:keplerianMotionSolver}

The 2D position of a celestial body of orbital elements, $\bm\mu = [a, e, i, \tau, \omega, \Omega, K]\transp$, projected on the sky plane at the epoch of observation $t$ is expressed as:
\begin{equation}
\bm\theta_t(\bm\mu)
= r_t
\begin{pmatrix}
    \, \cos\Omega\cos(\omega +\nu_t)-\sin{\Omega}\sin(\omega +\nu_t)\cos{i} \, \\
    \, \sin\Omega\cos(\omega +\nu_t)+\cos{\Omega}\sin(\omega +\nu_t)\cos{i} \,
\end{pmatrix} .
\label{eq:2DprojPositions}
\end{equation}
In Eq.~\eqref{eq:2DprojPositions}, $r_t$ is the apparent distance of the celestial body with respect to the host star at time, $t,$ such that:
\begin{equation}
\label{eq:trueSeparation}
    r_t = \frac{a \, (1-e^{2})}{1+e \cos{\nu_t}} = a \, (1-e\cos{E_t})\,,
\end{equation}
with $\nu_t$ the true anomaly at a time, $t,$ given by:
\begin{equation}
\label{eq:trueAnomaly}
    \nu_t = 2 \arctan{\left( \sqrt{\frac{1+e}{1-e}} \tan{\frac{E_t}{2}} \right)}\,,
\end{equation}
and $E_t$ the eccentric anomaly. Along with the mean anomaly, $M_t$, it defines the well-known Kepler's equation:
\begin{equation}
\label{eq:KeplerEquation}
    M_t = E_t - e \sin{E_t}\,.
\end{equation}
The mean anomaly, $M_t$, is a function of the epoch of periapsis passage and the orbital period (or the Kepler constant and the semi-major axis) it is therefore simply given by:
\begin{equation}
\label{eq:MeanAnomaly}
    M_t = \frac{2\pi \, (t - t_0)}{P} 
      = 2\pi\left(\frac{t}{P}-\tau\right)
      = 2\pi\left(\sqrt{\frac{K}{a^3}} \, t-\tau\right)\,.
\end{equation}
Computing the 2D projected positions of Eq.~\eqref{eq:2DprojPositions} given the orbital elements, $\bm\mu,$ amounts to find $E_t$ by solving Kepler's equation of Eq.~\eqref{eq:KeplerEquation}, which is transcendental and offers no analytical expression for $E_t$. The solution can however be approximated with numerical root-finding algorithms. For this, we use Brent's \texttt{fzero} method \citep{Brent1973}, which is derivative-free  and that guarantees good accuracy for a limited number of iterations. 

To initialize the search, a crude bracketing of the solution of Kepler's equation is given by $E_t \in [M_t-e, M_t+e]$. For elliptic orbits, the eccentricity is such that $0 \leq e < 1$ and $f(E_t) = E_t - M_t - e\sin(E_t)$ is a strictly non-decreasing function as $f'(E_t) = 1 - e\cos(E_t) \geq 1 - e > 0$. A narrower interval can be determined in that case, $E_t \in [M_t-e, M_t]$ if $M_t \in [\pi, 2\pi]$ or $E_t \in [M_t, M_t+e]$ if $M_t \in [0, \pi]$, which makes the search a bit faster. We use $M_t$ as initial guess of the solution.

\section{Derivatives of the projected positions with respect to the orbital elements}
\label{sec:projPosDerivWRTOrbElem}

This appendix details the analytical derivatives, $\partial \bm\theta_t/\partial \bm \mu,$ of the projected positions, $\V \theta_t$, with respect to the orbital elements, $\V \mu,$ needed to optimize locally the multi-epoch detection criterion (Eq.~\ref{eq:costfuncC}). Further details are given in  Sect. \ref{sec:LocalOptiRefinement}.

\subsection{Derivatives with respect to $i$, $\omega,$ and $\Omega$}

Given Eq.~\eqref{eq:2DprojPositions}, the derivatives of the apparent position $\V \theta_t$ with respect to the three orbital elements $i$, $\omega,$ and $\Omega$ are:
\begin{equation}
\frac{\partial\bm\theta_t}{\partial i}
= r_t
\begin{pmatrix}
    \, \sin \Omega \sin(\nu_t + \omega) \sin i\, \\
    \, - \cos \Omega \sin
      (\nu_t + \omega) \sin i  \,
\end{pmatrix}\,,
\vspace{2mm}
\end{equation}

\begin{equation}
\frac{\partial\bm\theta_t}{\partial \omega}
= -r_t
\begin{pmatrix}
    \, \cos{\Omega} \sin{(\nu_t+\omega)} + \sin{\Omega}\cos{(\nu_t + \omega)} \cos{i}\, \\
    \, \sin{\Omega} \sin{(\nu_t+\omega)} - \cos{\Omega}\cos{(\nu_t + \omega)} \cos{i}  \,
\end{pmatrix}\,,
\vspace{2mm}
\end{equation}

\begin{equation}
\frac{\partial\bm\theta_t}{\partial \Omega}
= r_t
\begin{pmatrix}
    \, - \sin{\Omega} \cos{(\nu_t+\omega)} - \cos{\Omega}\sin{(\nu_t + \omega)} \cos{i}\,   \\
    \,  \cos{\Omega} \cos{(\nu_t+\omega)} - \sin{\Omega}\sin{(\nu_t + \omega)} \cos{i}  \,
\end{pmatrix}\,.
\end{equation}

\subsection{Derivatives with respect to $a$, $e$ and $\tau$ and $K$}

Computing the derivatives of the apparent position $\V \theta_t$ with respect to the remaining orbital elements ($a$, $e$, $\tau$, and $K$) requires having the derivatives of the true anomaly $\nu_t$ and of the true separation $r_t$ with respect to $a$, $e$, $\tau$, and $K$. These derivatives require those of the eccentric anomaly $E_t$ with respect to $a$, $e$, $K$, $a$, and $\tau$. Indeed, for any $\beta$ in $\{ a, e, \tau, K \}$, it comes:
\begin{align}
\frac{\partial\bm\theta_t}{\partial \beta}
&=  \frac{\partial\bm\theta_t}{\partial r_t} \dfrac{\partial r_t}{\partial \beta} + \frac{\partial\bm\theta_t}{\partial \nu_t} \dfrac{\partial \nu_t}{\partial \beta} \\
&=
\begin{pmatrix}
    \, \dfrac{\partial r_t}{\partial \beta} \big( \cos{\Omega} \cos{(\nu_t+\omega)} - \sin{\Omega}\sin{(\nu_t + \omega)} \cos{i} \big) \, \\
    \, -r_t \big( \cos{\Omega} \sin{(\nu_t+\omega)} + \sin{\Omega}\cos{(\nu_t + \omega)} \cos{i} \big) \dfrac{\partial \nu_t}{\partial \beta} \, \\[0.5cm]
    \, \dfrac{\partial r_t}{\partial \beta} \big( \sin{\Omega} \cos{(\nu_t+\omega)} + \cos{\Omega}\sin{(\nu_t + \omega)} \cos{i} \big) \, \\
    \, -r_t \big( \sin{\Omega} \sin{(\nu_t+\omega)} - \cos{\Omega}\cos{(\nu_t + \omega)} \cos{i} \big) \dfrac{\partial \nu_t}{\partial \beta}  \, 
\end{pmatrix}. \notag
\end{align}

\subsubsection{Derivatives of the eccentric anomaly $E_t$}

Deriving Kepler's equation with respect to any parameter $\beta \in
\mathbb{R}$ yields:
\begin{equation}
    \frac{\partial M_t}{\partial \beta} = (1 - e \cos E_t)  \frac{\partial
   E_t}{\partial \beta} - \sin E_t  \frac{\partial e}{\partial \beta}\,,
\end{equation}
hence:
\begin{equation}
    \frac{\partial E_t}{\partial \beta} = \frac{1}{1 - e \cos E_t}  \left(
   \frac{\partial M_t}{\partial \beta} + \sin E_t  \frac{\partial e}{\partial
   \beta} \right)\,.
   \label{eq:dEt_dBeta}
\end{equation}
Substituting each $\beta \in \{ a, e, K, t, \tau \}$ in Eq.~\eqref{eq:dEt_dBeta} yields:
\begin{equation}
  \frac{\partial E_t}{\partial a} = \frac{- 3 \pi}{1 - e \cos E_t}
  \sqrt{\dfrac{K}{a^5}}\,,
\end{equation}
\begin{equation}
  \frac{\partial E_t}{\partial e} = \frac{\sin E_t}{1 - e \cos E_t}\,,
\end{equation}
\begin{equation}
  \frac{\partial E_t}{\partial \tau} = \frac{- 2 \pi}{1 - e \cos E_t }\,,
\end{equation}
\begin{equation}
  \frac{\partial E_t}{\partial K} = \dfrac{\pi t}{(1 - e \cos E_t) \sqrt{K
  a^3}}\,.
\end{equation}

\subsubsection{Derivatives of the true anomaly $\nu_t$}

The derivative of the true anomaly $\nu_t$ defined in Eq.~{\eqref{eq:trueAnomaly}} with
respect to the eccentric anomaly is:
\begin{equation}
  \frac{\partial \nu_t}{\partial E_t} = \frac{\sqrt{1 - e^2}}{1 - e \cos E_t}\,.
\end{equation}
Hence the derivative with respect to the eccentricity $e$ is:
\begin{align}
  \frac{\partial \nu_t}{\partial e} &= \frac{\sin E_t}{(1 - e \cos E_t)
  \sqrt{1 - e^2}} + \frac{\partial \nu_t}{\partial E_t}  \frac{\partial
  E_t}{\partial e} \notag\\
  &= \frac{\sin E_t}{1 - e \cos E_t}  \left( \frac{1}{\sqrt{1 - e^2}} +
  \frac{\sqrt{1 - e^2}}{1 - e \cos E_t} \right)\,.
\end{align}
The other derivatives are obtained by applying the chain rule:
\begin{equation}
  \frac{\partial \nu_t}{\partial a} = \frac{\partial \nu_t}{\partial E_t} 
  \frac{\partial E_t}{\partial a} = \frac{- 3 \pi t \sqrt{1 - e^2}}{(1 - e
  \cos E_t)^2 } \sqrt{\dfrac{K}{a^5}}\,,
\end{equation}
\begin{equation}
  \frac{\partial \nu_t}{\partial K} = \frac{\partial \nu_t}{\partial E_t} 
  \frac{\partial E_t}{\partial K} = \frac{\pi t \sqrt{1 - e^2}}{(1 - e \cos
  E_t)^2  \sqrt{K a^3} }\,,
\end{equation}
\begin{equation}
  \frac{\partial \nu_t}{\partial \tau} = \frac{\partial \nu_t}{\partial
  E_t}  \frac{\partial E_t}{\partial \tau} = \frac{- 2 \pi \sqrt{1 -
  e^2}}{(1 - e \cos E_t)^2 }\,.
\end{equation}

\subsubsection{Derivatives of the true separation, $r_t$}

The derivatives of the true separation, $r_t$, defined in
Eq.~{\eqref{eq:trueSeparation}} are obtained by applying the chain rule:
\begin{align}
     \frac{\partial r_t}{\partial a} &= 1 - e \cos E_t + a \, e \sin E_t 
     \dfrac{\partial E_t}{\partial a} \notag\\
     &= 1 - e \cos E_t - \dfrac{3 \pi \, t \, e \sin E_t}{1 - e \cos E_t}
     \sqrt{\dfrac{K}{a^3}}\,,
\end{align}
\begin{align}
     \frac{\partial r_t}{\partial e} &= - a \cos E_t + a \, e \sin E_t 
     \frac{\partial E_t}{\partial e} = - a \cos E_t + \frac{a \, e \sin^2 E_t }{1 - e \cos E_t} \notag \\
     &= \frac{e - \cos E_t}{1 - e \cos E_t},
\end{align}
\begin{align}
     \frac{\partial r_t}{\partial K} &= a \, e \sin E_t  \dfrac{\partial
     E_t}{\partial K}
     = - \dfrac{\pi \, t \, e \sin E_t}{(1 - e \cos E_t) \sqrt{K a}}\,,
\end{align}

\begin{align}
     \frac{\partial r_t}{\partial \tau} = \frac{\partial r_t}{\partial
     E_t}  \frac{\partial E_t}{\partial \tau}
     = \frac{- 2 \pi \, a \, e \sin E_t}{(1 - e \cos E_t)}\,.
\end{align}

\section{Data centering using satellite spots}
\label{sec:dataCent_satSpots}

In this appendix, we describe our data centering procedure coded in Python, using satellite spots, that has been delivered and integrated in the SPHERE Data Center\footnote{https://sphere.osug.fr/spip.php?rubrique16} (see Sect. \ref{subsec:pre_reduction}).

\subsection{Parametric model of a satellite spot}

We model a satellite spot at position $(x,y)$ by a 2D elliptical Gaussian pattern of parameter $\bm\xi = [A, x_0, y_0, \sigma_x, \sigma_y, \theta, c]\transp$ as:
\begin{align}
        g(x, y, \bm\xi) = A \exp\Bigg(
                      &-\bigg( \frac{\cos^2\theta}{2\sigma_x^2} + \frac{\sin^2\theta}{2\sigma_y^2} \bigg)  (x - x_0)^2 \notag \\
                      &- \bigg( \frac{\sin^2\theta}{2\sigma_x^2} + \frac{\cos^2\theta}{2\sigma_y^2} \bigg)(y - y_0)^2 \notag \\
                      - 2 \, &\bigg( \frac{\sin 2\theta}{4\sigma_x^2} - \frac{\sin 2\theta}{4\sigma_y^2} \bigg)(x - x_0)(y - y_0)
               \Bigg)
                      + c\,,
\end{align}
with $A$ as the amplitude, $(x_0, y_0)$ the center coordinates, $(\sigma_x, \sigma_y)$  as the $x$ and $y$ spreads, $\theta$ as the orientation of the pattern, and $c$ as a constant offset. For convenience, we denote the discrete 2D elliptical Gaussian function at 2D angular location $p$ as:
\begin{equation}
        g_p(\bm\xi) = g(x_p, y_p, \bm\xi)\,.
\end{equation}

\subsection{Minimization problem}

\begin{figure*}[t!]
    \centering
    
        \begin{tikzpicture}[scale=1]
                \node[inner sep=0pt, label={[align=center]below:\hspace{-0.7cm}Observed intensity $\{\bm r_k\}_{k=1:K}$}] (centerC) at (0,0) {\includegraphics[width=.2\textwidth]{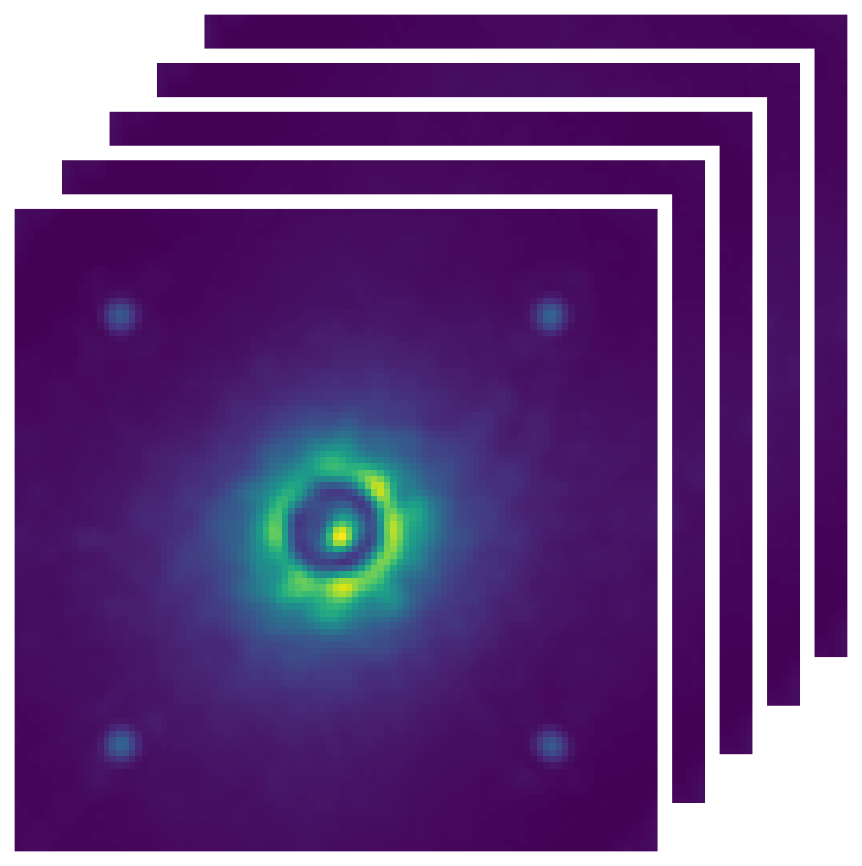}};
                \node[inner sep=0pt, label={[align=center]below: \hspace{-0.7cm}Extracted spots\\ \hspace{-0.7cm}$\big\{ \{ \bm d_{i,k} \}_{i=1:4} \big\}_{k=1:K}$}] (realW) at (5,0) {\includegraphics[width=.2\textwidth]{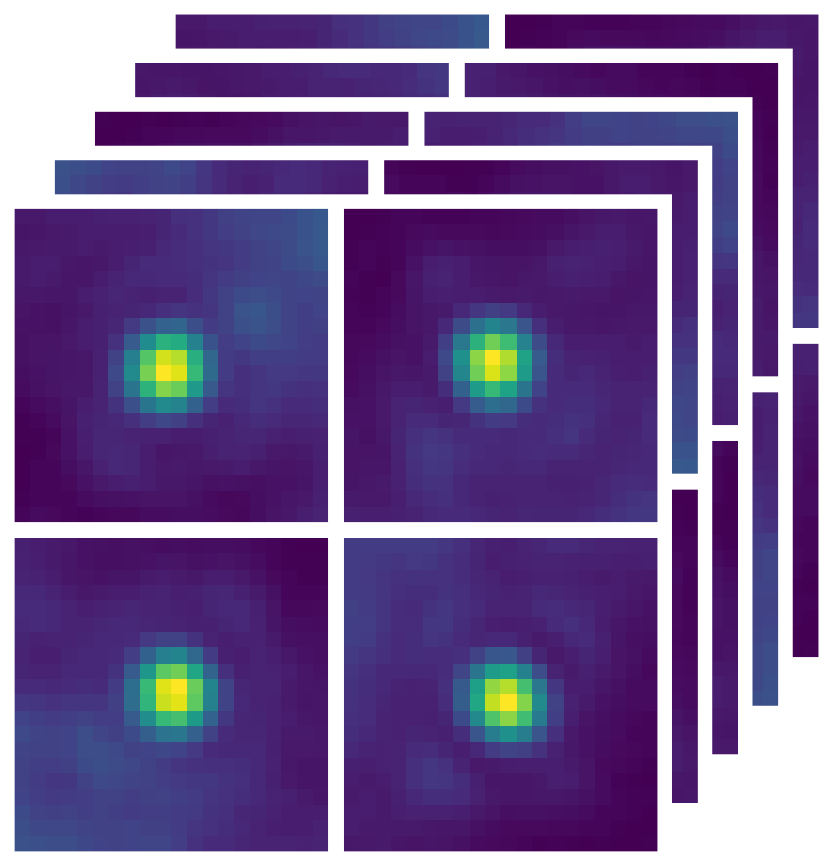}};
                \node[inner sep=0pt, label={[align=center]below: \hspace{-0.7cm}Fitted spots\\ \hspace{-0.7cm}$\big\{ \{ \bm g(\bm\xi)_{i,k} \}_{i=1:4} \big\}_{k=1:K}$}] (fittedW) at (10,0) {\includegraphics[width=.2\textwidth]{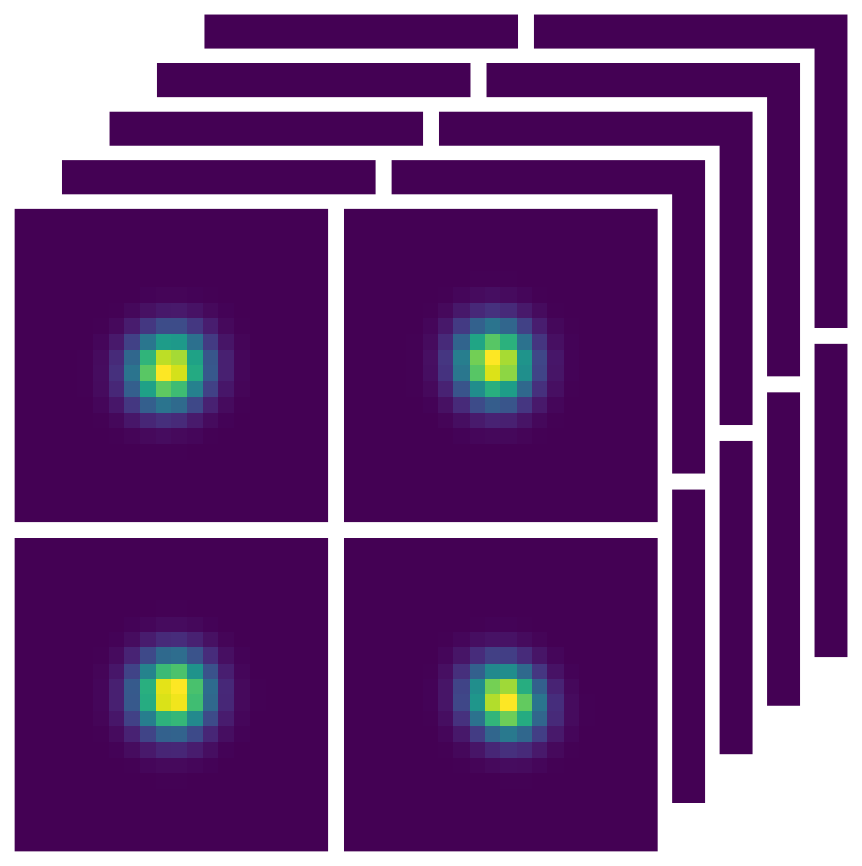}};
                \draw[->,thick] (centerC.east) -- (realW.west); 
                \draw[->,thick] (realW.east) -- (fittedW.west);
                
                \node[inner sep=0pt, align=center, rectangle, draw, minimum width=4cm, minimum height=2.5cm, rounded corners=0.25cm, line width=1.5pt, fill=black!5] (GaussParam) at (10,-4.5) { 2D Gaussian \\ parameters \\ \vspace{-0.2cm} \\ $\big\{ \{ \widehat{\bm\xi}_{i,k} \}_{i=1:4} \big\}_{k=1:K}$ \\ $\big\{ \{ \text{Cov}(\widehat{\bm\xi}_{i,k}) \}_{i=1:4} \big\}_{k=1:K}$};
                \draw[->,thick] ([yshift=0.5cm]fittedW.south east) to [out=-20,in=45] (GaussParam.north east);
                \node[inner sep=0pt, align=center, rectangle, draw, minimum width=2.85cm, minimum height=1.75cm, rounded corners=0.25cm, red, line width=1.5pt, fill=black!5] (StellarCenter) at (5,-4.5) {Rotation center \\ coordinates \\ $ \big\{(\widehat{x}_c,\widehat{y}_c)_k \big\}_{k=1:K}$};
                
                \draw[->,thick] (GaussParam.west) -- (StellarCenter.east); 
                
                
                \node[inner sep=0pt, label={[align=center]below:\hspace{-0.15cm}\textcolor{green!50!black}{Recentered data} \textcolor{green!50!black}{$\{\bm {r_{c_k}}\}_{k=1:K}$}}] (recenterC) at (0,-4.5) {\includegraphics[width=.2\textwidth]{images/centerCubes.png}};
                
                \draw[->,thick] (StellarCenter.west) -- (recenterC.east); 
        \end{tikzpicture}
    
    \caption{Schematic diagram of the proposed centering procedure using satellite spots with $K$ the total number of frames.}
    \label{fig:satelliteSpotsScheme}
    
\end{figure*}

At a given frame, $k,$ and spectral channel, $\ell$, satellite spots are considered separately. We denote by $\bm d = \big\{ d_{p} \big\}_{p=1:P}$ a 2D $P$-pixels sub-window containing a single  satellite spot extracted from the data, $\V r$. The 2D Gaussian model of parameter $\bm \xi$ is denoted by $\bm g(\bm\xi) = \big\{ g_{p}(\bm\xi) \big\}_{p=1:P}$. Finding the best estimator for the center coordinates $(x_0, y_0)$ of a satellite spot contained in data, $\bm d,$ amounts to minimizing the following nonlinear least-mean-square constrained problem:
\begin{equation}
    \label{eq:OptiProblemSatelliteSpots}
    \widehat{\bm\xi} =\min_{\bm\xi} \bigg\{ \sum_{p=1}^{P} \big( d_p - g_p(\bm\xi) \big)^2 : \, \bm\xi_l \leq \bm\xi \leq \bm\xi_u \bigg\}\,,
\end{equation}
where $\bm\xi_l$ and $\bm\xi_u$ are, respectively, the lower and upper bounds of the 2D Gaussian parameters $\bm \xi$. We do not account for bad pixels as our procedure operates at a late stage of the pipeline where they are already taken into account and corrected.
To solve Eq.~\eqref{eq:OptiProblemSatelliteSpots}, we use a Subspace Trust region Interior Reflective algorithm (STIR, \cite{branch1999subspace}), which is known for its robustness and for being particularly suitable for large sparse problems with bound constrains.
A schematic view of the proposed centering procedure using the satellite spots is given in Fig.~\ref{fig:satelliteSpotsScheme}.

\subsection{Implementation details of the fitting procedure}

The satellite spots are extracted in 30 pixels-wide squares (i.e., $P=900$) centered on their theoretical positions ($\simeq 14\,\lambda / D$ away from the theoretical rotation center). For IRDIS data, if a satellite fit does not converge\footnote{This situation occur when the satellite spots are highly dominated by residual stellar leakages.}, a median spatial filtering is applied to the extracted sub-images, $\V d,$ of the satellite spots and the fit is reprocessed on them. For IFS data, the nuisance component is stronger so the median spatial filter is always applied. A circular mask is also systematically applied to the extracted sub-images, $\V d,$ of the satellite spots to reduce the effects of stellar leakages in the corners. If some satellite spots are not fitted correctly, the rotation center is estimated from the remaining ones. Given the information on the spots orientation (''\rotatebox[origin=c]{45}{+}'' or ''+'' shape) and on the width of the spectral band (dual band or broad band) retrieved from the dataset header, some parameters are fixed in the optimization procedure (e.g., $\theta$, $\sigma_x = \sigma_y$) to speed up the convergence to the solution.

\subsubsection{Application on SPHERE/IRDIS datasets}

\begin{figure*}[t!]
    \centering
    \begin{subfigure}[t]{0.49\textwidth}
        \centering
        \includegraphics[width=\linewidth]{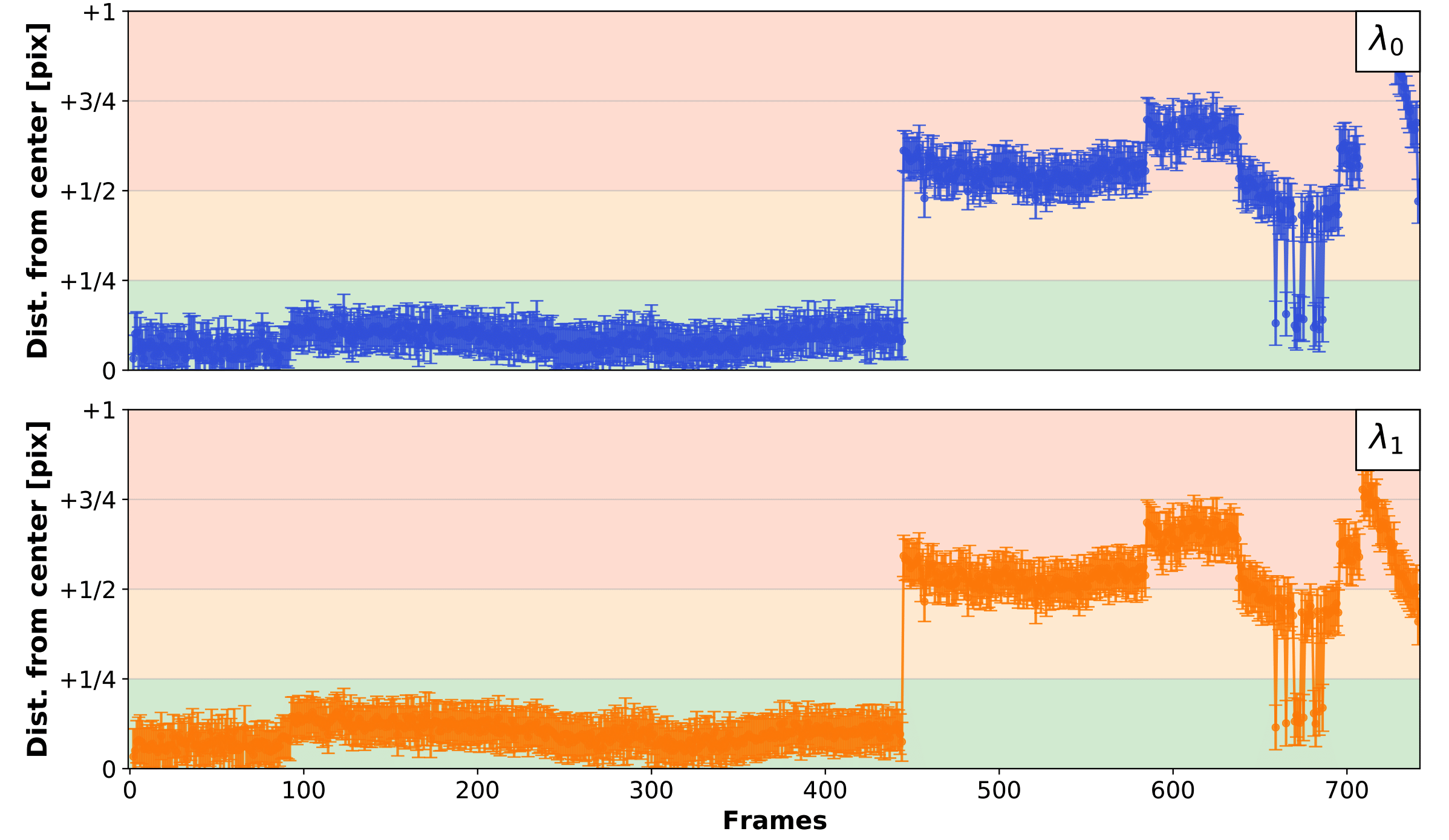}
        \caption{HR 8799 (2015-07-30) in DB J23,}
        \label{fig:distanceFromCenter_2015-07-29}
    \end{subfigure}
    \hfill
    \begin{subfigure}[t]{0.49\textwidth}
        \centering
        \includegraphics[width=\linewidth]{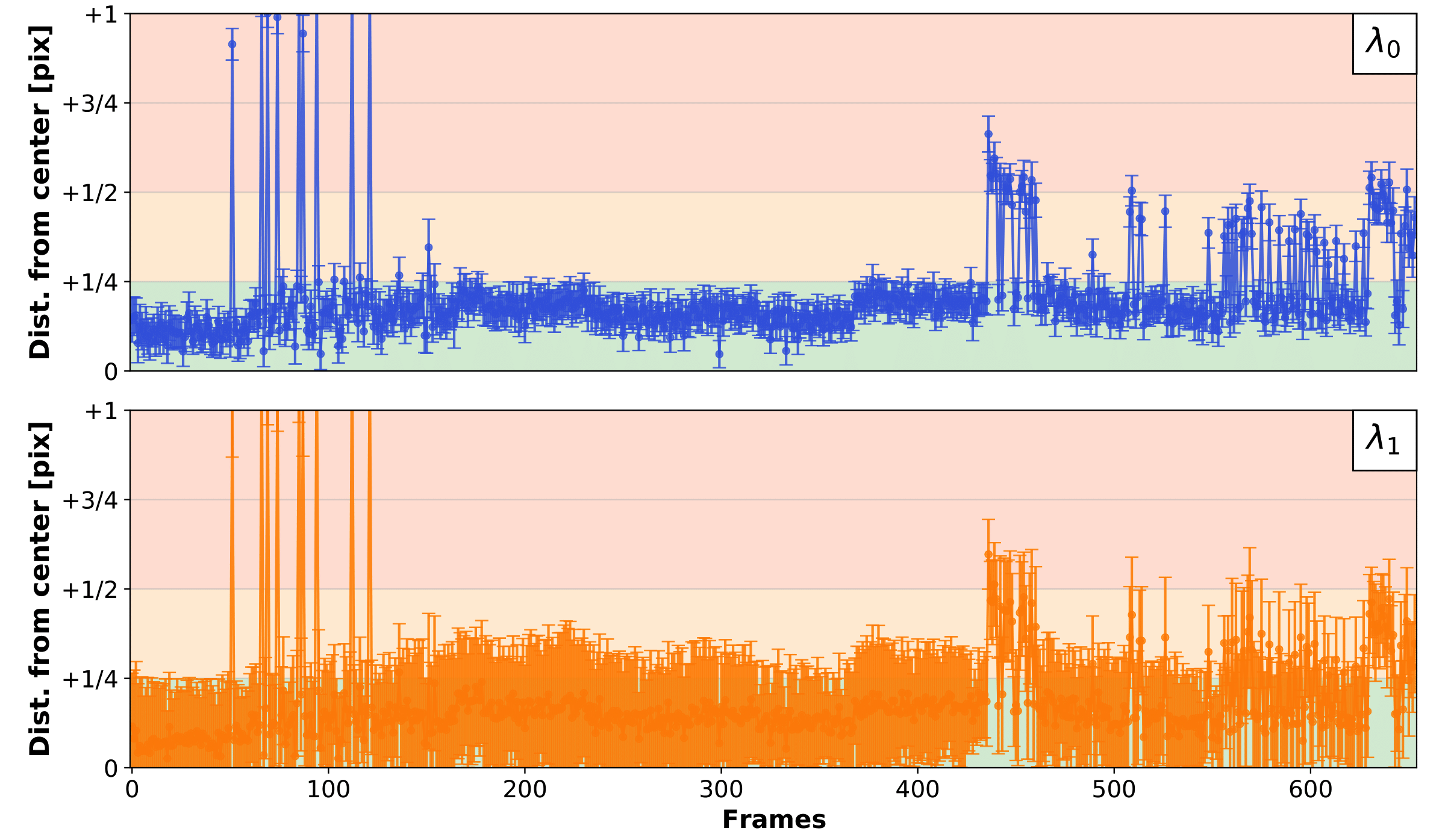}
        \caption{HR 8799 (2015-07-31) in DB K23.}
        \label{fig:distanceFromCenter_2015-07-30}
    \end{subfigure}
    \caption{Distance between the measured center of each frame and the theoretical star center for two HR 8799 observations of 2015-07-30 and 2015-07-31. See Table \ref{tab:dataset_logs} for observation logs.}
    \label{fig:distanceFromCenter}
\end{figure*}

We illustrate the importance of a proper centering with two datasets from HR 8799 acquired with the IRDIS instrument: an observation of 2015-07-30 in the J2-J3 spectral band, and an other of 2015-07-31 in K1-K2. More information regarding these observations can be found in Table \ref{tab:dataset_logs}. We applied our centering method to both datasets, and we measured for each frames the distance of their rotation center with respect to the theoretical rotation center. The obtained results are given in Figs.~\ref{fig:distanceFromCenter_2015-07-29} and \ref{fig:distanceFromCenter_2015-07-29}. In addition to these measurements, the gain in terms of S/N of detection after post-processing with the \PACO algorithm of the datasets before and after centering are given in Tables \ref{tab:gainSNRcenteringProcedure_2015-07-29} and \ref{tab:gainSNRcenteringProcedure_2015-07-30}). For the observations of 2015-07-30, about a third of the ASDI sequence is offset by half a pixel, which strongly impacts the overall S/N of the four sources. Indeed, the centering procedure leads to a gain in S/N ranging from $4.3\,\%$ to $15.7\,\%$ in the two spectral channels. The observation of 2015-07-31 displays smaller, but yet non-negligible, gains in the overall S/N with the recentered dataset ($0.8\,\%$ to $2.9\,\%$) due to the fewer number of frames impacted by a significant shift.

\begin{table*}[t!]
    \hspace{-4mm}
    \caption{Comparison of the S/N of detection of the four known exoplanets HR 8799 b, c, d and e.}
    \begin{subtable}[h]{0.45\textwidth}
        \centering
        \begin{tabular}{ccccccc}
        \toprule
         \multirow{3}{*}{\textbf{Planet}} & \multicolumn{6}{c}{$\bm{\mathcal{S}/\mathcal{N}}_{t,\ell}$} \\
         & \multicolumn{3}{c}{$\ell=1$} & \multicolumn{3}{c}{$\ell=2$} \\ 
         & Old & New & \textbf{Gain (\%)} & Old & New & \textbf{Gain (\%)} \\ \toprule
         b & 46.7 & 48.8 & \bm{$+4.3$} & 90.7 & 94.8 & \bm{$+4.3$} \\
         c & 30.9 & 35.1 & \bm{$+12.0$} & 49.7 & 54.4 & \bm{$+8.6$}\\
         d & 7.0 & 8.3 & \bm{$+15.7$} & 17.7 & 20.2 & \bm{$+12.4$}\\
         e & 13.3 & 15.7 & \bm{$+15.3$} & 31.2 & 34.4 & \bm{$+9.0$} \\
        \bottomrule
        \end{tabular}
        \caption{HR 8799 (2015-07-30)}
        \label{tab:gainSNRcenteringProcedure_2015-07-29}
    \end{subtable}
    \hspace{10mm}
    \begin{subtable}[h]{0.45\textwidth}
        \centering
        \begin{tabular}{ccccccc}
        \toprule
         \multirow{3}{*}{\textbf{Planet}} & \multicolumn{6}{c}{$\bm{\mathcal{S}/\mathcal{N}}_{t,\ell}$} \\
         & \multicolumn{3}{c}{$\ell=1$} & \multicolumn{3}{c}{$\ell=2$} \\ 
         & Old & New & \textbf{Gain (\%)} & Old & New & \textbf{Gain (\%)} \\ \toprule
         b & 112.0 & 115.4 & \bm{$+2.9$} & 89.0 & 90.6 & \bm{$+0.8$} \\
         c & 91.1 & 92.9 & \bm{$+1.9$} & 95.4 & 96.2 & \bm{$+1.8$}\\
         d & 95.4 & 96.2 & \bm{$+2.6$} & 77.7 & 78.6 & \bm{$+1.1$}\\
         e & 74.7 & 76.7 & \bm{$+2.5$} & 39.6 & 39.9 & \bm{$+0.8$} \\
        \bottomrule
        \end{tabular}
        \caption{HR 8799 (2015-07-31)}
        \label{tab:gainSNRcenteringProcedure_2015-07-30}
     \end{subtable}
     \label{tab:gainSNRcenteringProcedure}
     \tablefoot{The SNR maps were computed with \PACO before (denoted "Old") and after (denoted "New") centering the IRDIS HR 8799 datasets of 2015-07-30 and 2015-07-31, see Table \ref{tab:dataset_logs} for observation logs.}
\end{table*}

\section{Details of the individual (mono-epoch) S/N maps}
\label{sec:details_ind_snr}

In this section, we give a local visualization of the maps produced by \PACO on single-epoch datasets (i.e., without \PACOME processing) for the numerical experiments conducted in Sect. \ref{subsec:results_ss_data} with injected sources, and for the analysis of HR 8799 archival data performed in Sect. \ref{subsec:results_real_data}. 

Comparison between (i) Figs. \ref{fig:indivSNRinjections_part1}-\ref{fig:indivSNRinjections_part2}, and (ii) Fig. \ref{fig:costFuncMapInjection} as well as Table \ref{tab:optimalOrbitalElementsAndInjections} shows that injected sources (2), (3), and (4) are never detected at a single-epoch (S/N below the $5\sigma$ confidence level on each individual epoch), while they are detected at a combined multi-epoch (S/N between 9.6 and 10.5). A comparison between (i) Figs.~\ref{fig:indivSNR_HR_part1}-\ref{fig:indivSNR_HR_part2} and (ii) Fig. \ref{fig:costFuncMapHR8799bcde}, as well as Table \ref{tab:optimalOrbitalElementsHR8799}, shows that the four known exoplanets are detected at all single-epoch (mean mono-epoch S/N ranging from 23.8 to 57.0), while they are detected at a combined multi-epoch S/N lying between 231.0 and 529.9.
For the first case, the S/N gain brought by \PACOME is very close to the theoretical gain in $\sqrt{T}$ that is expected when combining optimally $T$ (independent) datasets. For the second case, the gain does not scale as $\sqrt{T}$ but, again, this is expected as the data is very heterogeneous in quality (see Sect.~\ref{subsec:results_real_data} for further information).

\begin{figure*}[t!]
    \centering
    \begin{subfigure}[t]{0.49\textwidth}
        \centering
        \includegraphics[width=\linewidth]{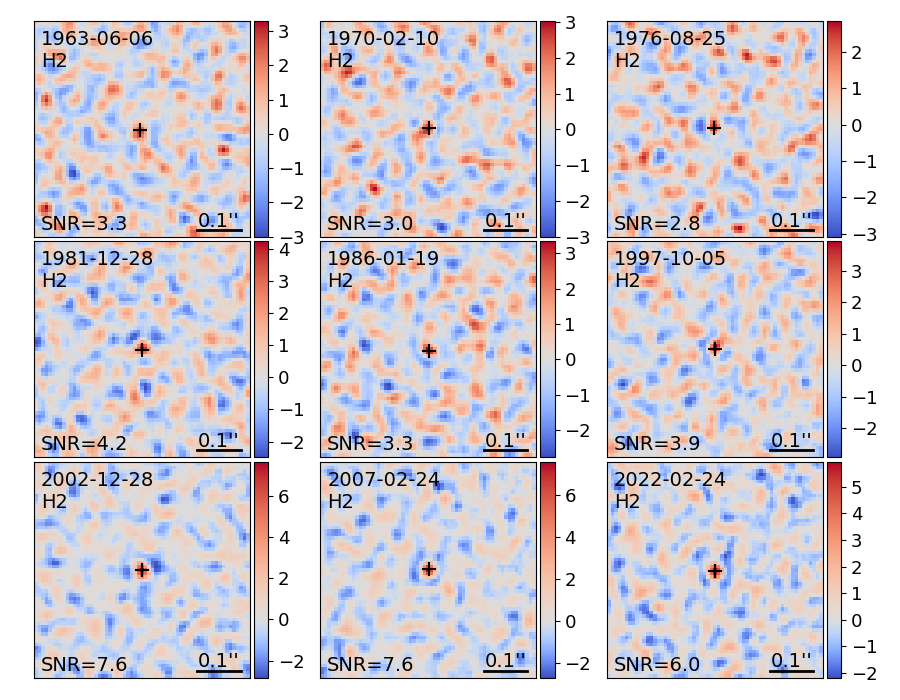}
        \caption{Source (1) / H2 band,}
        \label{fig:indivSNRinjections_b_lam1}
    \end{subfigure}
    \hfill
    \begin{subfigure}[t]{0.49\textwidth}
        \centering
        \includegraphics[width=\linewidth]{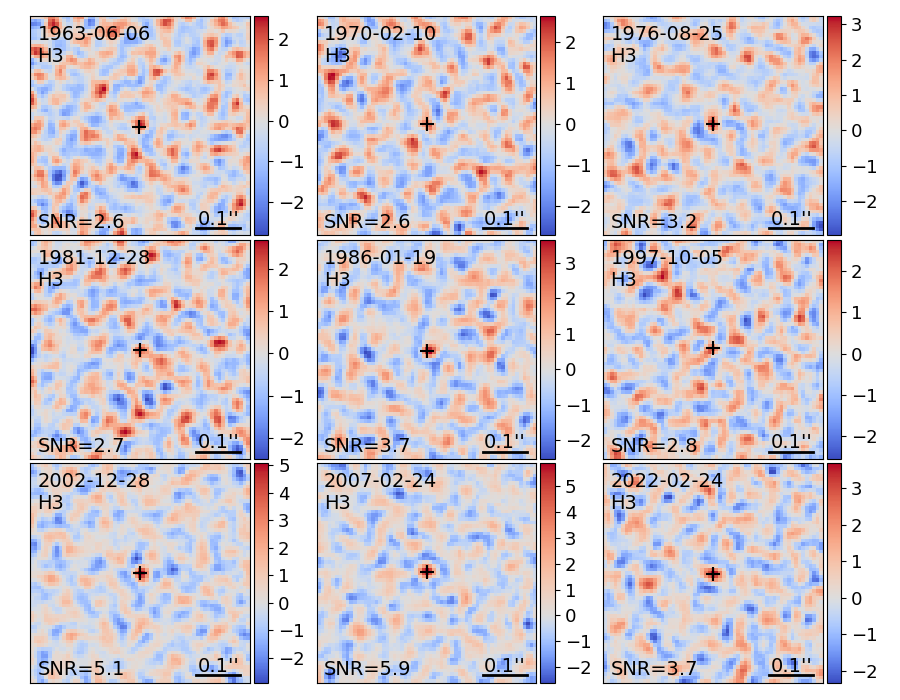}
        \caption{Source (1) / H3 band,}
        \label{fig:indivSNRinjections_b_lam2}
    \end{subfigure}
    \medskip
    \vspace{1mm}
    \begin{subfigure}[t]{0.49\textwidth}
        \centering
        \includegraphics[width=\linewidth]{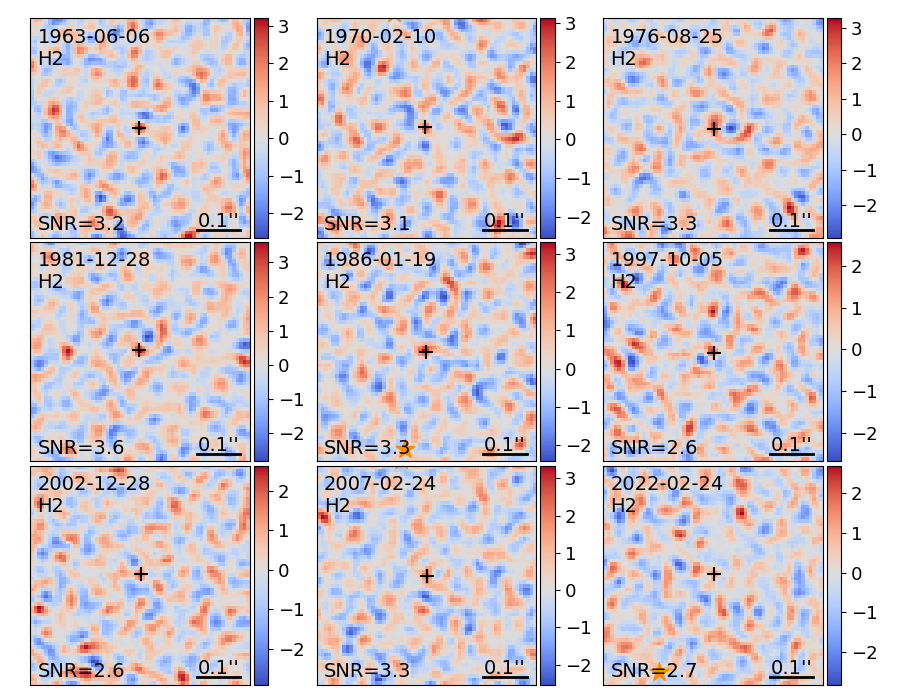}
        \caption{Source (2) / H2 band,}
        \label{fig:indivSNRinjections_c_lam1}
    \end{subfigure}
    \hfill
    \begin{subfigure}[t]{0.49\textwidth}
        \centering
        \includegraphics[width=\linewidth]{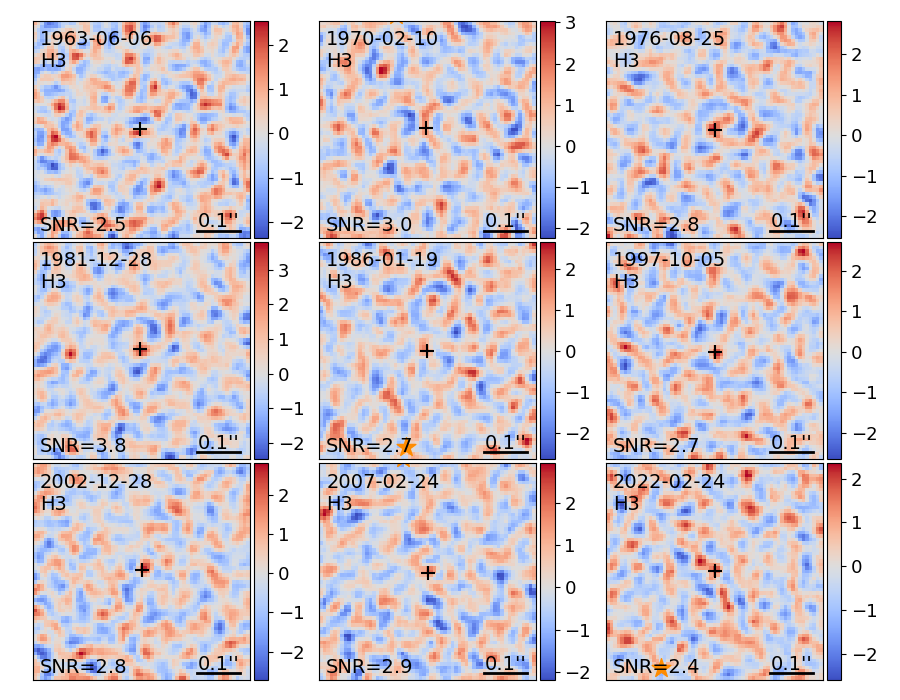}
        \caption{Source (2) / H3 band,}
        \label{fig:indivSNRinjections_c_lam2}
    \end{subfigure}
    \medskip
    \vspace{1mm}
    \begin{subfigure}[t]{0.49\textwidth}
        \centering
        \includegraphics[width=\linewidth]{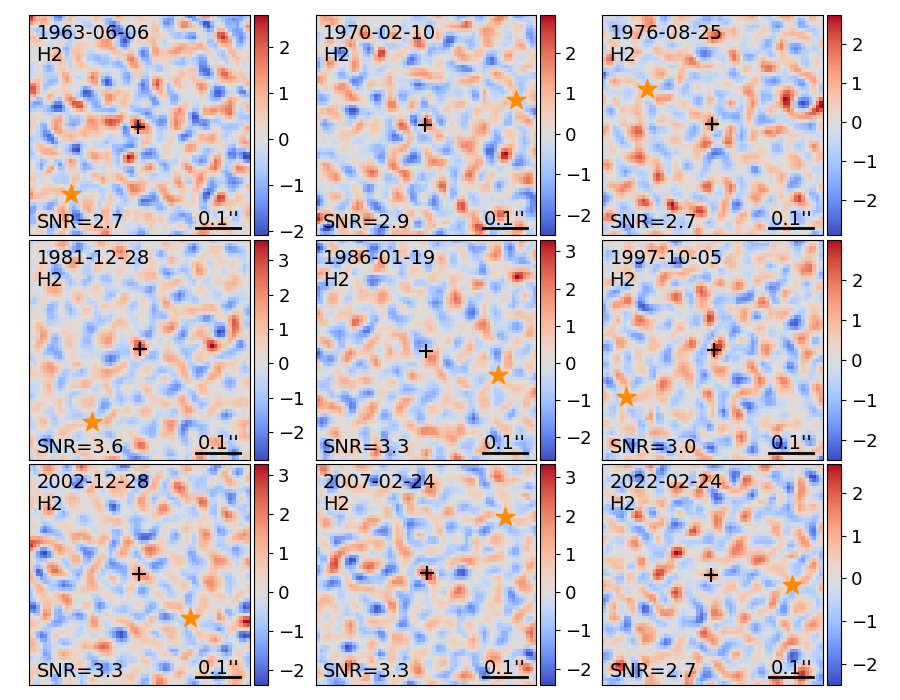}
        \caption{Source (3) / H2 band,}
        \label{fig:indivSNRinjections_d_lam1}
    \end{subfigure}
    \hfill
    \begin{subfigure}[t]{0.49\textwidth}
        \centering
        \includegraphics[width=\linewidth]{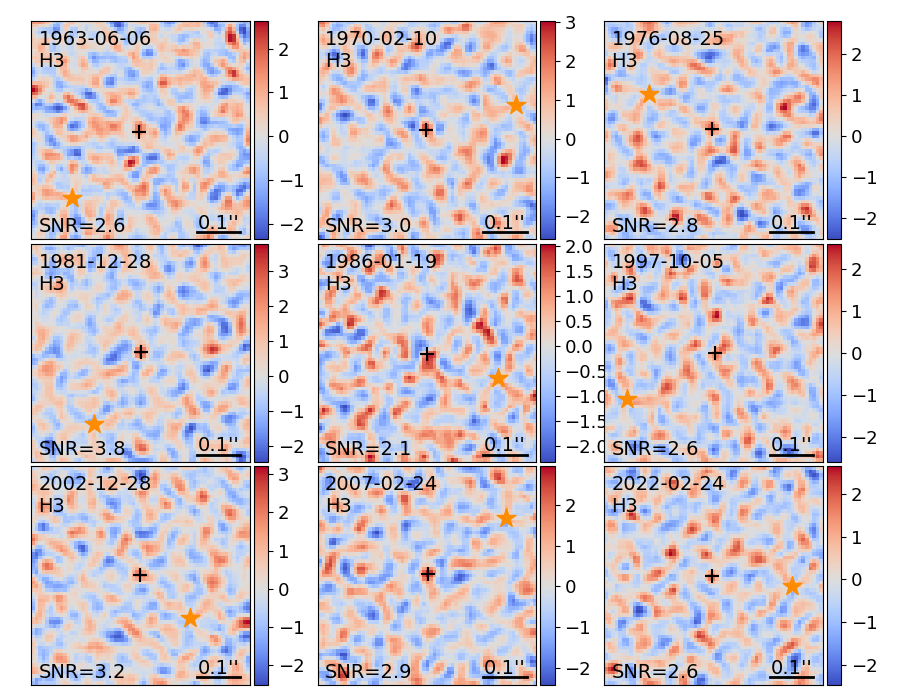}
        \caption{Source (3) / H3 band.}
        \label{fig:indivSNRinjections_d_lam2}
    \end{subfigure}
    \caption{Individual (mono-epoch) $\SNR_{t,\ell}$ maps produced by \PACO around the optimal solutions found by \PACOME for the four injected sources (1), (2), (3), and (4) considered in the semi-synthetic benchmark of Sect. \ref{subsec:results_ss_data}. The black cross indicates the source location found by \PACOME. The dynamic of the color bars is adapted to the minimum and maximum values of the displayed ROIs.}
    \label{fig:indivSNRinjections_part1}
\end{figure*}

\begin{figure*}[t!]
    \begin{subfigure}[t]{0.49\textwidth}
        \centering
        \includegraphics[width=\linewidth]{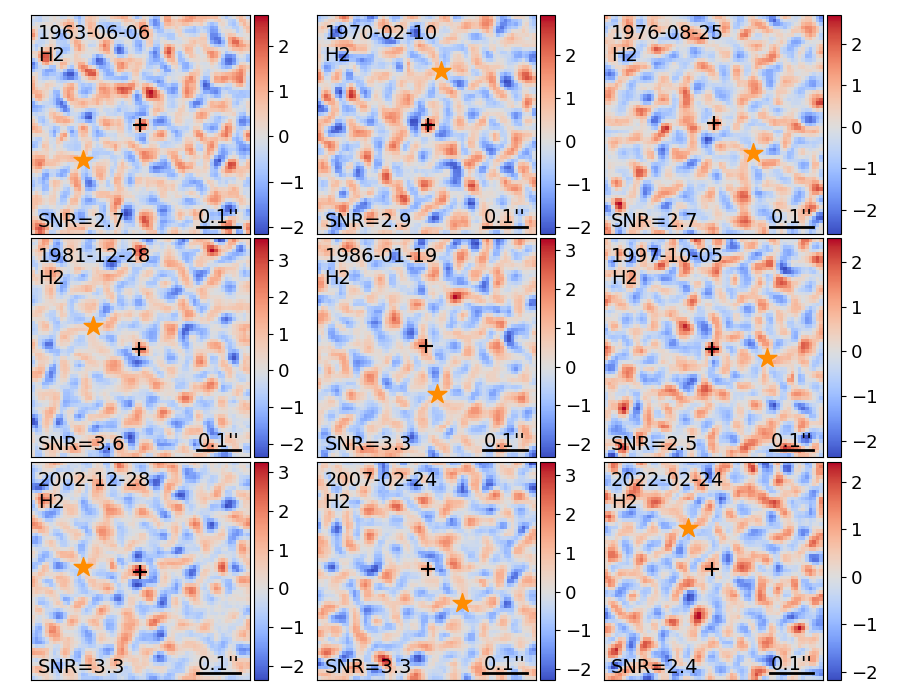}
        \caption{Source (4) / H2 band}
        \label{fig:indivSNRinjections_e_lam1}
    \end{subfigure}
    \hfill
    \begin{subfigure}[t]{0.49\textwidth}
        \centering
        \includegraphics[width=\linewidth]{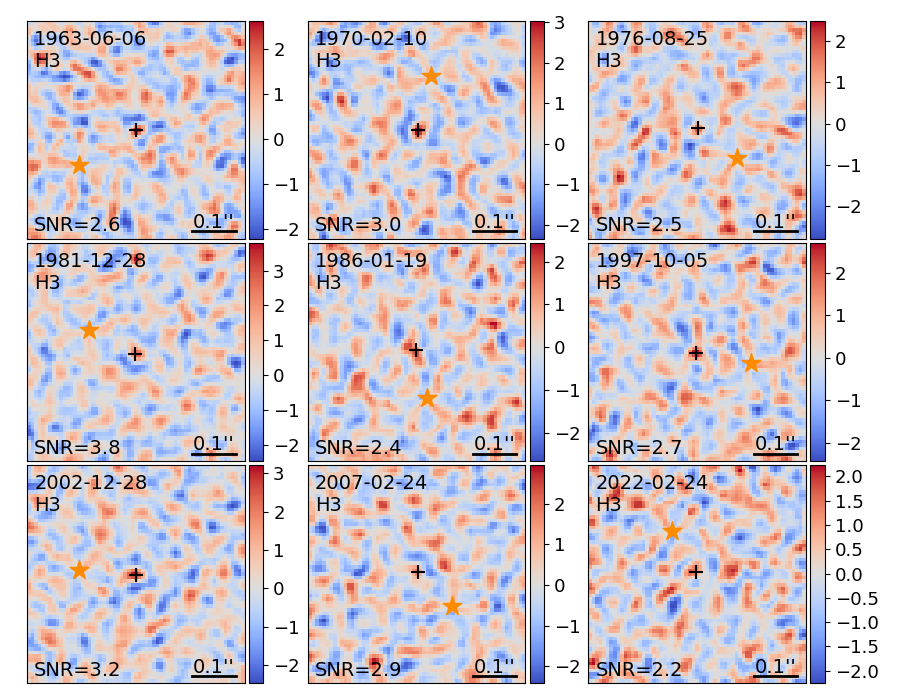}
        \caption{Source (4) / H3 band}
        \label{fig:indivSNRinjections_e_lam2_part2}
    \end{subfigure}
    \caption{Continuation of Fig. \ref{fig:indivSNRinjections_part1} for injected source (4).}
    \label{fig:indivSNRinjections_part2}
\end{figure*}


\begin{figure*}[t!]
    \centering
    \begin{subfigure}[t]{0.49\textwidth}
        \centering
        \includegraphics[width=\linewidth]{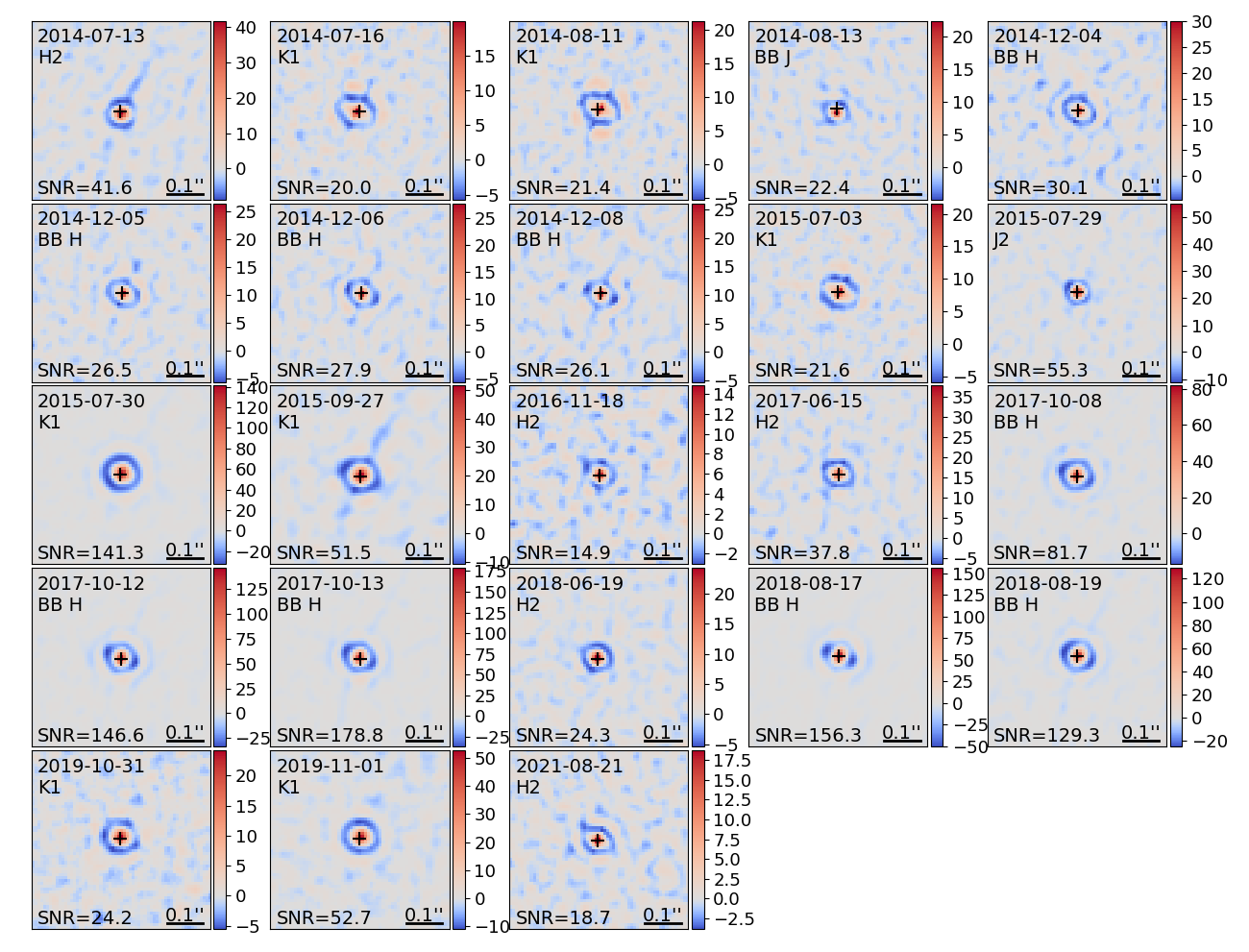}
        \caption{HR 8799b / $\ell=1,$}
        \label{fig:indivSNR_HR_b_lam1}
    \end{subfigure}
    \hfill
    \begin{subfigure}[t]{0.49\textwidth}
        \centering
        \includegraphics[width=\linewidth]{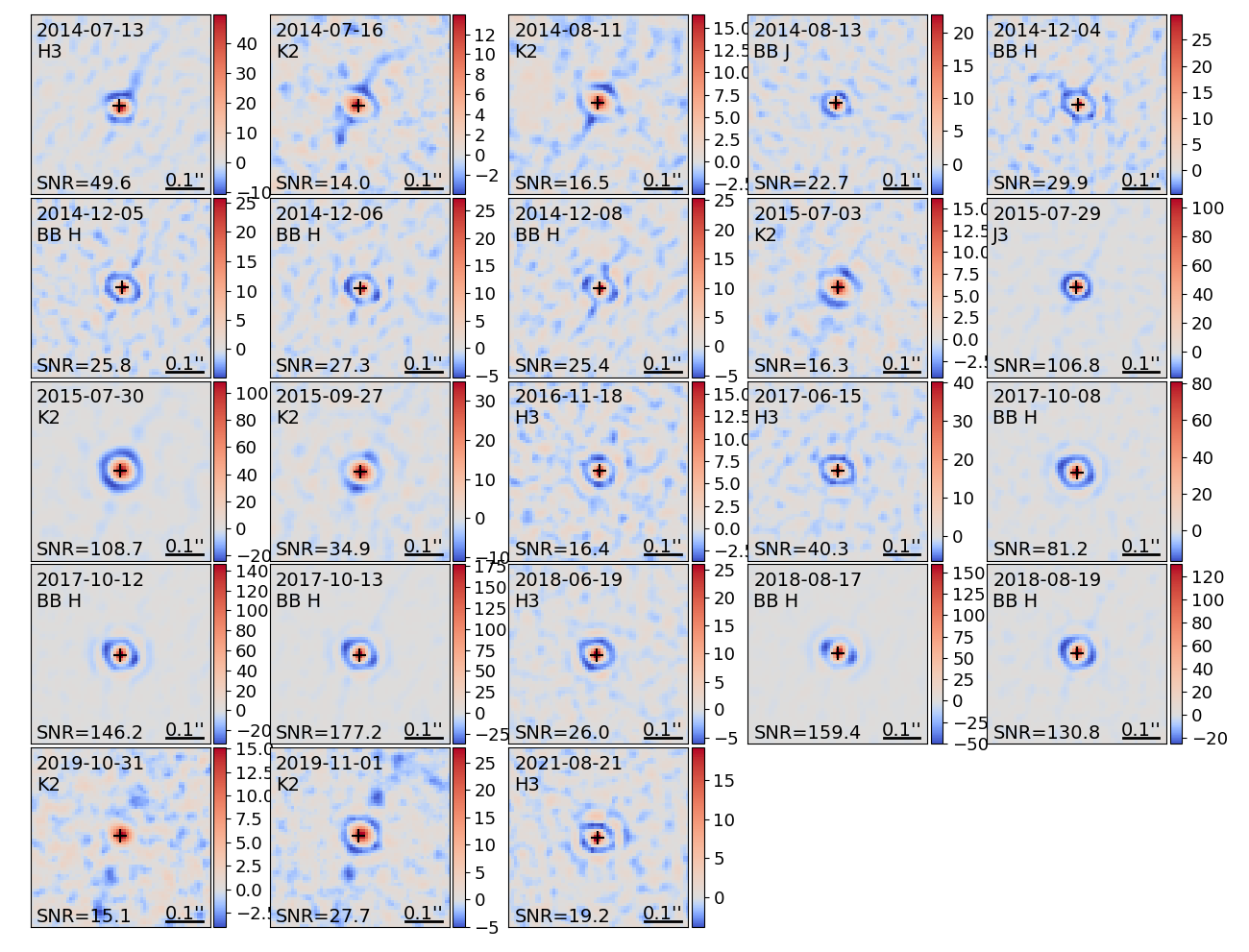}
        \caption{HR 8799b / $\ell=2,$}
        \label{fig:indivSNR_HR_b_lam2}
    \end{subfigure}
    \medskip
    \vspace{1mm}
    \begin{subfigure}[t]{0.49\textwidth}
        \centering
        \includegraphics[width=\linewidth]{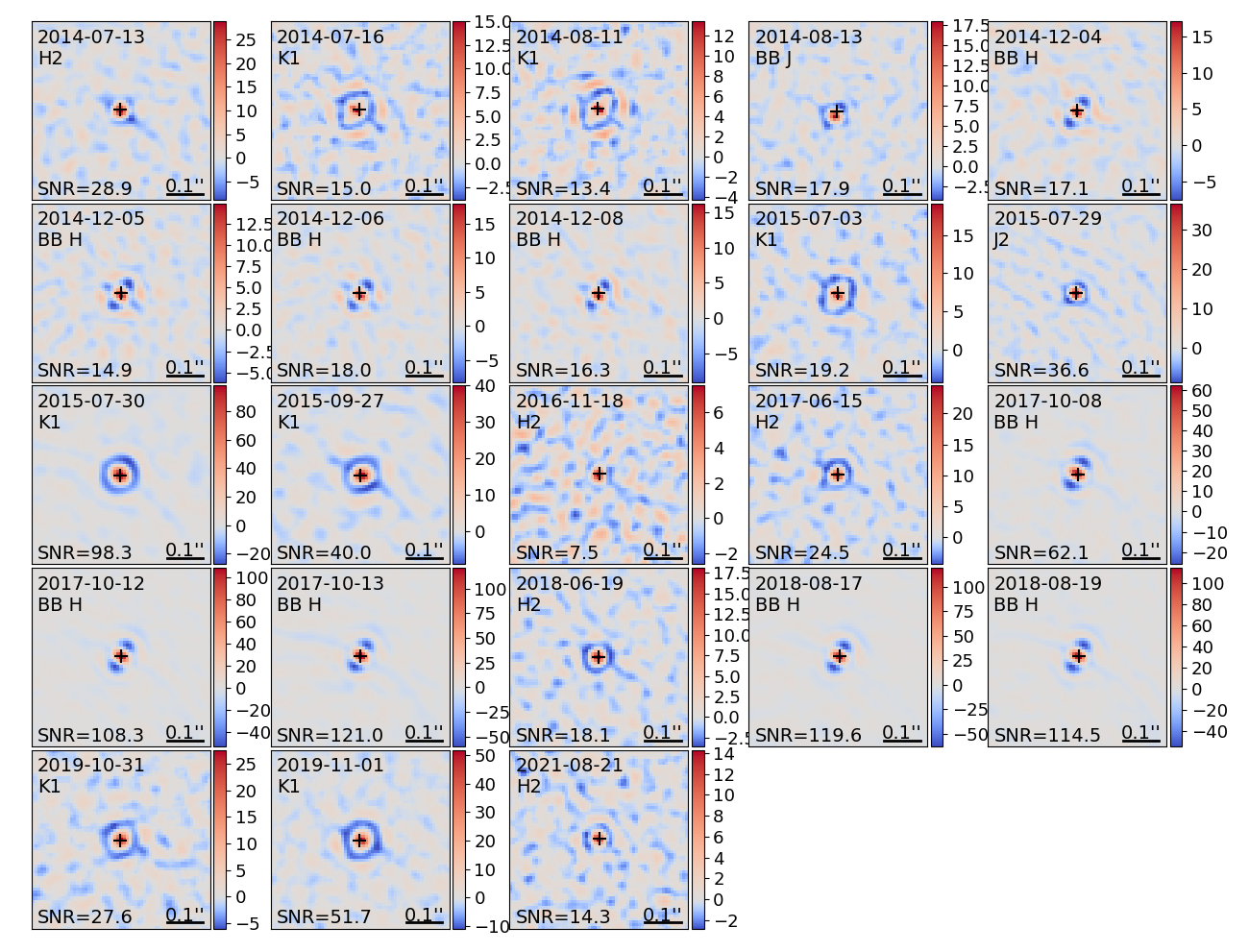}
        \caption{HR 8799c / $\ell=1,$}
        \label{fig:indivSNR_HR_c_lam1}
    \end{subfigure}
    \hfill
    \begin{subfigure}[t]{0.49\textwidth}
        \centering
        \includegraphics[width=\linewidth]{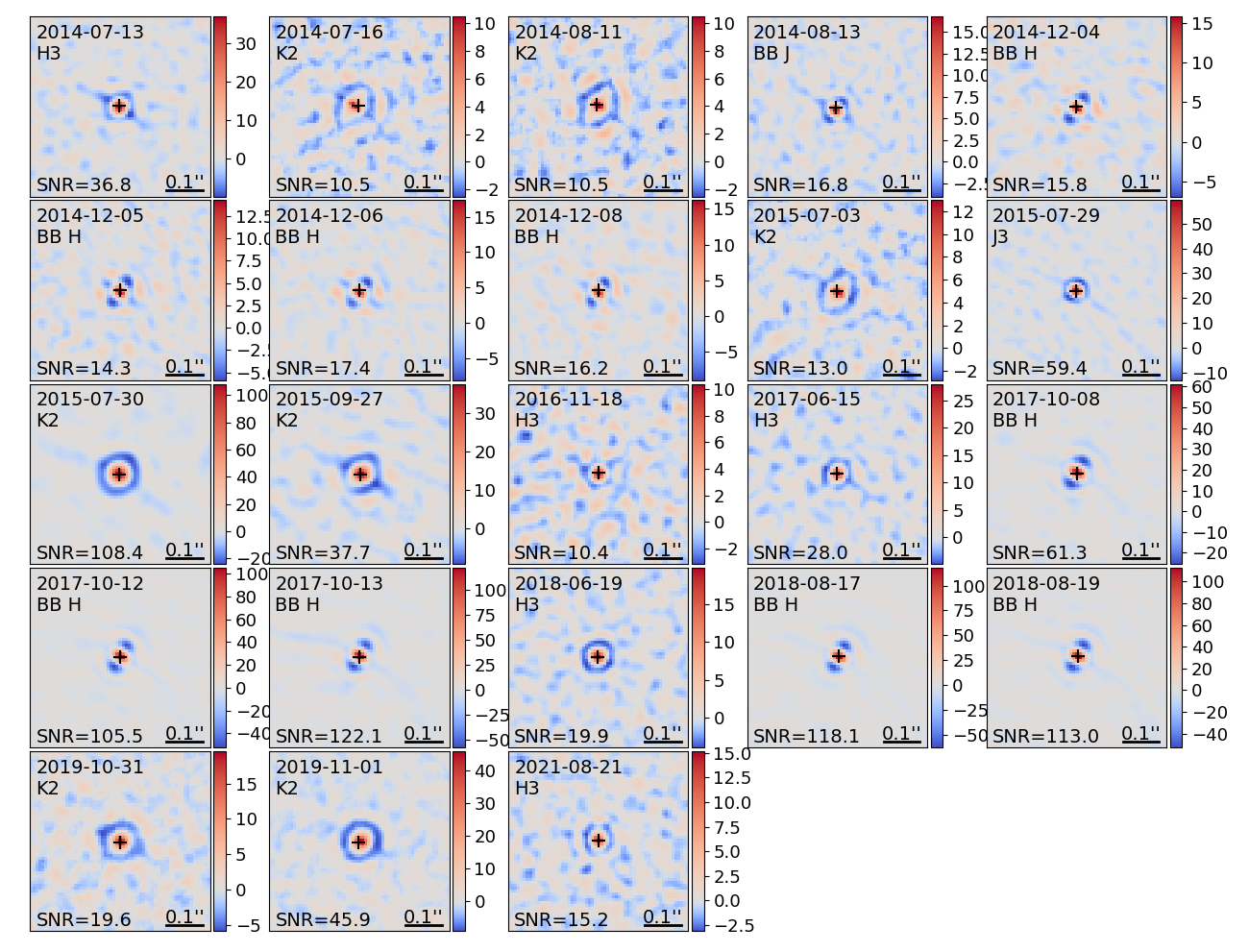}
        \caption{HR 8799c / $\ell=2.$}
        \label{fig:indivSNR_HR_c_lam2}
    \end{subfigure}
    \caption{Mono-epoch $\SNR_{t,\ell}$ maps produced by \PACO around the \PACOMEs optimal solutions for the four planets of HR 8799.}
    \label{fig:indivSNR_HR_part1}
\end{figure*}

\begin{figure*}[t!]
    \begin{subfigure}[t]{0.49\textwidth}
        \centering
        \includegraphics[width=\linewidth]{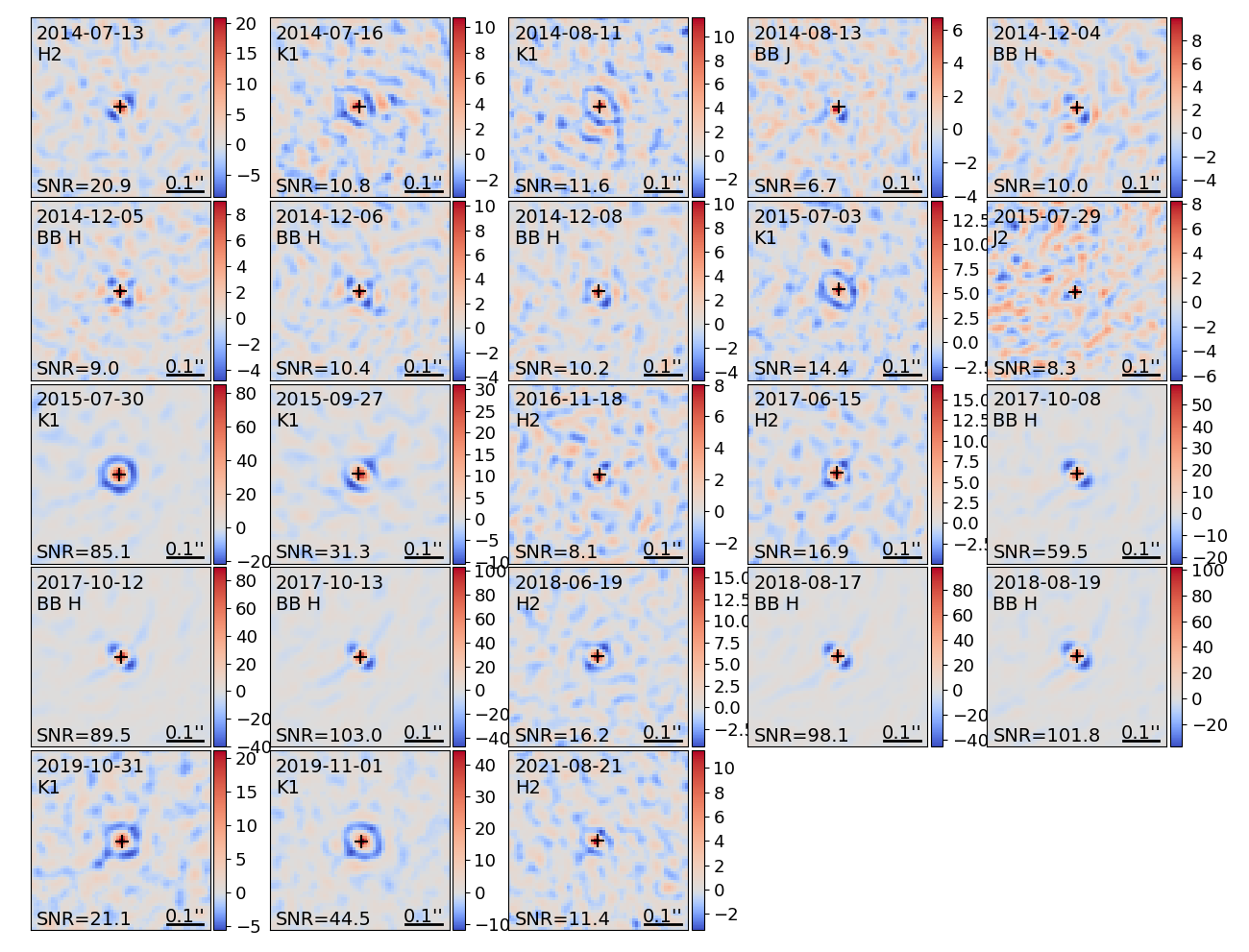}
        \caption{HR 8799d / $\ell=1,$}
        \label{indivSNR_HR_d_lam1}
    \end{subfigure}
    \hfill
    \begin{subfigure}[t]{0.49\textwidth}
        \centering
        \includegraphics[width=\linewidth]{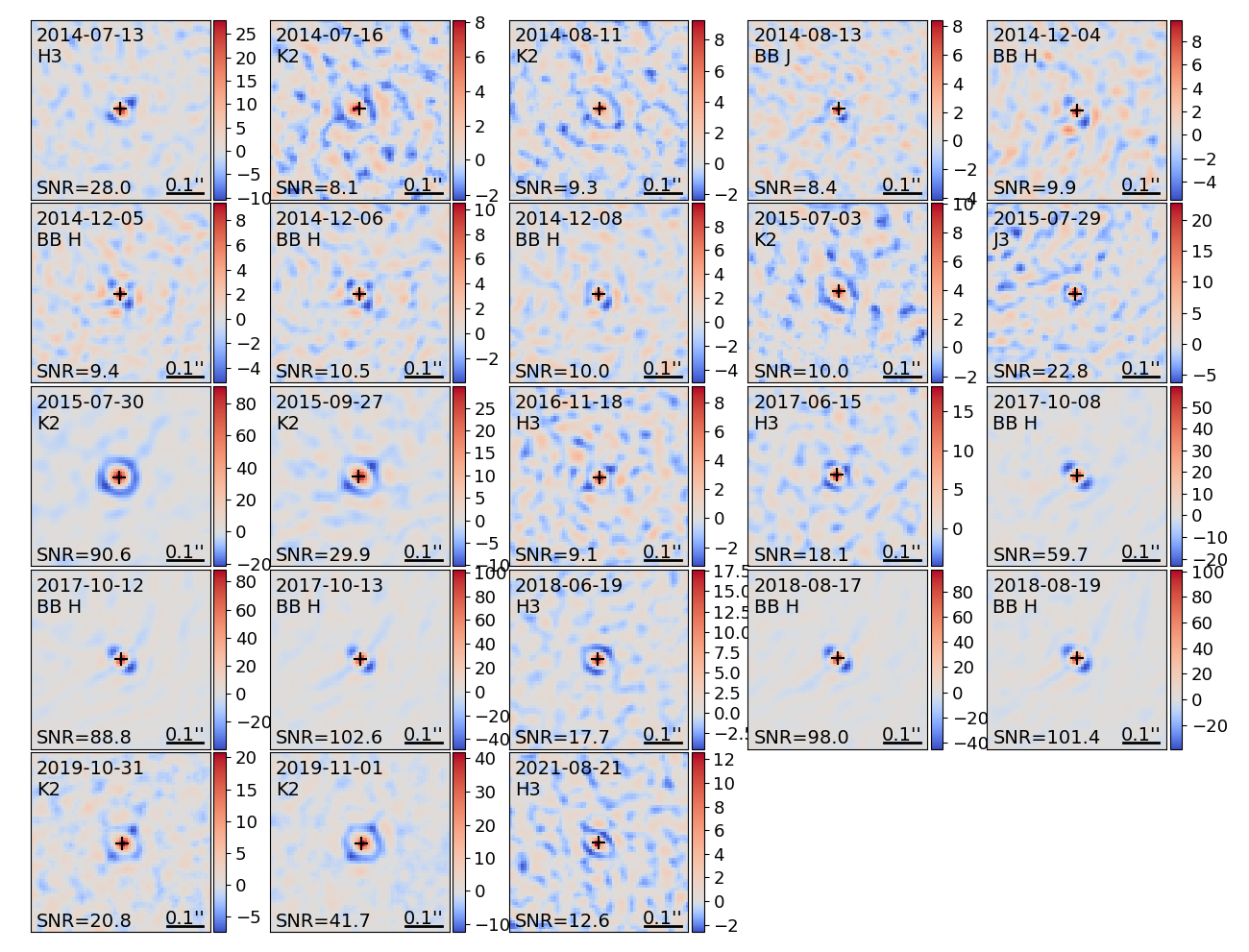}
        \caption{HR 8799d / $\ell=2,$}
        \label{indivSNR_HR_d_lam2}
    \end{subfigure}
    \vspace{1mm}
    \begin{subfigure}[t]{0.49\textwidth}
        \centering
        \includegraphics[width=\linewidth]{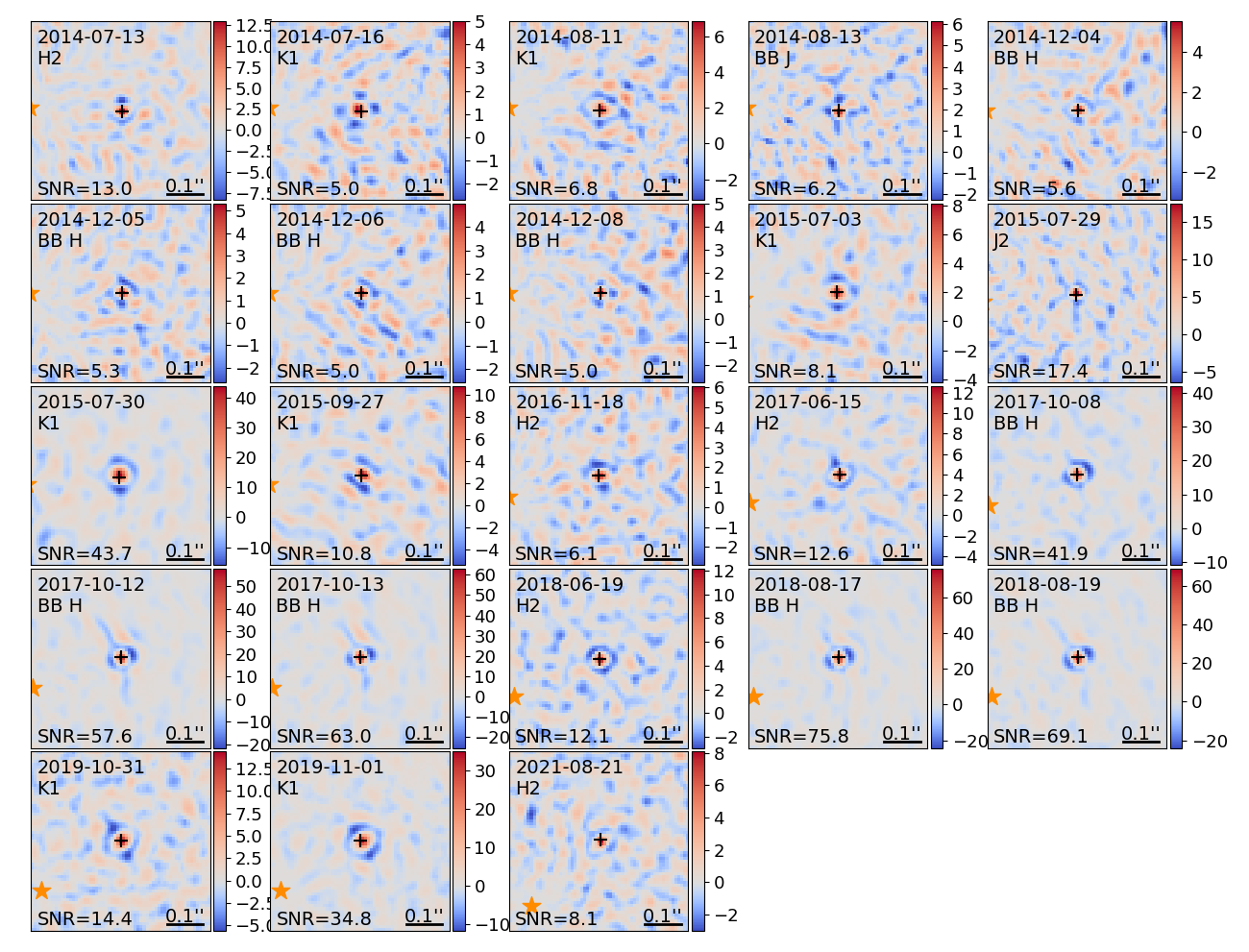}
        \caption{HR 8799e / $\ell=1,$}
        \label{indivSNR_HR_e_lam1}
    \end{subfigure}
    \hfill
    \begin{subfigure}[t]{0.49\textwidth}
        \centering
        \includegraphics[width=\linewidth]{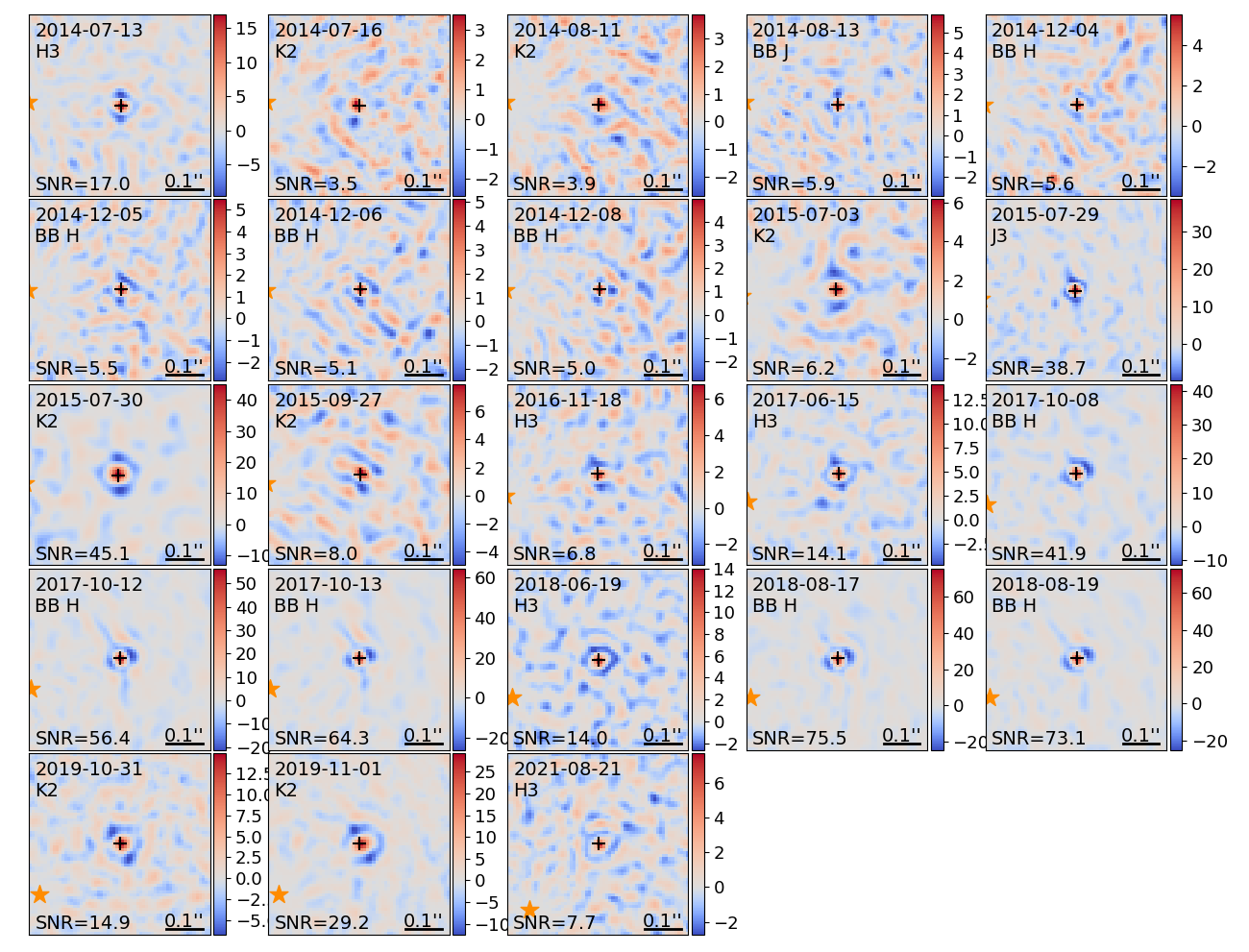}
        \caption{HR 8799e / $\ell=2.$}
        \label{fig:indivSNR_HR_d_lam2}
    \end{subfigure}
    \caption{Continuation of Fig. \ref{fig:indivSNR_HR_part1} for HR 8799.}
    \label{fig:indivSNR_HR_part2}
\end{figure*}
\end{appendix}

\end{document}